\documentclass[
aip,
amsmath,amssymb,preprint,floatfix
]{revtex4-1}

\usepackage{physics}
\usepackage{verbatim}
\usepackage{natbib}
\usepackage{placeins}
\usepackage{adjustbox}
\usepackage{tikz}
\usepackage{soul}
\usepackage{subcaption}
\usepackage{cleveref}
\usepackage{graphicx}
\usepackage{float}
\usepackage{xcolor}

\Crefname{equation}{Eq.}{Eqs.}
\Crefname{figure}{Fig.}{Figs.}
\DeclareMathAlphabet{\mathpzc}{OT1}{pzc}{m}{it}
\DeclareMathAlphabet\mathbfcal{OMS}{cmsy}{b}{n}
\begin{document}
\title{The electron cyclotron drift instability: a comparison of particle-in-cell and continuum Vlasov simulations}
\author{Arash Tavassoli}
	\email{art562@usask.ca}
	\author{Mina Papahn Zadeh}
	\author{Andrei Smolyakov}
	\author{Magdi Shoucri}
	\affiliation{Department of Physics and Engineering Physics, University of Saskatchewan,  Saskatoon,  Saskatchewan, Canada S7N 5E2}
	\author{Raymond J. Spiteri}
	\affiliation{Department of Computer Science, University of Saskatchewan,  Saskatoon, Saskatchewan, Canada S7N 5C9}
	\date{\today}

\begin{abstract}
    The linear and nonlinear characteristics of the electron cyclotron drift instability (ECDI) have been studied through the particle-in-cell (PIC) and continuum Vlasov simulation methods in connection with the effects of the azimuthal length (in the $E \cross B$ direction) on the simulations. Simulation results for a long azimuthal length (17.82 cm $= 627\;v_d/\omega_{ce}$, where $\omega_{ce}$ is the electron cyclotron frequency and $v_d$ is the $E\times B$  drift of the electrons) are reported, for which a high resolution is achieved in Fourier space. For simulations with a long azimuthal length,  the linear growth rates of the PIC simulations show a considerable discrepancy with the theory, whereas the linear growth rate of the Vlasov simulations remains close to the theory.  In the nonlinear regime, the inverse cascade is shown in both PIC and Vlasov simulations with a sufficiently large azimuthal length. In simulations with a short azimuthal length, however, the inverse cascade is barely observed. Instead, the PIC simulations with a short azimuthal length (0.5625 cm $=19.8\;v_d/\omega_{ce}$) show an essentially continuous nonlinear dispersion, similar to what is predicted by the ion-sound turbulence theory. It is shown that, in the PIC and Vlasov simulations, the inverse cascade coincides with the formation and merging of  electron structures  in phase space. This process, however, terminates differently in the PIC simulations compared with the Vlasov simulations. Larger amplitudes of ECDI fluctuations are  observed in the PIC simulations compared with the Vlasov simulations, leading to an intensified electron heating and anomalous current. This suggests that the statistical noise of PIC simulations might contribute to the extreme electron heating that has been observed in previous studies.
\end{abstract}

\maketitle

\section{Introduction}
In recent years, theoretical and experimental studies of the electron cyclotron drift instability (ECDI) have received considerable  attention \cite{boeuf2018b,charoy2021interaction,sengupta2020mode,asadi2019numerical,hara2020cross,mandal2020cross,janhunen2018nonlinear,janhunen2018evolution,tavassoli2022nonlinear}. The ECDI is a leading candidate to explain  fluctuations that increase the electron conductivity across the magnetic field, well above the predictions of the collisional transport theory, in particular, for   Hall thrusters, where the electron anomalous current strongly affects  the discharge and thrust characteristics \cite{boeuf2017tutorial}. A similar instability occurs in space plasma conditions \cite{muschietti2006electron,muschietti2013microturbulence}. Therefore,  understanding the mechanism of the instability and the related anomalous current due to ECDI is important  for the operation of Hall thrusters and space physics.   

The linear regime of the ECDI is well studied from the dispersion equation \cite{gary1993theory}. The linear spectrum of unstable ECDI modes for purely perpendicular propagation is characterized by discrete and sharp bands of the growth rate that result from the interaction of the ion mode with the Doppler-shifted electron cyclotron resonances \cite{lampe1972theory}. When the wave vector along the magnetic field  is finite, the linear spectrum of the ECDI becomes smoother and approaches the spectrum  of unmagnetized ion-sound instability \cite{cavalier2013hall,janhunen2018evolution}. It has been argued \cite{lafleur2016theory1,lafleur2018anomalous} that even for propagation purely perpendicular to the magnetic field,  the ECDI can be described as the ion-sound instability.   It has been suggested \cite{lampe1971nonlinear,lampe1972theory} that the  nonlinear transition to the ion-sound regime for  perpendicular propagation may occur similarly to the  transition from discrete Bernstein modes with a finite magnetic field to the Landau damped modes in the limit of the vanishing magnetic field, $B \rightarrow 0$, Bernstein--Landau paradox \cite{sukhorukov1997bernstein}. Such a transition  has been debated, however, in  other studies, where the effects of the magnetic field have been emphasized \cite{forslund1972anomalous,janhunen2018nonlinear,janhunen2018evolution,tavassoli2022nonlinear}.

Numerical simulations have been making  major contributions to the  understanding of the ECDI in the nonlinear regimes. The nonlinear  anomalous current as a result of the ECDI has been demonstrated in many numerical simulations \cite{adam2004study,janhunen2018nonlinear,janhunen2018evolution,tavassoli2022nonlinear,lafleur2016theory1,lafleur2016theory2,lafleur2018anomalous,charoy20192d,villafana20212d,ducrocq2006high,BoeufPoP2018}. In the simulations, the anomalous current is often accompanied by fast electron heating \cite{janhunen2018nonlinear,lafleur2016theory1}. The origin of this heating is not well understood, but Refs. \onlinecite{adam2004study,heron2013anomalous,stasiewicz2020stochastic} suggest it is the result of stochastization of electron trajectories   \cite{karney1977stochastic,karney1978stochastic,karney1979stochastic}. Nevertheless, the heating rates in the PIC simulations of Refs.~\onlinecite{janhunen2018nonlinear,janhunen2018evolution,lafleur2016theory1} exceed  experimentally observed values\cite{stasiewicz2020stochastic}.  Another characteristic of the ECDI demonstrated by simulations is the flow of energy from the short-wavelength modes to the long-wavelength modes in the nonlinear regime (the inverse cascade) \cite{janhunen2018nonlinear,tavassoli2022nonlinear}, which is believed to have an important role in the generation of the anomalous current.

Despite many insights provided by numerical simulations, there are still many challenges to be addressed. Most simulations of the ECDI are done using the particle-in-cell (PIC) method, which is known to suffer from statistical noise. This noise is a result of the limited number of markers (macroparticles) that are used to sample the phase space, and it only decreases as $1/\sqrt{N_{ppc}}$, where $N_{ppc}$ is the number of macroparticles per cell. It is well known that the noise of PIC simulations can undermine the accuracy of physical results such as the linear growth rates \cite{tavassoli2021role}, heating and transport rates \cite{nevins2005discrete,holod2007statistical}, and the threshold of instabilities \cite{palodhi2019counterstreaming}. The noise can also lead to a ``numerical relaxation" towards Maxwellian distributions in PIC simulations \cite{birdsall2004plasma,turner2006kinetic}. This effect can potentially introduce  numerical transport in the simulations, thus distorting the physics. The effect of the $N_{ppc}$ on the ion density, electric field, and anomalous current has been investigated
in the 2D ECDI simulations of Refs.~\onlinecite{charoy20192d,villafana20212d}. In both studies, it is concluded that numerical convergence happens at $N_{ppc} \approx 100-250$. In the 1D PIC simulations of Ref.~\onlinecite{lafleur2016theory1}, however, it is found that neither increasing $N_{ppc}$ nor using the "quiet start" initialization have a significant effect on the anomalous mobility in ECDI. 

Another challenge for the simulation of ECDI is related to the effects of the simulation box length on the physics of the linear and nonlinear regimes. The limited length in the direction of the $E\times B$ drift (the azimuthal length) results in the discrete spectrum of the modes that can be excited; i.e., only the modes with an azimuthal wave vector $k_x=2\pi m/L$ (where $L$ is the azimuthal length and $m=1,2,\ldots$) are permitted in the simulation. In this work, we refer to these modes as ``box resonant modes" or simply ``resonant modes".

Because of the computational costs, many studies use a relatively short azimuthal length of about 0.5 cm to 2 cm \cite{villafana20212d,charoy20192d,croes2018effect,lafleur2018anomalous,croes20172d,janhunen2018evolution}. By increasing  $L$, the number of box resonant modes, and therefore the resolution of the simulation in Fourier space, increases. In the linear regime of ECDI simulations, resolving the steep variation of the  growth rates in Fourier space demands a particularly high resolution in this space (see e.g.,~\Cref{fig:resonant_modes}). Also in the nonlinear regime, the generation of the long-wavelength waves due to inverse cascade can be affected by a limited azimuthal length.
Previous studies of the ECDI have mainly relied on convergence tests and benchmarking against other PIC codes for validating the results of the PIC simulations. Despite their merits, such an approach would not reveal any systematic problem that might be present in the PIC approach itself. Therefore, benchmarking against alternative kinetic simulation methods seems prudent. The continuum Vlasov is an alternative simulation method to the PIC method. The continuum Vlasov method is known to be free of statistical noise, and for this reason, benchmarking PIC results against Vlasov results is a potentially useful experiment for investigating the effect of the noise in PIC simulations. 

In Ref.~\onlinecite{tavassoli2022nonlinear}, we presented one of the first Vlasov simulations of the ECDI. The results of these simulations demonstrated some of the characteristics of the ECDI such as nonlinear transitions in the fluctuation profiles, inverse cascade, nonlinear heating, and anomalous transport. It was also shown that the transition to ion-sound theory is not likely to exist in the nonlinear regime of the simulation. Some of the similarities and discrepancies of the PIC simulations with the Vlasov simulations were  also discussed.

In this work, we present a detailed comparison between the PIC and Vlasov simulations of the ECDI, with special emphasis on the effect of azimuthal length. In the linear regime, we have measured the linear growth rates from simulations and compared them with the theoretical growth rates from the dispersion relation. It is shown that, for an intermediate length ($L=156.8/k_0$, where $k_0\equiv v_d/\omega_{ce}$,  $\omega_{ce}$ is the electron cyclotron frequency and $v_d$ is the $E\times B$  drift of the electrons) and as long as the $N_{ppc}$ is sufficiently large, the linear growth rates in both the PIC and Vlasov simulations are fairly consistent with theory. This conclusion, however, does not hold for the simulations with a long azimuthal length ($L=627/k_0$). For this length, many new resonant modes are resolved, some of which (especially the low-growth-rate modes) show a much better consistency with theory in the Vlasov simulation than the PIC simulation even with $N_{ppc}$ as large  as $N_{ppc}=10^4$. In the nonlinear regime, the inverse cascade is shown in both PIC and Vlasov simulations for a sufficiently large $L$. For small $L$ ($L=19.8/k_0$), the inverse cascade is not clearly observed, and the spectrum of PIC simulation becomes mostly similar to  the ion-sound turbulence theory of Ref.~\onlinecite{lampe1972theory}. This similarity is not seen in the Vlasov simulation, likely because of the lack of the additional effect of the statistical noise. We also show that the inverse cascade coincides with the formation of some structures in the particular positions of the electron phase space. The subsequent merging of these structures is likely to explain the inverse cascade. Although this merging is observed in both PIC and Vlasov simulations, it terminates differently in them. The intensity of the nonlinear fluctuations in the PIC and Vlasov simulations is also compared. It is shown that the electrostatic energy, the electron temperature, and the anomalous current are generally larger in the PIC simulations than in the Vlasov simulations. The effect of the $L$ on these quantities is also studied. It is shown that all these quantities increase with a significant increase in $L$. Moreover, a small variation of $L$ can lead to a significant change in these quantities in the nonlinear regime, especially in the PIC simulations.

The remainder of the paper is organized as follows. In \cref{sec:setup}, the physical and numerical setups of the ECDI problem are illustrated. In \cref{sec:linear_regime}, the linear regime of the ECDI is discussed, and the growth rates of the PIC and Vlasov simulations are compared. In \cref{sec:eigenspectrums}, the eigenspectra of the simulations are compared, and the inverse cascade is discussed. In \cref{sec:bunching}, the formation of the electron structures and its possible relation to the inverse cascade are discussed. In \cref{sec:es_heating}, the electrostatic energy and the electron temperature in the PIC and Vlasov simulations are compared. The effect of the azimuthal length on these results is also discussed. In \cref{sec:anomalous}, the  analysis is repeated for the anomalous current. Finally, in \cref{sec:conclusion}, the results are discussed, and conclusions are given.  

\section{Problem setup and numerical methods} \label{sec:setup}

In our setup, a constant magnetic field ($B_0$) and a constant electric field ($E_0$) are applied in the direction of $\hat{y}$ and $\hat{z}$, respectively. Therefore, the $E\times B$ drift velocity of the electrons is $v_d=-\frac{E_0}{B_0}\hat{x}$. The ions are taken to be unmagnetized and unaffected by $E_0$, a common assumption  in ECDI studies. \Cref{table:par} shows the parameters used in the simulations. We note that the values of the physical parameters are chosen as in Refs.~\onlinecite{janhunen2018nonlinear,janhunen2018evolution,tavassoli2022nonlinear}. These values are close to the typical operation regime of an SPT100 Hall thruster \cite{boeuf2017tutorial}. With these parameters, the  Debye length is $\lambda_D\equiv \sqrt{\epsilon_0T_{e0}/n_0e^2}=7.43\times 10^{-3}$ cm, the electron thermal velocity is $v_{te}\equiv \sqrt{T_{e0}/m_e}=1.329 \times 10^6$ m/s, the electron cyclotron frequency is $\omega_{ce}\equiv eB_0/m_e=3.5$ rad/ns, and the electron Larmor radius is $\rho_e\equiv v_{te}/\omega_{ce}=0.377$ mm.  The initial condition for all simulations is a Maxwellian distribution, constructed using a random number generator in the PIC simulations. Due to the low level of noise  in the Vlasov simulations, a perturbation is commonly used to excite the instability. In the Vlasov simulations, the initial density of electrons and ions is perturbed as $n_i=n_e=n_0(1+\epsilon \sin \frac{2\pi}{L}x)$, where $\epsilon=1.38\times 10^{-4}$.  {We also tried a case where $\epsilon$ was 100 times larger ($\epsilon=1.38\times 10^{-2}$), and no significant change was observed in the results.}

In all simulations, we have used a periodic boundary condition in the spatial subspace.  {The reason for this choice is to avoid any difficulty with regard to sheath formation at the boundaries and also to be consistent with the cylindrical geometry of the Hall thruster.\cite{boeuf2017tutorial}}. In the Vlasov simulations, the boundary condition in the velocity subspace is open.  {We have used four lengths of the domain in the azimuthal direction: $L=627/k_0=17.824$ cm, $L=158.4/k_0=4.502$ cm, $L=39.6/k_0=1.125$ cm, and $L=19.8/k_0=0.5625$ cm. We note that azimuthal lengths $L=39.6/k_0$ and $L=19.8/k_0$ are not sufficient to properly capture the physics of the problem; however, we have confirmed that even for the smallest length $L$ ($L=19.8/k_0$), five cyclotron peaks are ``resolved" (see \Cref{fig:resonant_modes}). Therefore, these two cases are shown as cases that might be tempting to use to reduce the computational cost of simulations while resolving the sharp cyclotron peaks of the ECDI but are in fact not well designed for investigating this problem.}

The numerical method used for the Vlasov simulation is the semi-Lagrangian scheme \cite{shoucri2008numerical}. In this method, the Vlasov equation is split into three sub-equations, which are then integrated using the method of characteristics with cubic spline interpolation \cite{cheng1976integration,cheng1977integration}. The three sub-equations are the advection equations in the $x$, $v_x$, and $v_z$ directions, respectively. Because the ions are unmagnetized and unaffected by $E_0$, their advection equation in the $v_z$ direction is trivial. In the Vlasov simulations, the Gauss law is solved, self-consistently, using the Fast Fourier transform (FFT). 
The PIC code used is ``EDIPIC", which is also used in Refs.~\onlinecite{tavassoli2021role,janhunen2018nonlinear,SmolyakovPPR2020}. The direct-implicit method is used for advancing the particles, and the finite difference method is used for solving the Poisson equation \cite{sydorenko_2006}.  The numerical parameters used in the simulations are listed in \Cref{table:par}. In the PIC simulations, we observed that if we use a time step the same as what is listed in \cref{table:par} for Vlasov simulations, the energy conservation is greatly violated. Therefore, the time step that is used for the PIC simulations is much smaller than the one used for the Vlasov simulations.  {On one hand, the requirement of a small time step for the PIC simulations might not be surprising. In the PIC simulations, it is generally recommended that the time step be small enough that a few particles are displaced more than a cell in one time step. On the other hand, in our PIC simulations, the extensive electron heating makes satisfying this criterion difficult (see e.g.~\Cref{fig:avg_Txe_PIC_VL}). This criterion does not however apply to the semi-Lagrangian Vlasov simulations.} In all simulations, the energy was generally conserved to within 1 percent.  In the Vlasov simulation with $L=156.8/k_0$, time steps in the range of $5.6$ ps to $11.2$ ps were tried, and no significant change was observed in the results.  {For $L=156.8/k_0$, the Vlasov simulation took about 10 days, and the PIC simulations with $N_{ppc}=10^4$ took about 6 days, both on 32 processors. These times scale linearly with $L$.} In addition, we verified the convergence of Vlasov simulations in terms of the grid resolution of the velocity subspace (which is not shown here).

\begin{table}[htbp]
\begin{tabular}{ccc}
Parameter                                                                      & Symbol    & Value (s)                     \\ \hline
Magnetic field                                                                 & $B_0$     & 200 G                         \\
Electric field                                                                 & $E_0$     & 200 V/cm                      \\
Ion mass                                                                       & $m_i$     & 131.293 u                     \\
Electron temperature                                                           & $T_{e0}$  & 10 eV                         \\
Ion temperature                                                                & $T_{i0}$  & 0.2 eV                        \\
Density                                                                        & $n_0$     & $10^{17}\;m^{-3}$             \\
PIC time step                                                                  & --        & 0.56 ps                       \\
Vlasov time step                                                               & --        & 5.6 ps to 11.2 ps             \\
Spatial cell size                                                              & --        & $0.28\;\lambda_D$             \\
\begin{tabular}[c]{@{}c@{}}Electron velocity cell size\\ (Vlasov)\end{tabular} & --        & $0.054\;v_{te}$               \\
\begin{tabular}[c]{@{}c@{}}Ion velocity cell size\\ (Vlasov)\end{tabular}      & --        & $10^{-4}\;v_{te}$            
\end{tabular}
\caption{The parameters used in the simulations of ECDI.\label{table:par}}
\end{table}

\section{Linear regime of ECDI and effect of azimuthal length on the linear growth rates of PIC and Vlasov simulations}\label{sec:linear_regime}

\begin{figure}[htbp]
\centering
\captionsetup[subfigure]{labelformat=empty}
\subcaptionbox{\label{fig:resonant_modes}}{\includegraphics[width=0.49\textwidth]{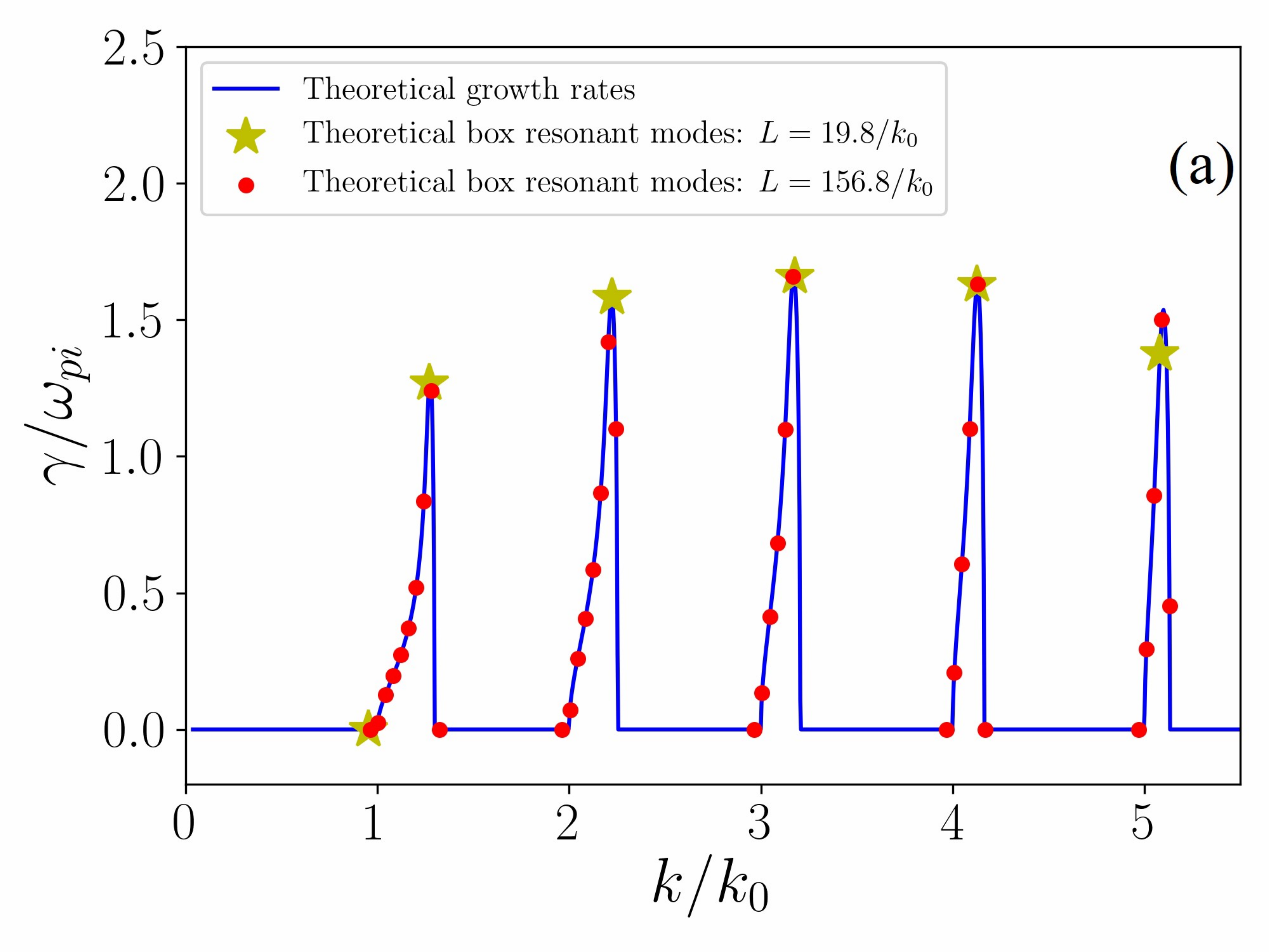}}
\subcaptionbox{\label{fig:PIC_short_growths}}{\includegraphics[width=0.49\textwidth]{{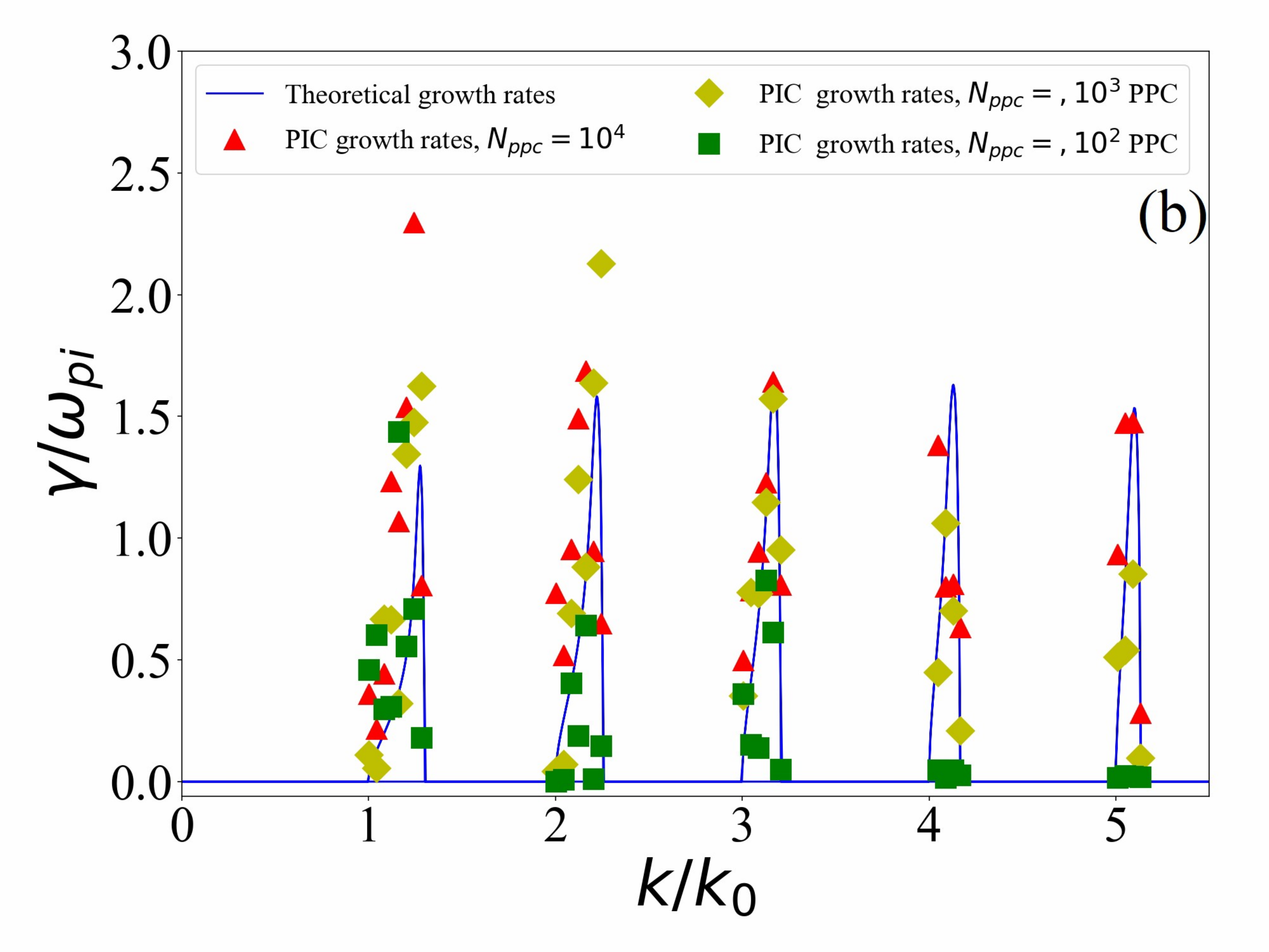}}}
\subcaptionbox{\label{fig:Vlasov_short_growths}}{\includegraphics[width=0.49\textwidth]{{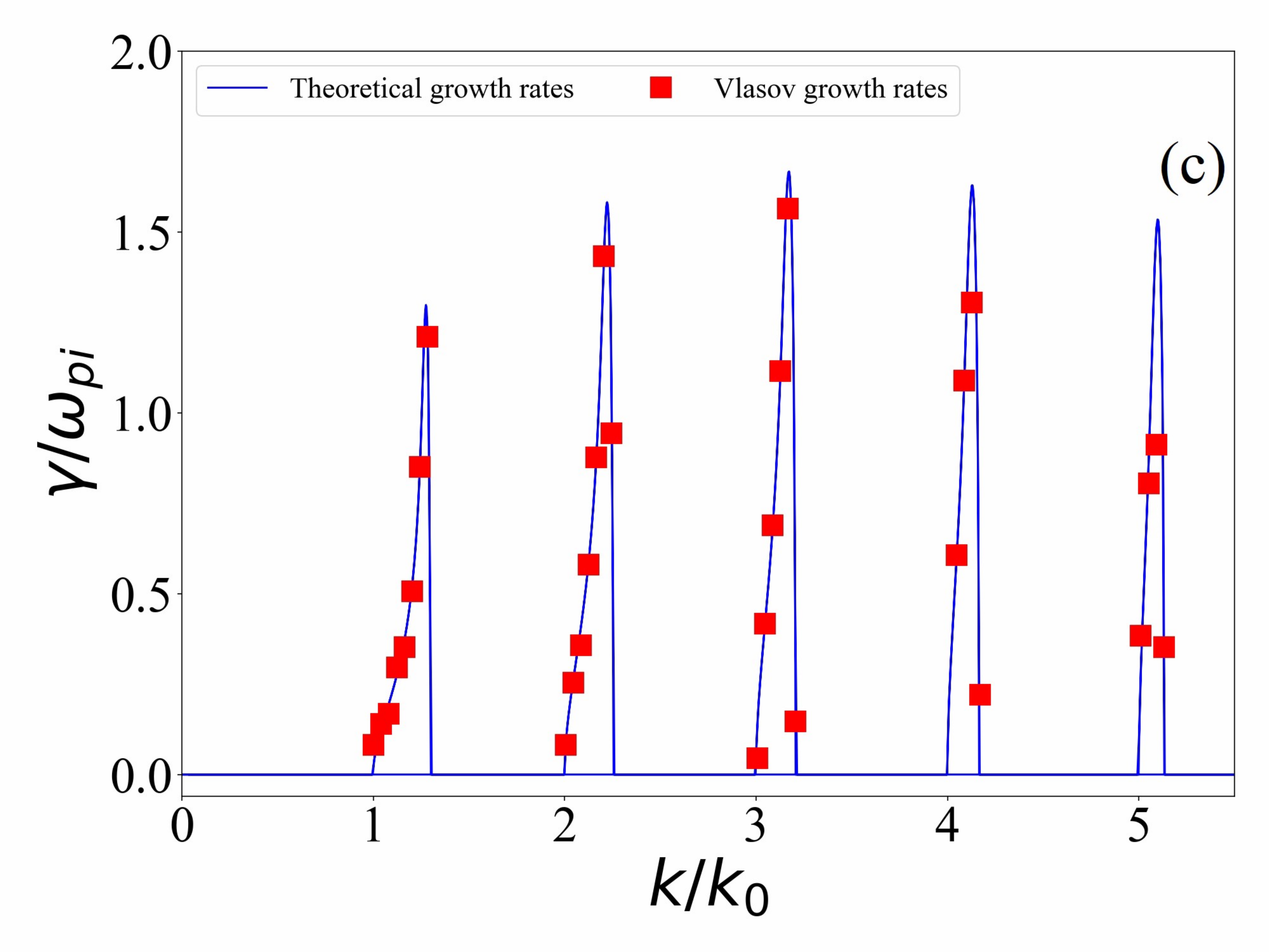}}}
\caption{a) The box resonant modes for two different azimuthal lengths. b) The comparison of the measured and theoretical growth rates for PIC simulations; $L=156.8/k_0$. c) The comparison of the measured and theoretical growth rates for the Vlasov simulation; $L=156.8/k_0$.}
\end{figure}

 The linear dispersion relation of the ECDI is $1+\chi_i+\chi_e=0$, where $\chi_i$ and $\chi_e$ are the ion and electron susceptibilities, respectively. The cold ion susceptibility is $\chi_i=-\frac{\omega_{pi}^2}{\omega ^2}$, where $\omega_{pi}\equiv\sqrt{n_0e^2/m_i\epsilon_0}$, $n_0$ is the plasma density, $e$ is the electron charge, and $m_i$ is the ion mass. The two-dimensional electron susceptibility is
\begin{gather}
    	\chi_e=\frac{1}{\lambda_{D}^2k^2}{\Bigg [} 1+\frac{
    		(\omega-k_xv_d)e^{-\lambda}}{\sqrt{2}\abs{k_y}v_{te}}\sum_{\ell=-\infty}^{\infty}I_\ell(\lambda)Z(\xi _\ell){\Bigg ]},
    	\label{eq:chie}
\end{gather}
where $k\equiv \sqrt{k_x^2+k_y^2}$ is the magnitude of wave vector, $I_\ell$ is the Bessel function of the second kind of order $\ell$, $Z$ is the plasma dispersion function, $\lambda\equiv (k_xv_{te}/\omega_{ce})^2$, and $\xi_l\equiv (\omega-k_xv_d+l\omega_{ce})/\sqrt{2}\abs{k_y}v_{te}$.
 In the ECDI, the overlapping resonances of Bernstein and ion-sound modes lead to a resonance condition 
 \begin{equation}
  \omega-n\omega_{ce}- k_xv_d=0, 
  \label{eq:resonance}
 \end{equation}
 where  $k_x$ is the wave vector in the direction of $E \times B$ drift, and $n$ is an integer. When $\omega\approx 0$, the resonances occur at $k_x\approx n k_0$.  {In this study, we refer to these modes as ``cyclotron peaks".} For strictly perpendicular propagation ($k_y=0$), the unstable growth rates form  a set of discrete narrow-band modes near these peaks (see, e.g.,~\Cref{fig:resonant_modes}).
 
\begin{figure}[htbp]
\centering
\captionsetup[subfigure]{labelformat=empty}
\subcaptionbox{\label{fig:PIC10000_t_Ek}}{\includegraphics[width=0.49\linewidth]{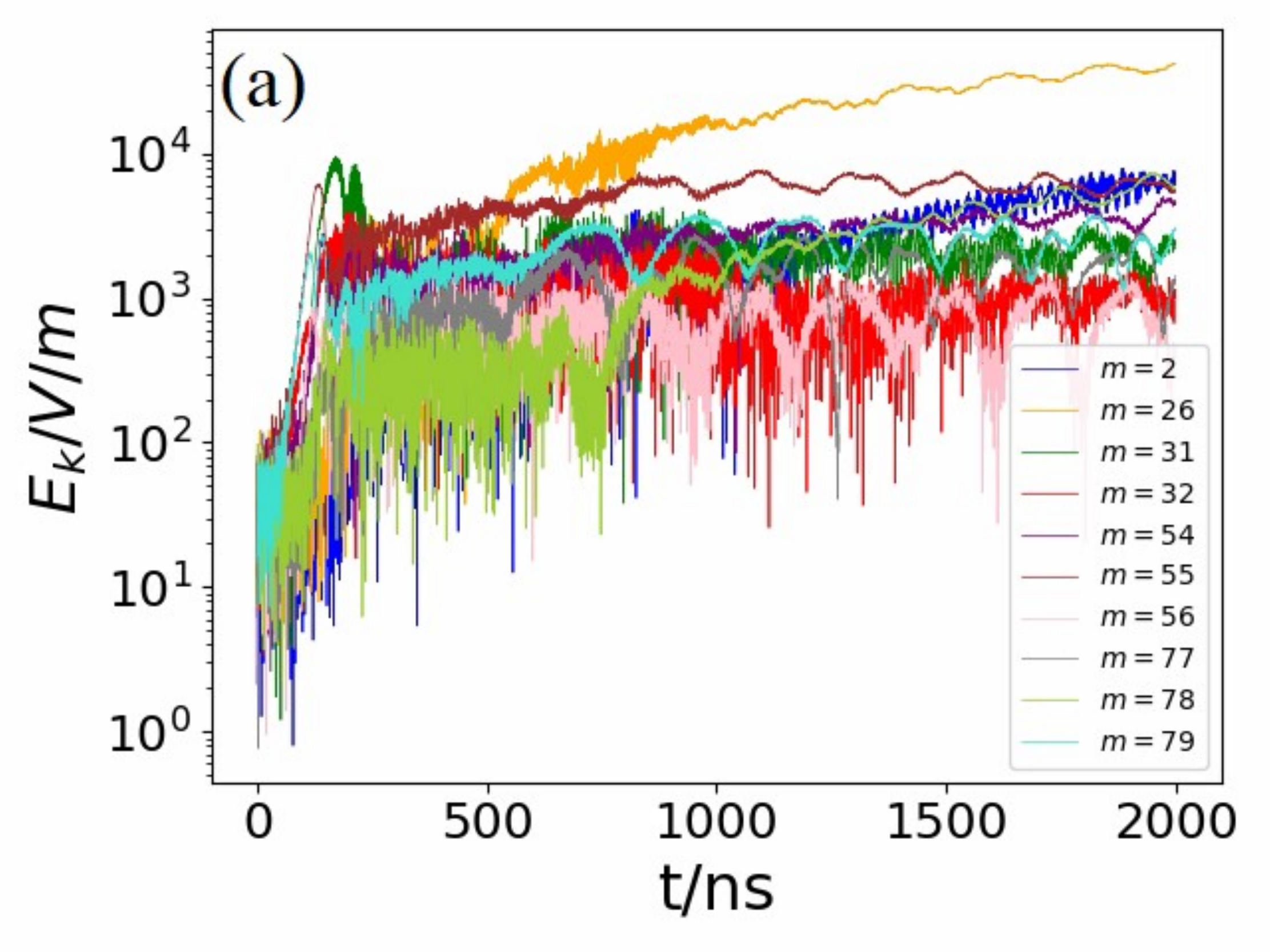}}
\subcaptionbox{\label{fig:vlasov_t_Ek}}{\includegraphics[width=0.49\linewidth]{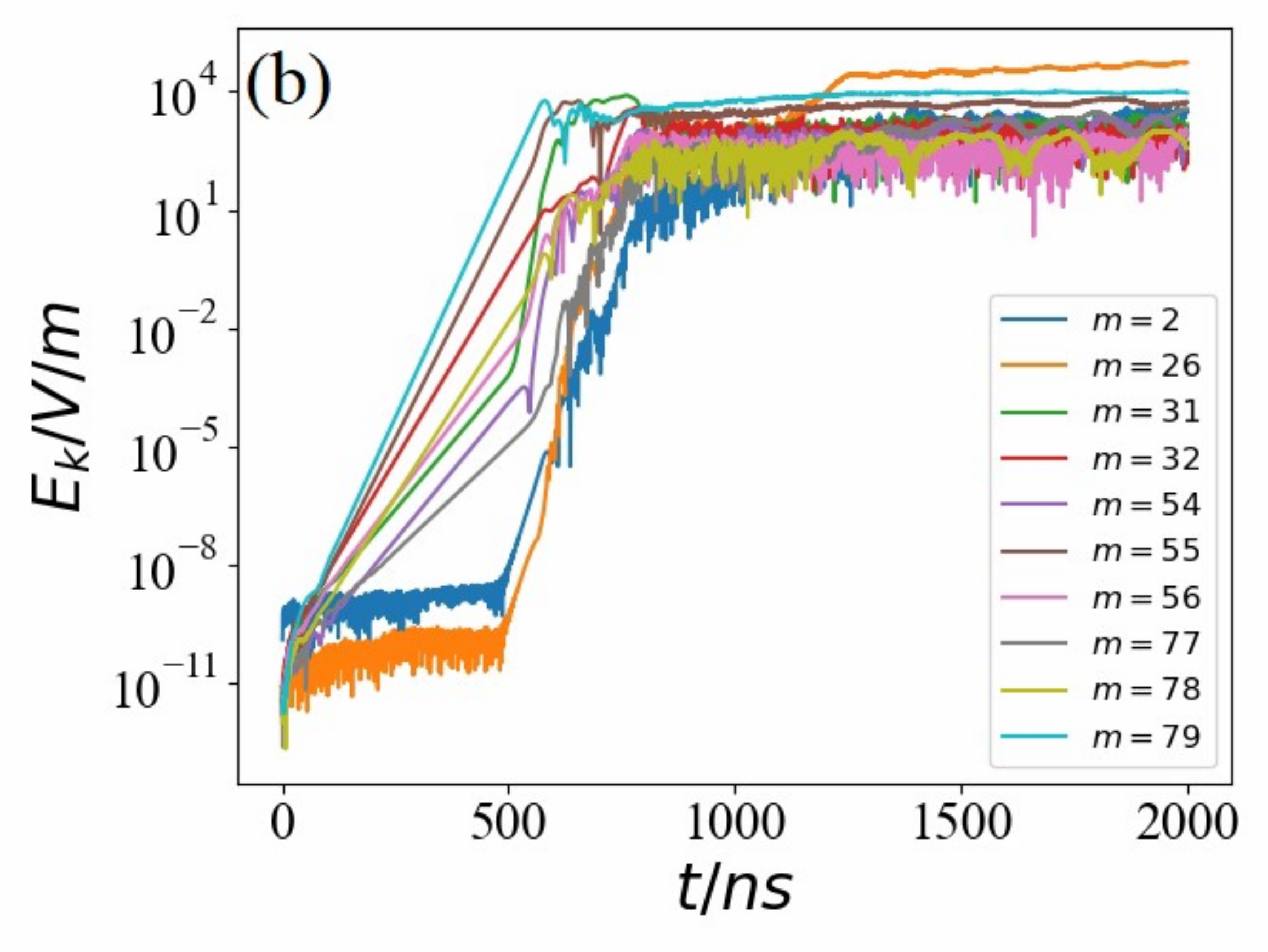}}
\subcaptionbox{\label{fig:Ep_PIC_VL_Log}}{\includegraphics[width=0.49\linewidth]{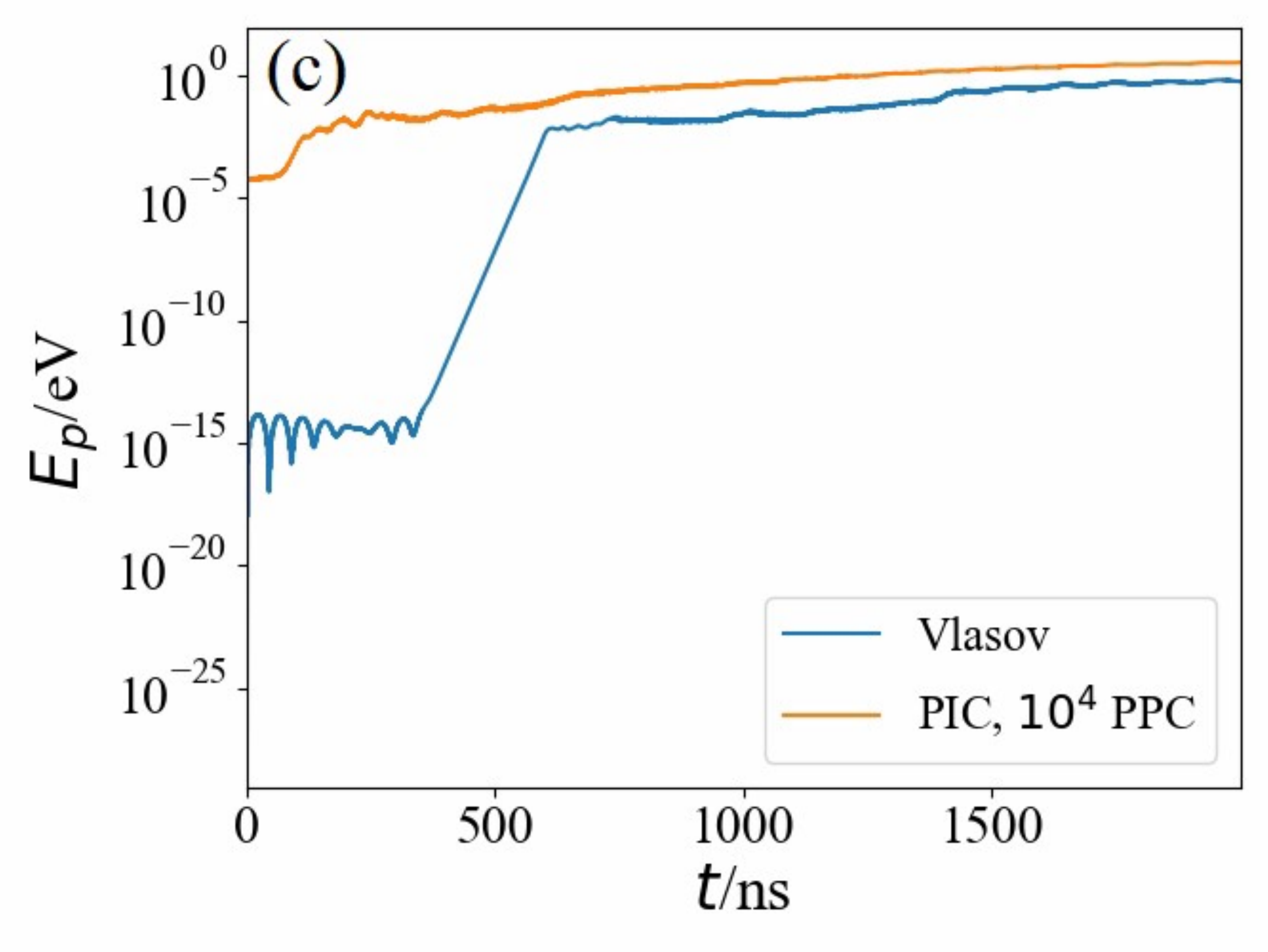}}
\caption{The growth of individual resonant modes of the electric field, using $L=156.8/k_0$. a) PIC simulation ($N_{ppc}=$ $10^4$) and b) Vlasov simulation. The total electrostatic energies are shown in (c). \label{fig:t_Ek_PIC_Vl}}
\end{figure}

In this study, we solve the dispersion relation with a  method described in Ref.~\onlinecite{cavalier2013hall}. As a test of the simulation results, we also measure the linear growth rates through simulations and compare them with the solution of the dispersion relation (the theoretical growth rates).  {For all azimuthal lengths, we ensured that at least five cyclotron peaks are resolved in the simulation. This point is made clear for $L=19.8/k_0$ and $L=156.8/k_0$ in \Cref{fig:resonant_modes}, where the resonant modes and their theoretical growth rates are shown.} \Cref{fig:PIC_short_growths,fig:Vlasov_short_growths} shows the comparison of the theoretical growth rates and the growth rates measured with different simulation methods when the azimuthal length is $L=156.8/k_0$. We can see that, for the PIC simulation with $N_{ppc}=10^3$ or $N_{ppc}=10^4$, and the Vlasov simulation, the measured growth rates decently agree with the theoretical values, despite a few outliers that exist in the three simulations. For the PIC simulation with $N_{ppc}=10^2$, however, the growth rates can significantly deviate from the theory. This deviation increases with $k$ and is particularly clear for higher cyclotron peaks such as $k\approx 4k_0$ and $k\approx 5k_0$. For these peaks, we basically see no growth rate in the PIC simulation with $N_{ppc}=10^2$. The linear growth rates are measured from the data by fitting a line in the linear growth region, as for example can be seen clearly in \Cref{fig:vlasov_t_Ek}. However, in all of our PIC simulations, the linear growth is subject to numerical noise that causes spurious oscillations (see \Cref{fig:PIC10000_t_Ek}). Therefore, the measurement of growth rates in PIC simulations is inevitably subject to errors that are much higher than those of the Vlasov simulations. The total electrostatic energy in the PIC and Vlasov simulations is also shown in \Cref{fig:Ep_PIC_VL_Log}.

\begin{figure}[htbp]
\centering
\captionsetup[subfigure]{labelformat=empty}
\subcaptionbox{\label{fig:PIC_long_growths}}{\includegraphics[width=0.49\textwidth]{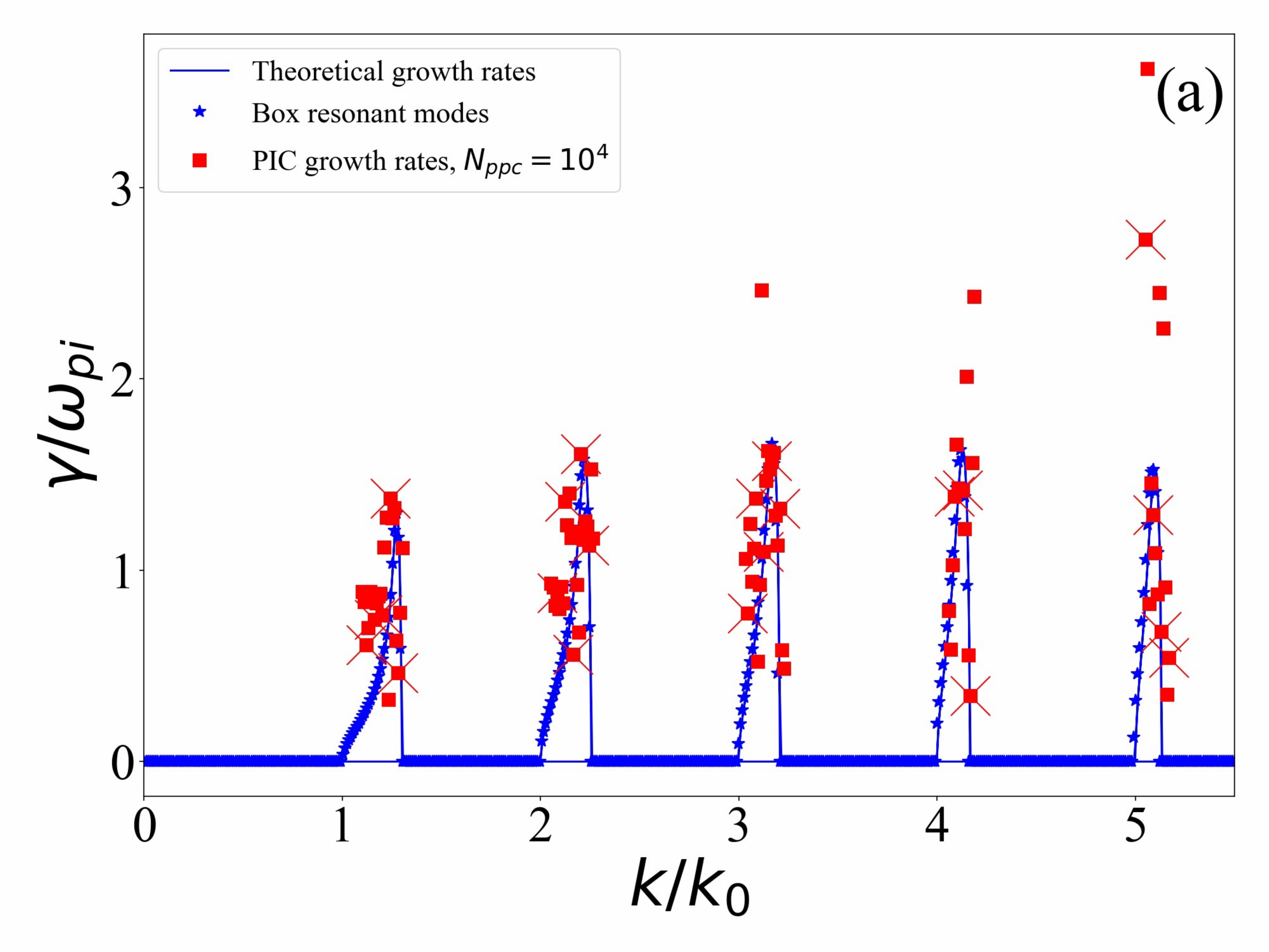}}
\subcaptionbox{\label{fig:VL_long_growths}}{\includegraphics[width=0.49\textwidth]{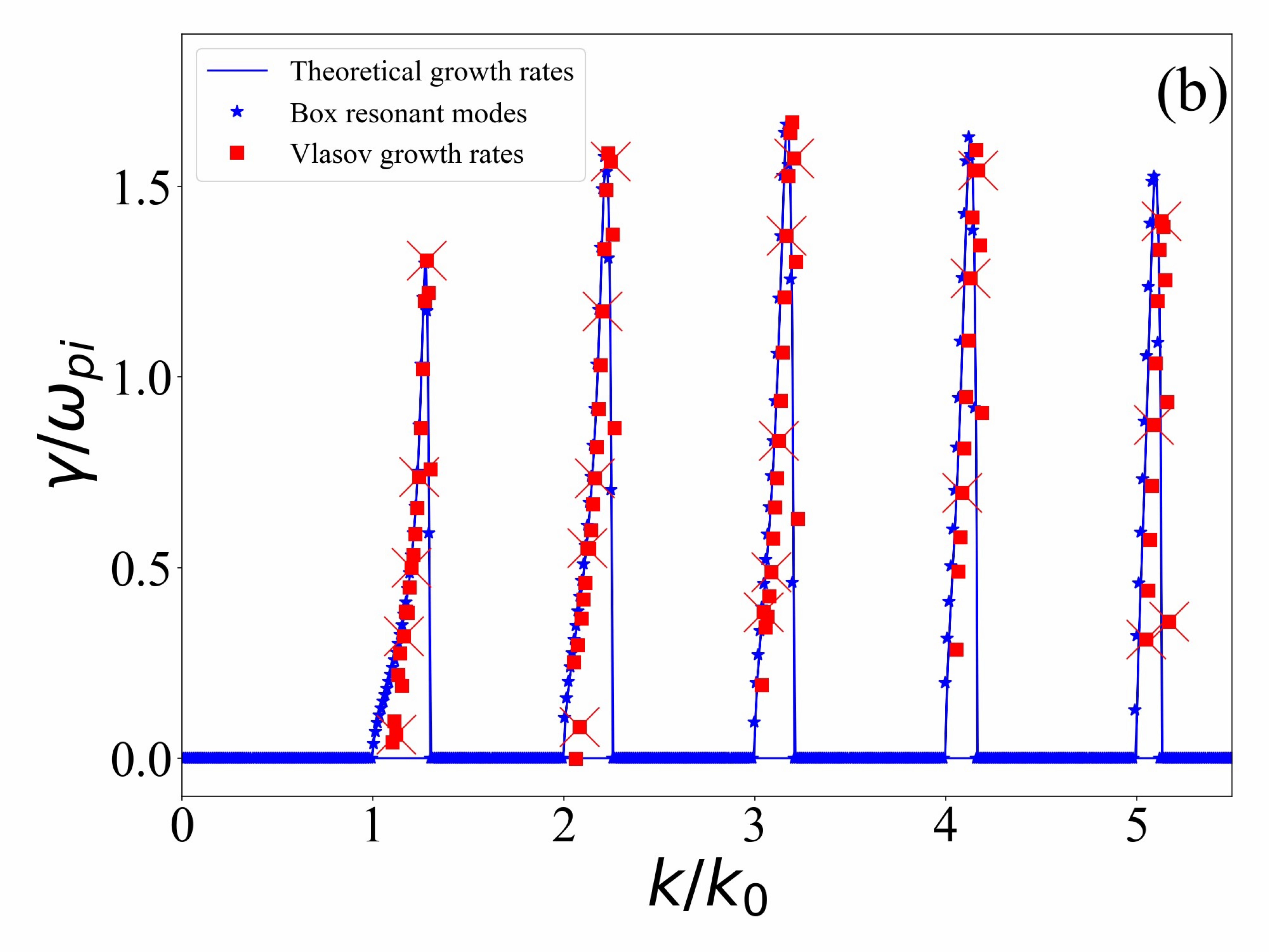}}
\caption{The growth-rates for a) PIC ($N_{ppc}=10^4$) and b) Vlasov, both with $L=627/k_0$. The $\times$ sign marks the box resonant modes of this simulation that also existed in the case of $L=156.8/k_0$ cm.}
\end{figure}

\Cref{fig:VL_long_growths} shows the theoretical and measured growth rates in a Vlasov simulation with azimuthal length $L = 627/k_0$. We see that, in comparison with the case of $L=156.8/k_0$, many new resonant modes exist in the simulation.  In the simulation with $L=627/k_0$, the growth rates of the resonant modes of the $L=156.8/k_0$ also show a better agreement with the theory. This effect is likely due to the reduced aliasing effects in the simulation with a longer azimuthal length. The results of the PIC simulation with $L=627/k_0$ are also presented in \Cref{fig:PIC_long_growths}. Similar to the results of the Vlasov simulation, we see that many new resonant modes are captured by the simulation. However, the growth rates of some of the new resonant modes show a significant discrepancy with the theoretical results. This discrepancy is more noticeable for the low-growth-rate resonant modes, where the measured growth rates are much larger than the theory. These large growth rates are likely an effect of the PIC noise, as also shown in the study of the linear regime of the Buneman instability \cite{tavassoli2021role}.

\section{Nonlinear spectra, backward waves, and the role of the azimuthal length.}\label{sec:eigenspectrums}

\begin{figure}[htbp]
\centering
\captionsetup[subfigure]{labelformat=empty}
\subcaptionbox{\label{fig:Ek_L17_PIC_1e4}}{\includegraphics[width=0.41\linewidth]{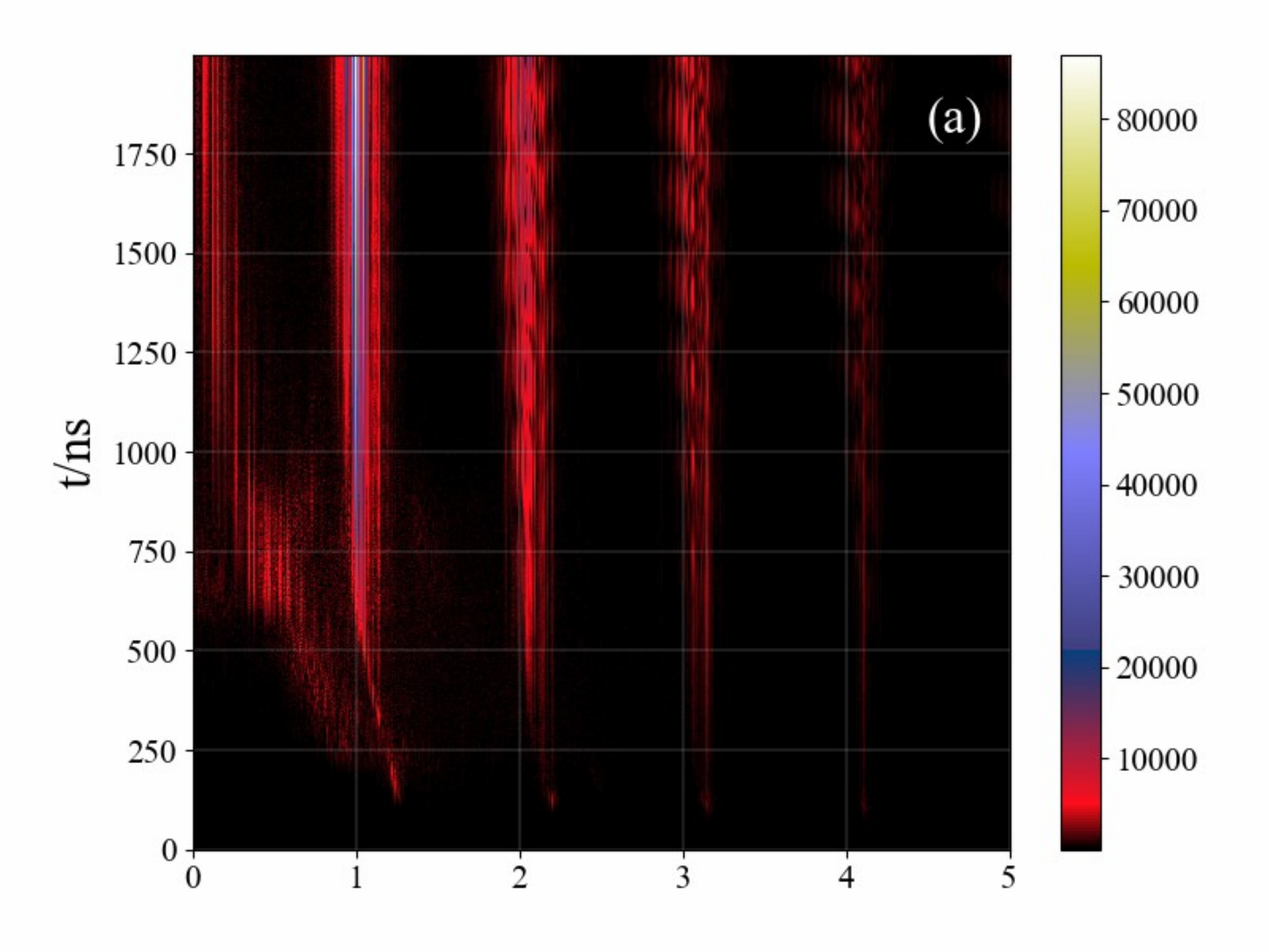}}
\vspace{-1.4cm}
\subcaptionbox{\label{fig:Ek_L17Vl}}{\includegraphics[width=0.41\linewidth]{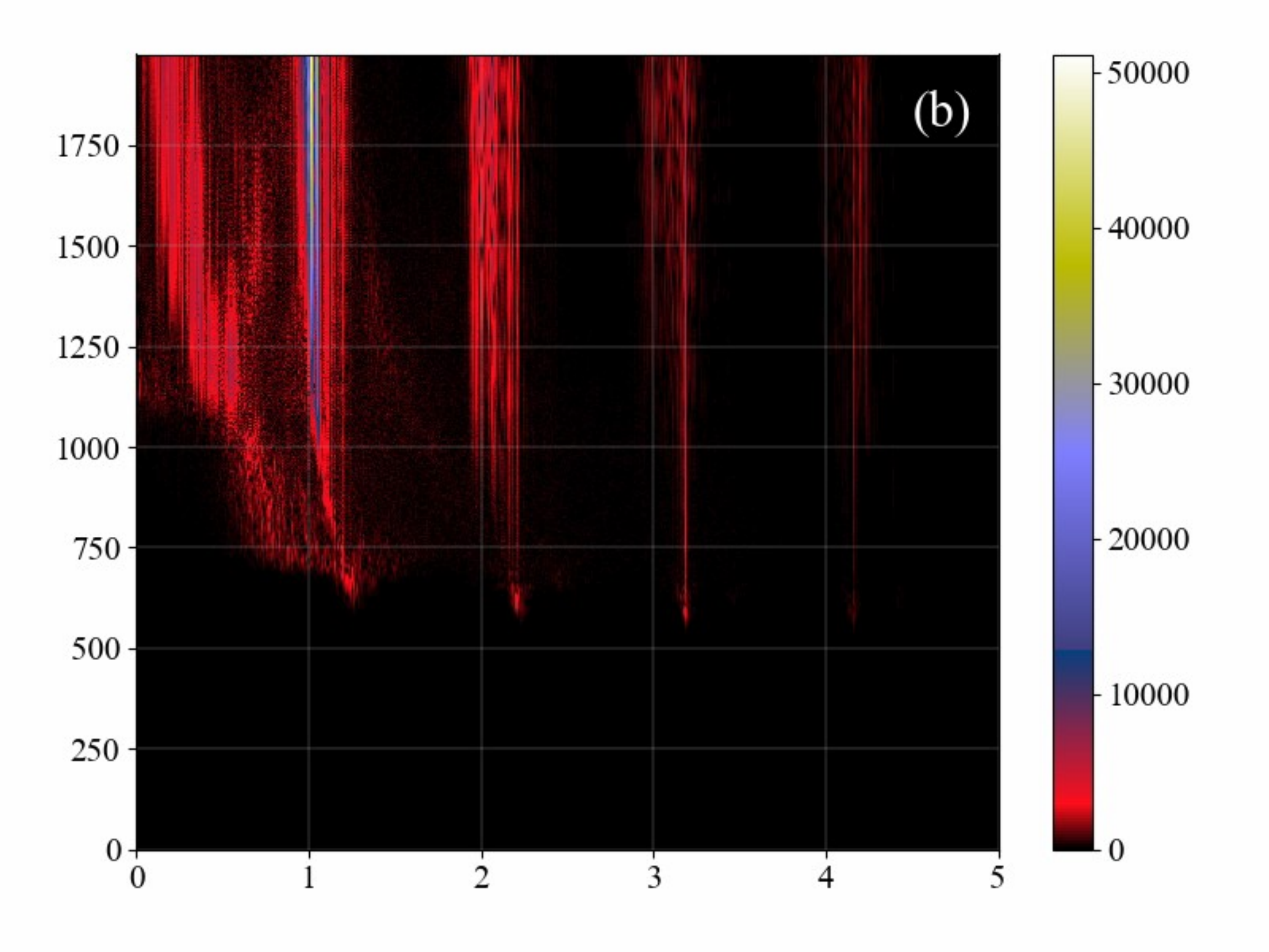}}
\vspace{-1.4cm}
\subcaptionbox{\label{fig:Ek_L4_PIC_1e4}}{\includegraphics[width=0.41\linewidth]{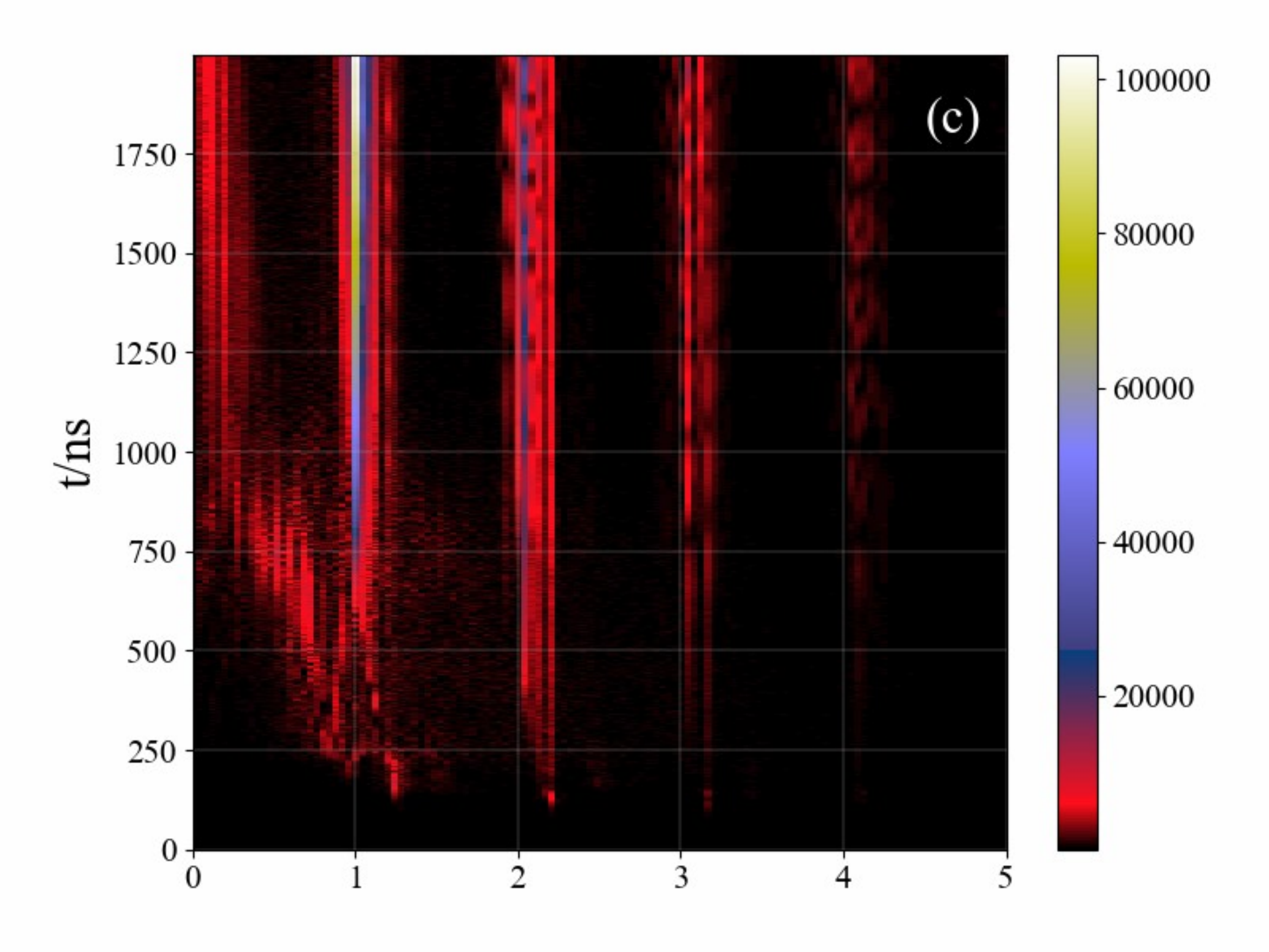}}
\subcaptionbox{\label{fig:Ek_L4Vl}}{\includegraphics[width=0.41\linewidth]{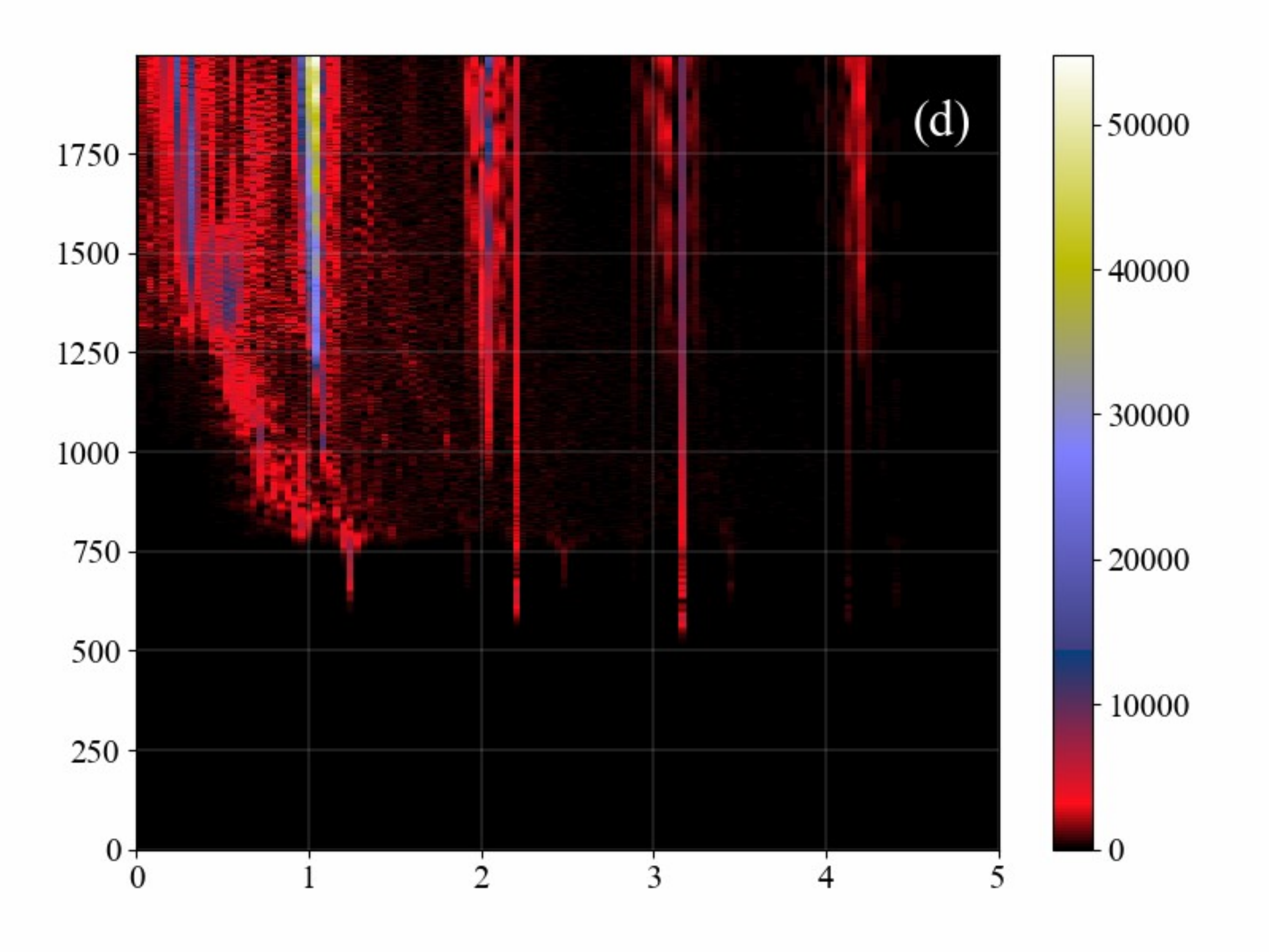}}
\vspace{-1.4cm}
\subcaptionbox{\label{fig:Ek_L1_PIC_1e4}}{\includegraphics[width=0.41\linewidth]{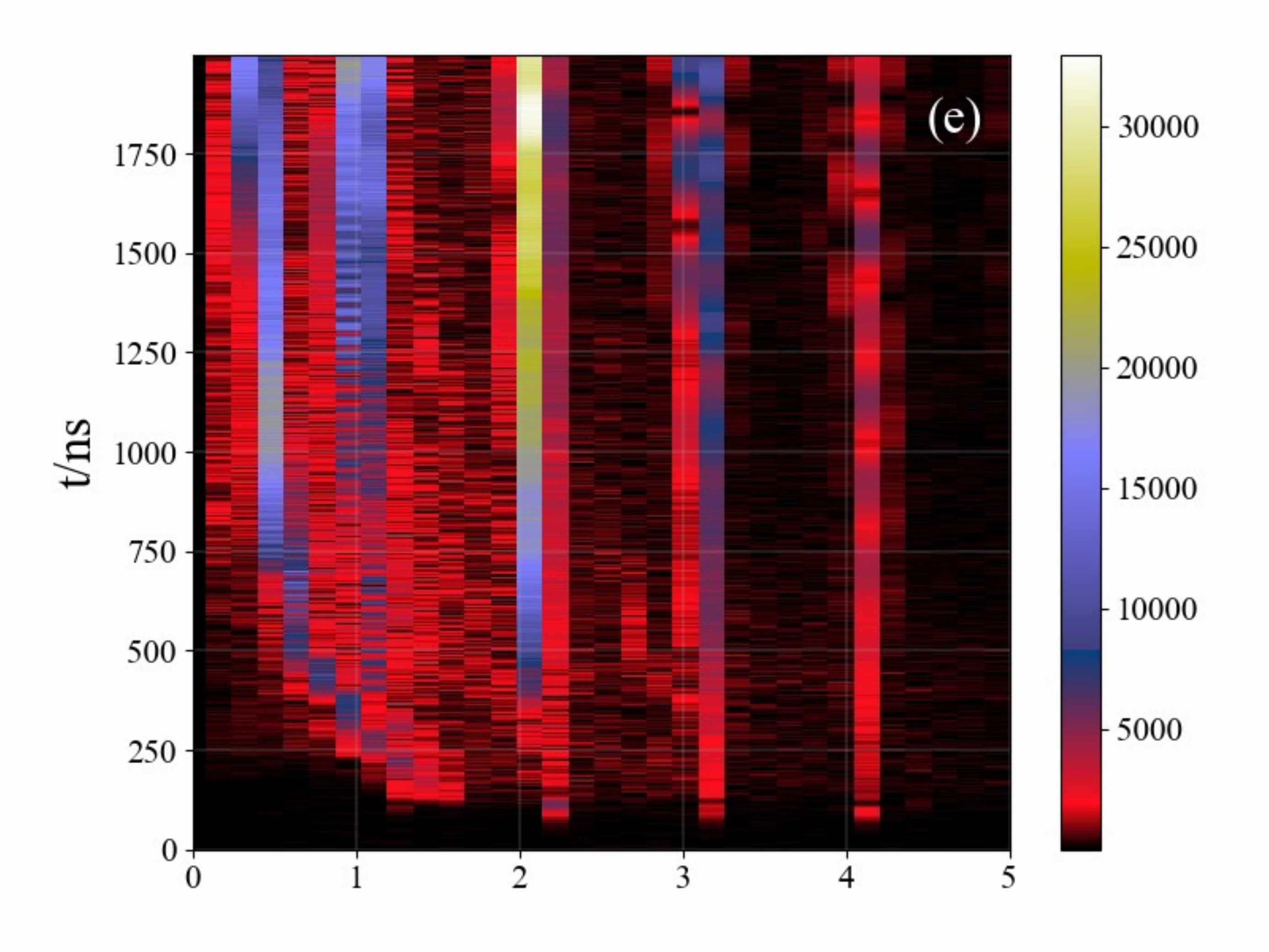}}
\subcaptionbox{\label{fig:Ek_L1Vl}}{\includegraphics[width=0.41\linewidth]{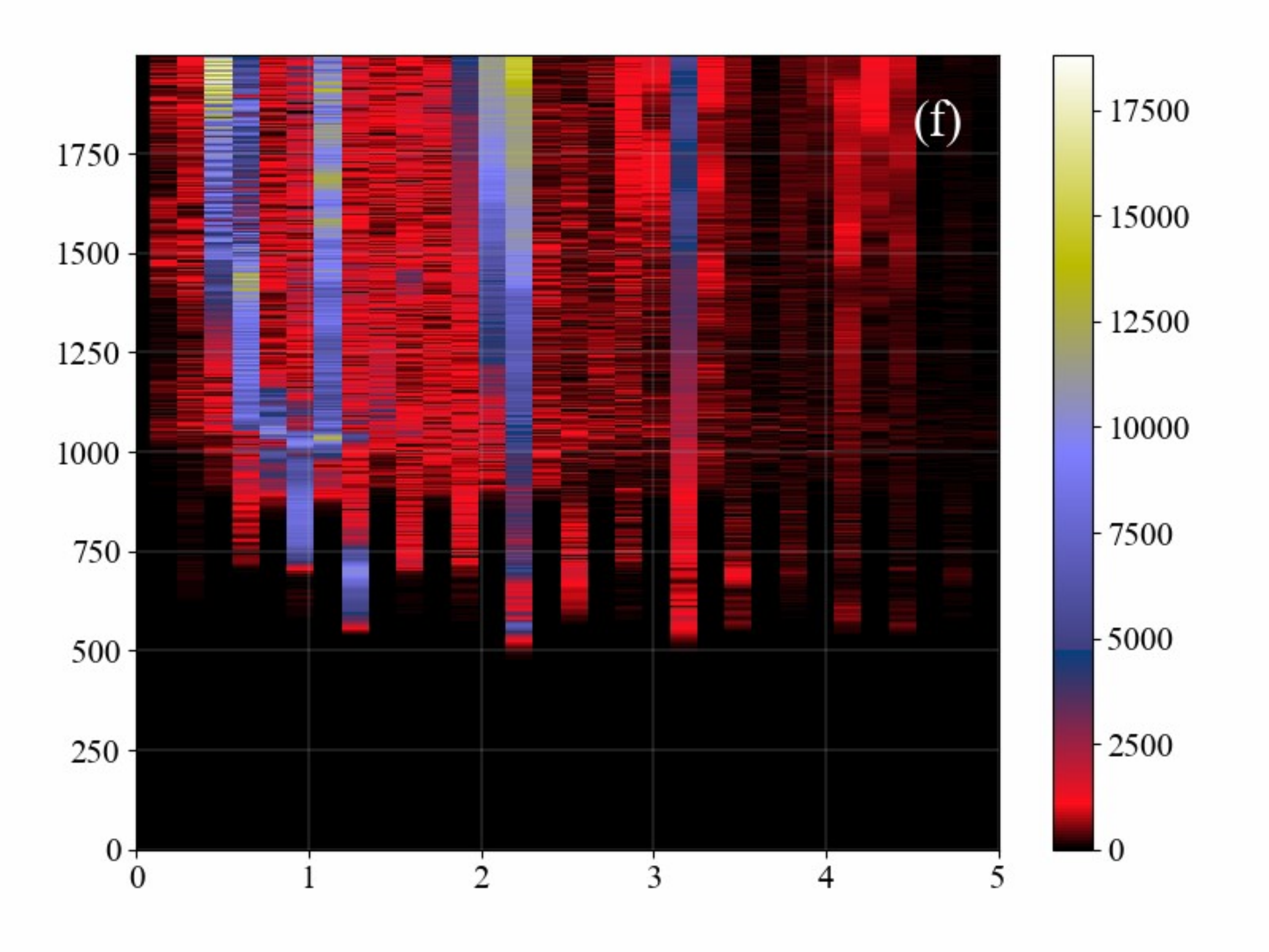}}
\vspace{-1.4cm}
\subcaptionbox{\label{fig:Ek_L005_PIC_1e4}}{\includegraphics[width=0.41\linewidth]{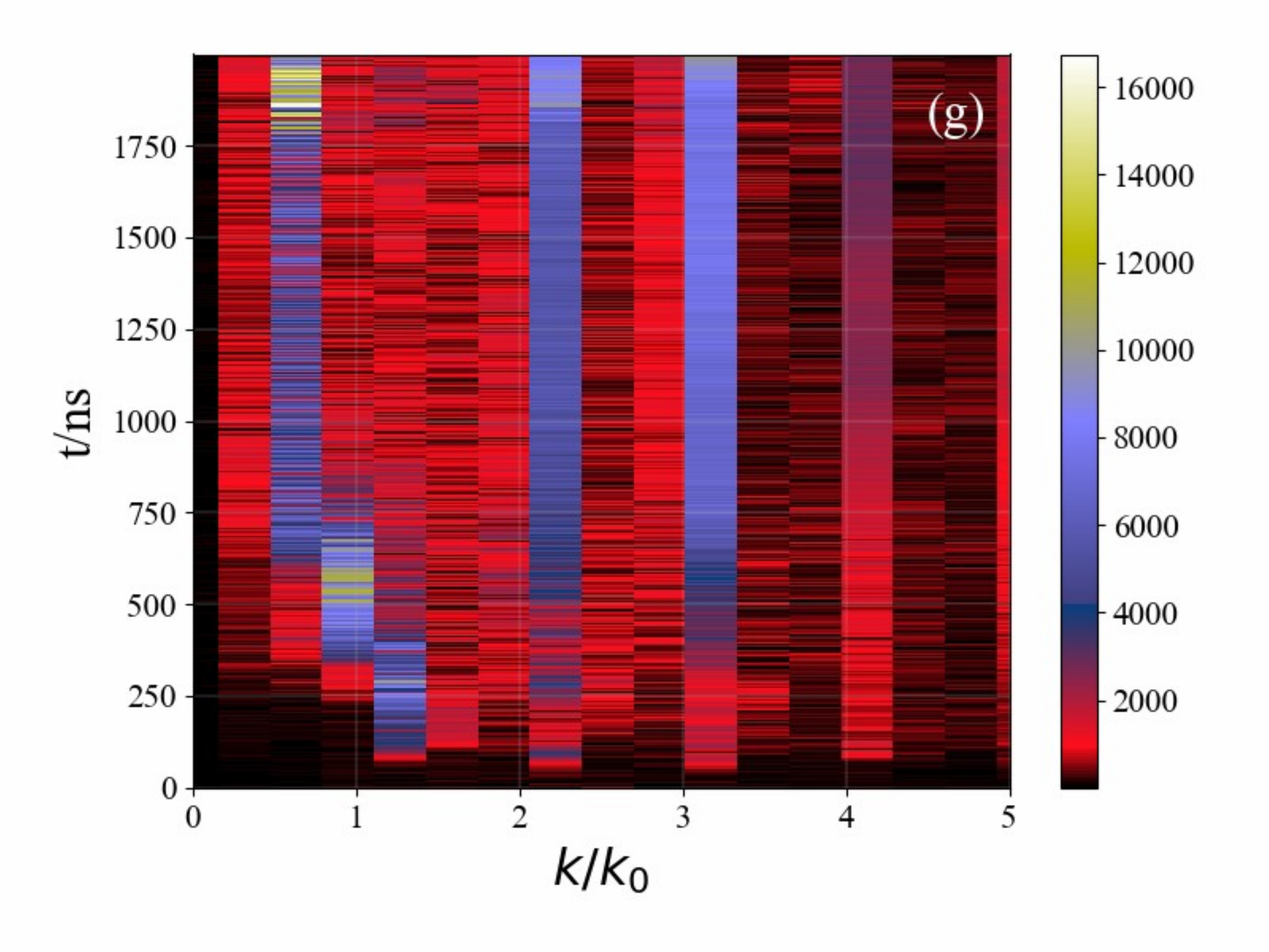}}
\subcaptionbox{\label{fig:Ek_L005Vl}}{\includegraphics[width=0.41\linewidth]{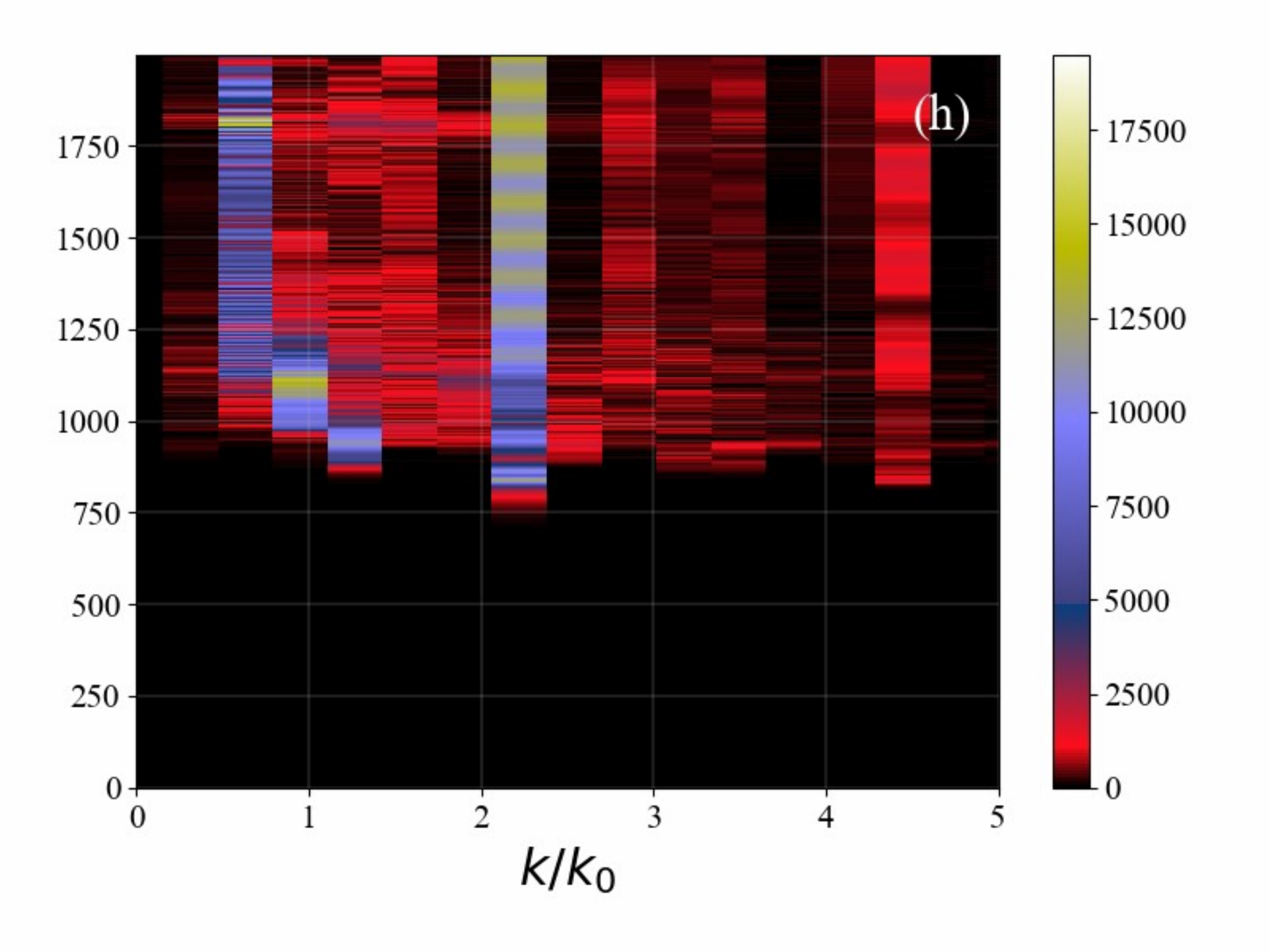}}

\caption{The Fourier spectrum of the electric field. (a), (c), (e),  and (g) are the PIC results ($N_{ppc}=10^4$). (b), (d), (f), and (h) are the Vlasov results. $L=627/k_0$,  $156.8/k_0$, $39.6/k_0$, and $19.8/k_0$ in the first, second, third, and fourth rows respectively. \label{fig:Ek}}  
\end{figure}

\Cref{fig:Ek} shows the nonlinear spectrum of the electric field in the PIC and Vlasov simulations for different azimuthal lengths. In all cases, the amplitude of the high-intensity modes is much lower in the Vlasov simulations than in the PIC simulations.  {In the simulations with $L=627/k_0$ and $L=156.8/k_0$,  we see that although the modes with $k\lesssim k_0$ do not grow significantly in the linear regime, they gradually grow in the nonlinear regime until a wavelength of about the size of the system appears. In this study, we refer to this process as the inverse cascade. This process happens along with the growth of other cyclotron modes, especially between $k_0$ to $5k_0$.} We note that the modes shown in \Cref{fig:Ek} occur in the nonlinear regime. Therefore, comparison of their growth with the linear growth rates is not easy because of the nonlinear effects (see \Cref{fig:t_Ek_PIC_Vl}).  The inverse cascade is, however,  less clear in the simulations with $L=39.6/k_0$ and is barely seen in the simulations with $L=19.8/k_0$.  In \Cref{fig:Ek_L17Vl,fig:Ek_L17_PIC_1e4,fig:Ek_L4Vl,fig:Ek_L4_PIC_1e4}, we can see that for $L=156.8/k_0$ cm and $L=627/k_0$, a mode with $k\approx k_0$ is eventually dominant in the late nonlinear regime. However, \Cref{fig:Ek_L1Vl,fig:Ek_L1_PIC_1e4} show that this is not true when the azimuthal length decreases to $L=39.6/k_0$ cm. In this case, a mode with $k\approx 2k_0$ in the PIC simulation and two modes with $k\approx 2k_0$ and $k\approx 0.5k_0$ in the Vlasov simulation have the highest amplitude at about $t=2000$ ns. It is also seen that, although the discrete cyclotron peaks are still visible,  there are also many finite-amplitude modes that fill the spaces between them. This suggests that decreasing the azimuthal length can make the Fourier spectrum smoother, similar to what is suggested in the theory of transition to ion-sound turbulence\cite{lampe1971nonlinear,lampe1972theory}. The process of smoothing is even more clear in \Cref{fig:Ek_L005Vl,fig:Ek_L005_PIC_1e4}, where the azimuthal length is reduced to $L=19.8/k_0$ cm.

\begin{figure}[htbp]
\centering
\captionsetup[subfigure]{labelformat=empty}
\subcaptionbox{\label{fig:Ekw_L17_PIC_1e4}}{\includegraphics[width=0.41\linewidth]{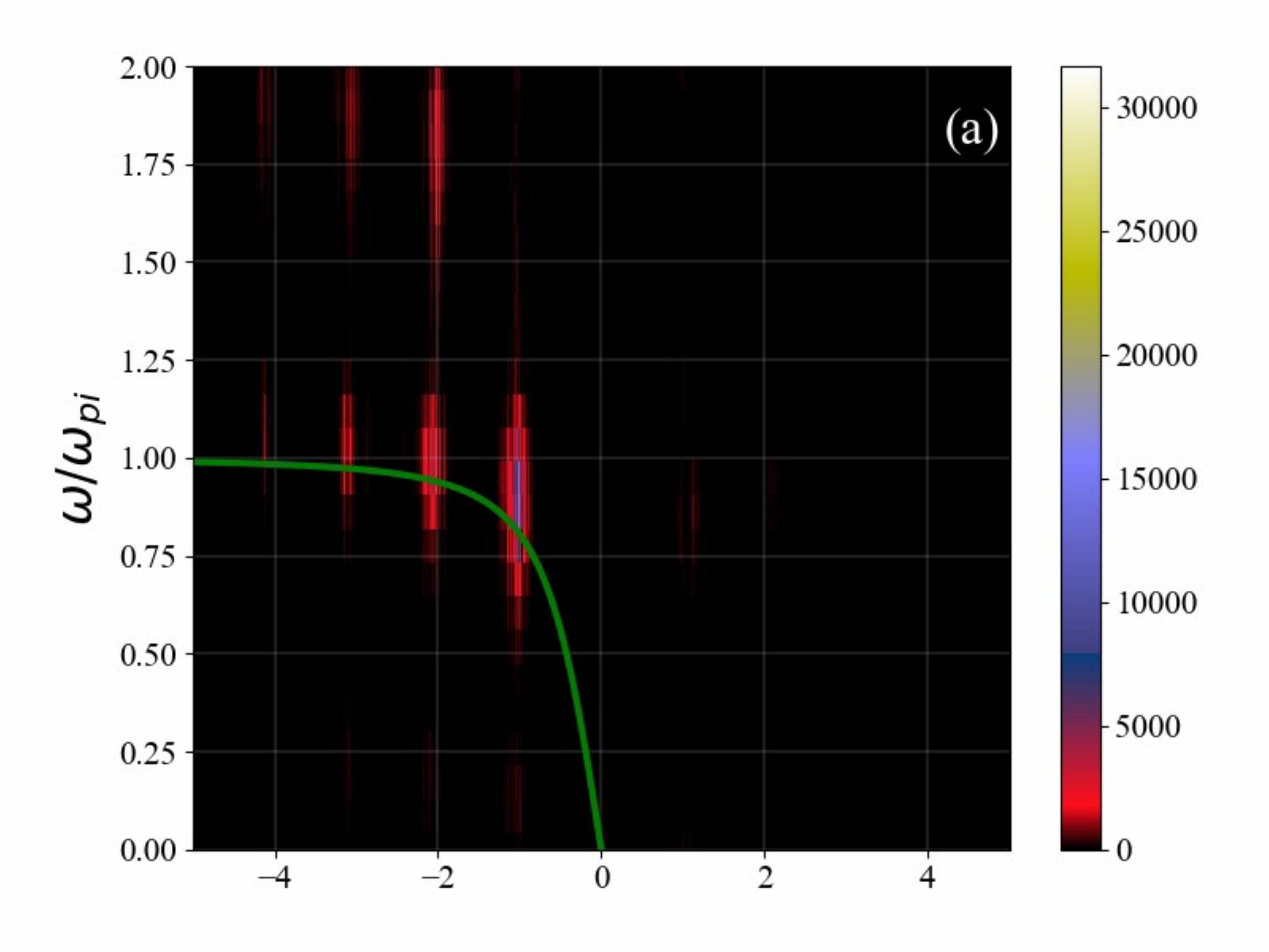}}
\vspace{-1.4cm}
\subcaptionbox{\label{fig:Ekw_L17Vl}}{\includegraphics[width=0.41\linewidth]{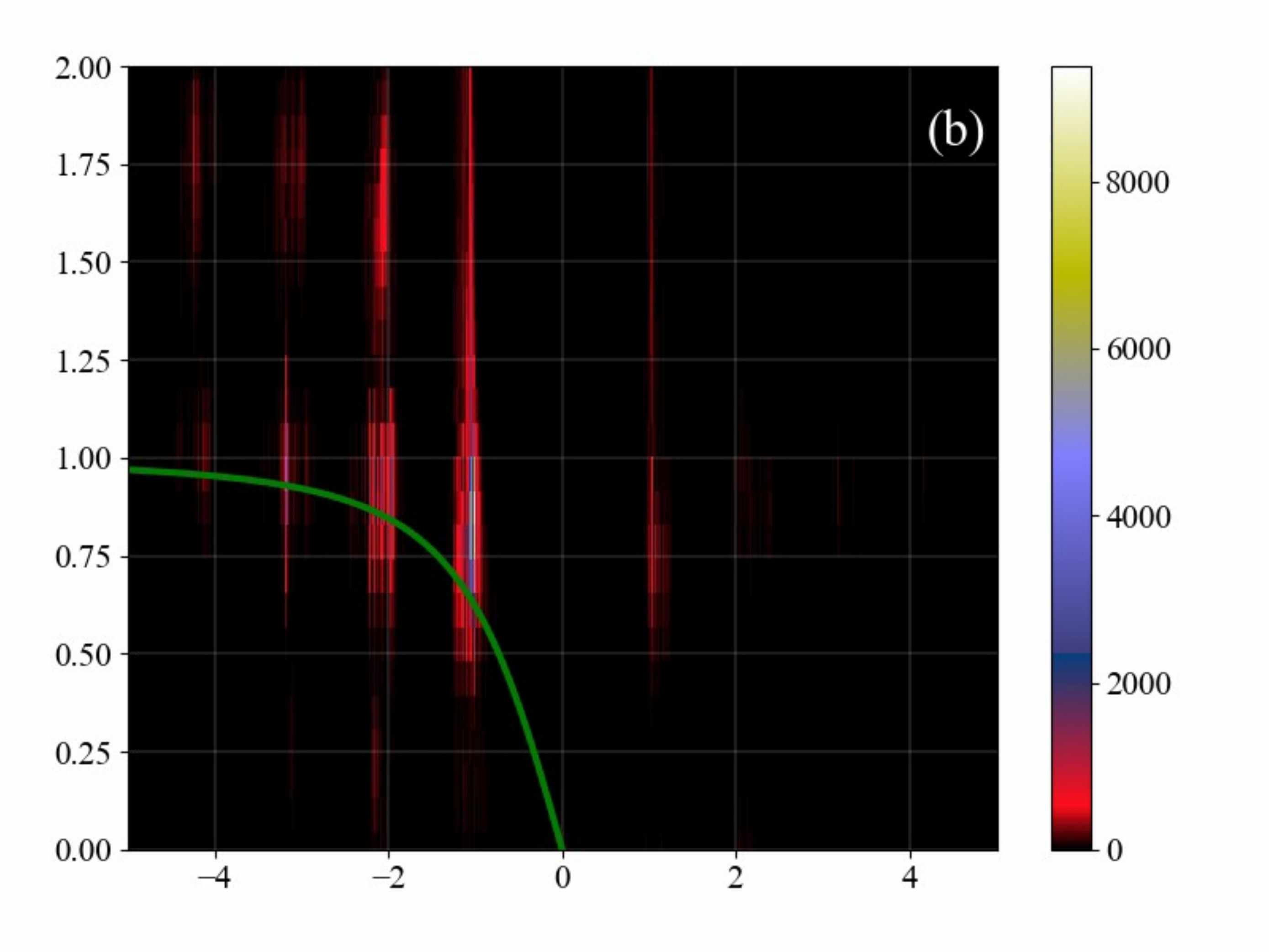}}
\vspace{-1.4cm}
\subcaptionbox{\label{fig:Ekw_L4_PIC_1e4}}{\includegraphics[width=0.41\linewidth]{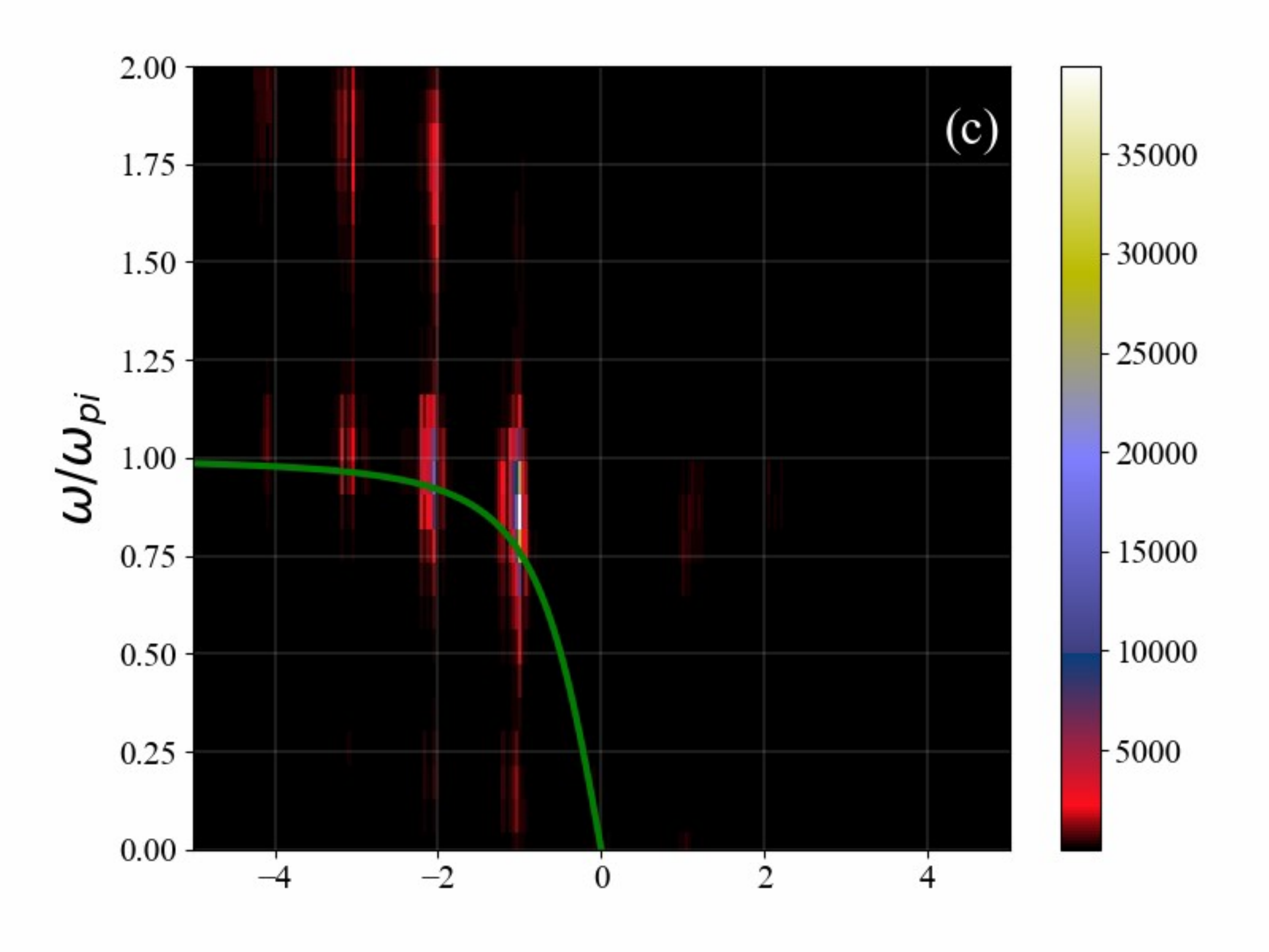}}
\subcaptionbox{\label{fig:Ekw_L4Vl}}{\includegraphics[width=0.41\linewidth]{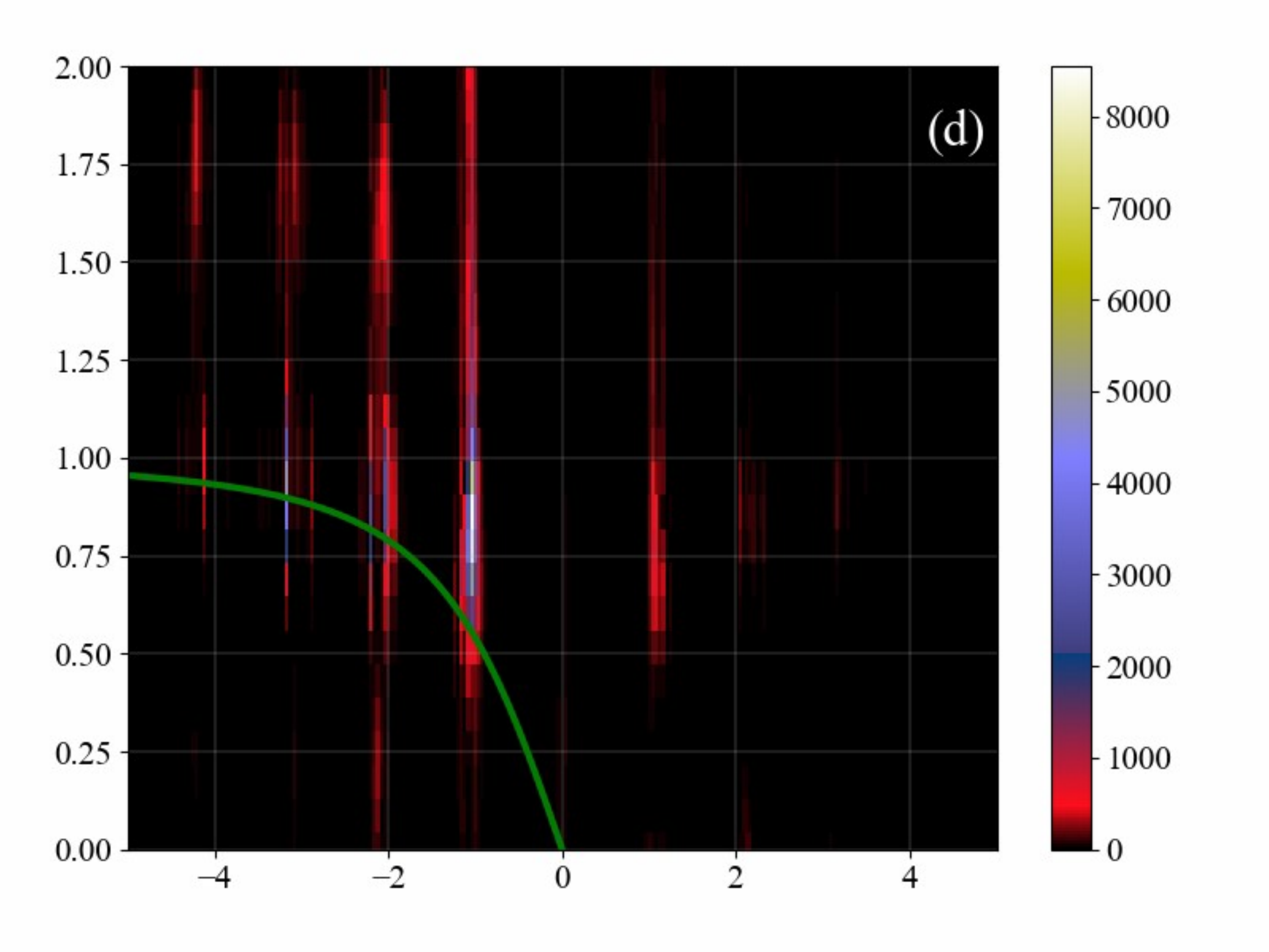}}
\vspace{-1.4cm}
\subcaptionbox{\label{fig:Ekw_L1_PIC_1e4}}{\includegraphics[width=0.41\linewidth]{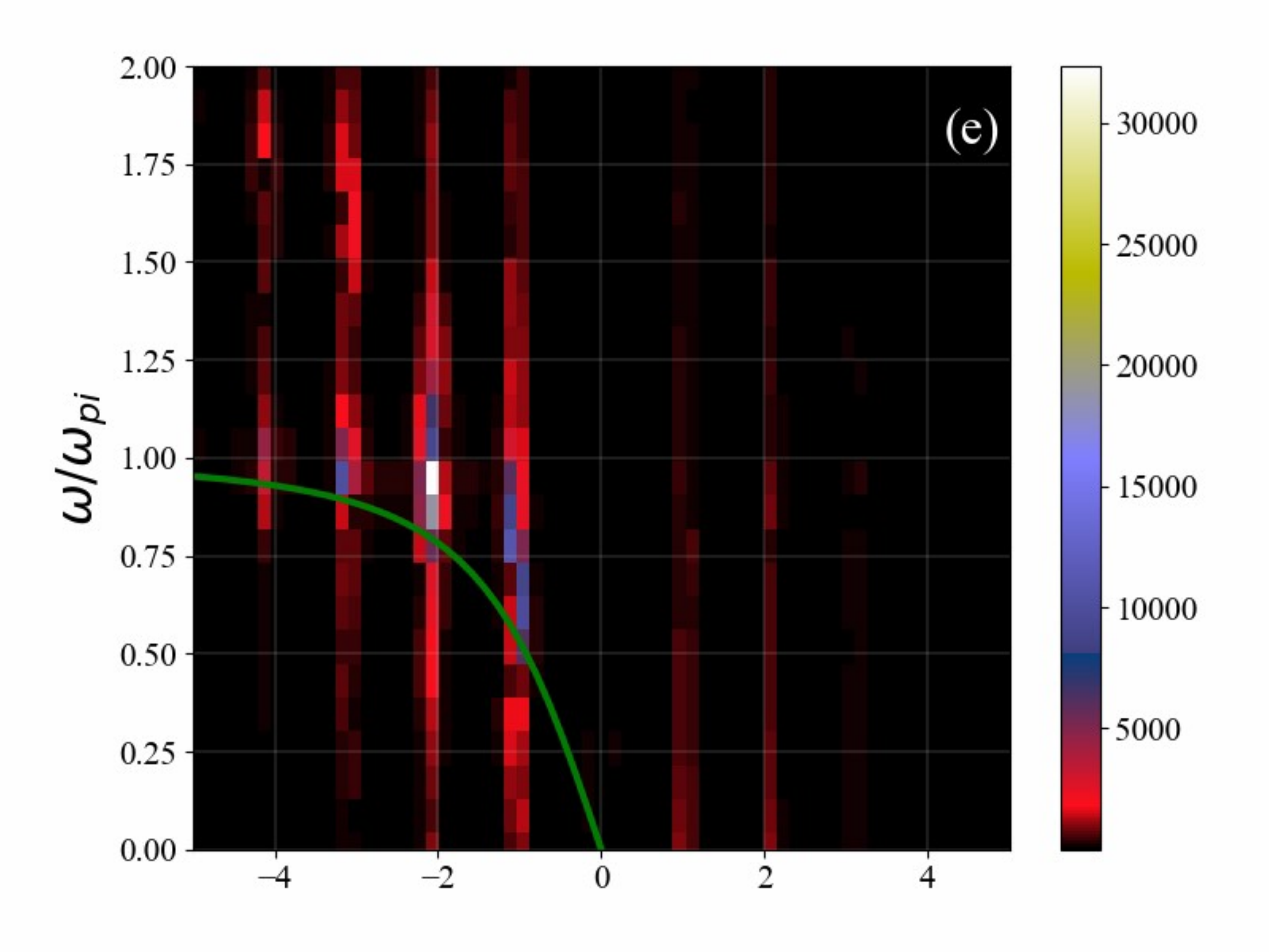}}
\subcaptionbox{\label{fig:Ekw_L1Vl}}{\includegraphics[width=0.41\linewidth]{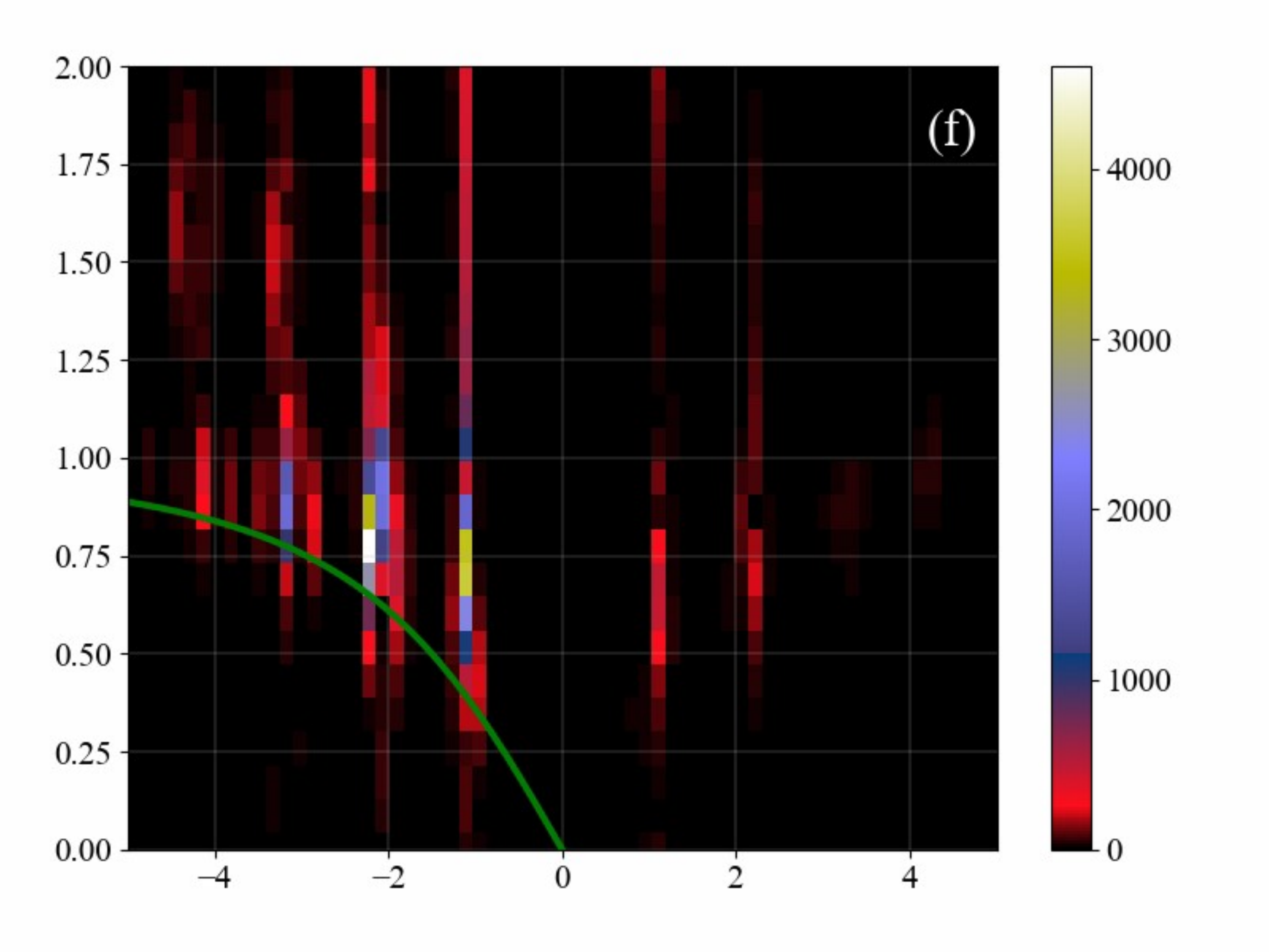}}
\subcaptionbox{\label{fig:Ekw_L005_PIC_1e4}}{\includegraphics[width=0.41\linewidth]{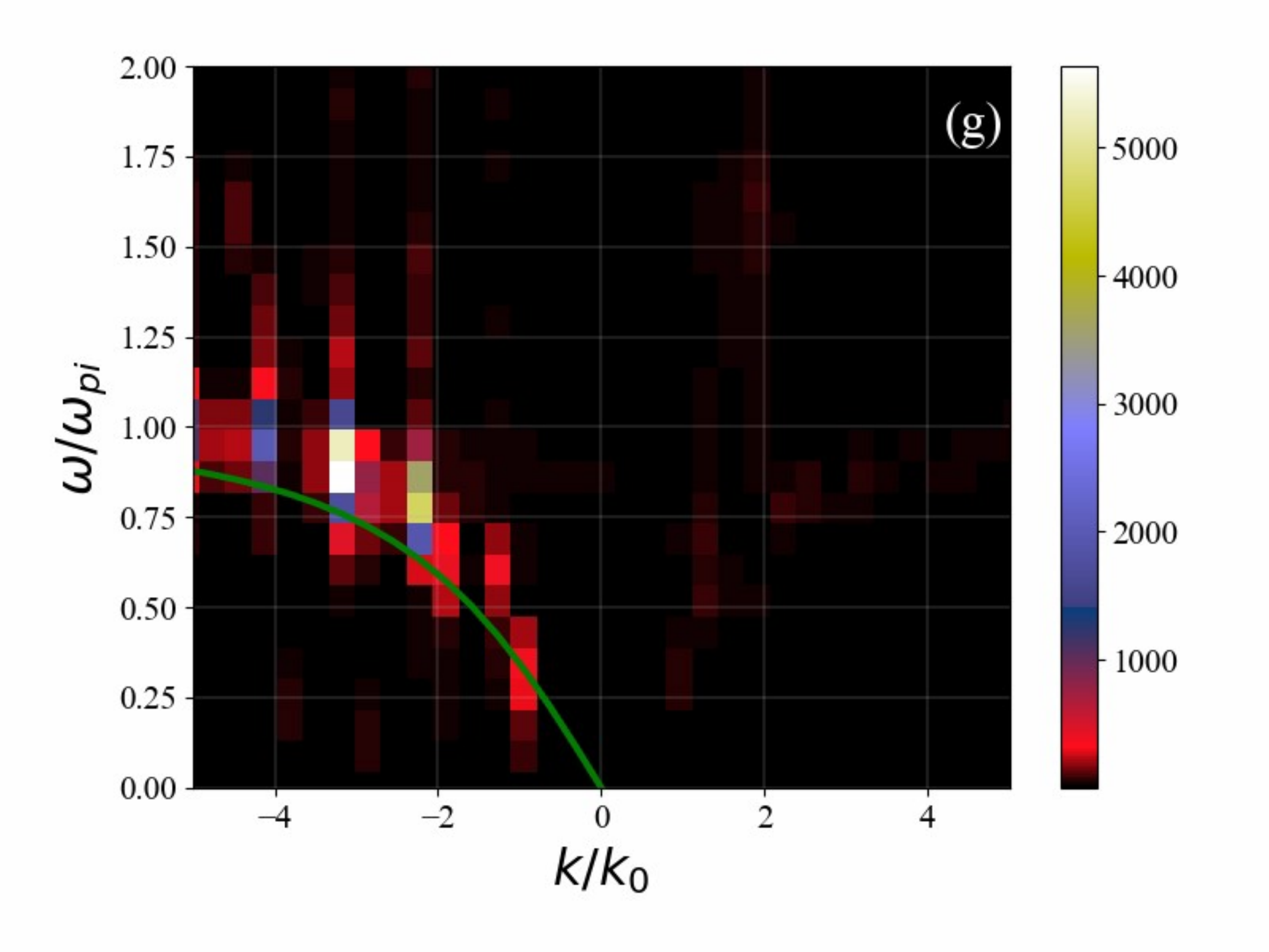}}
\subcaptionbox{\label{fig:Ekw_L005Vl}}{\includegraphics[width=0.41\linewidth]{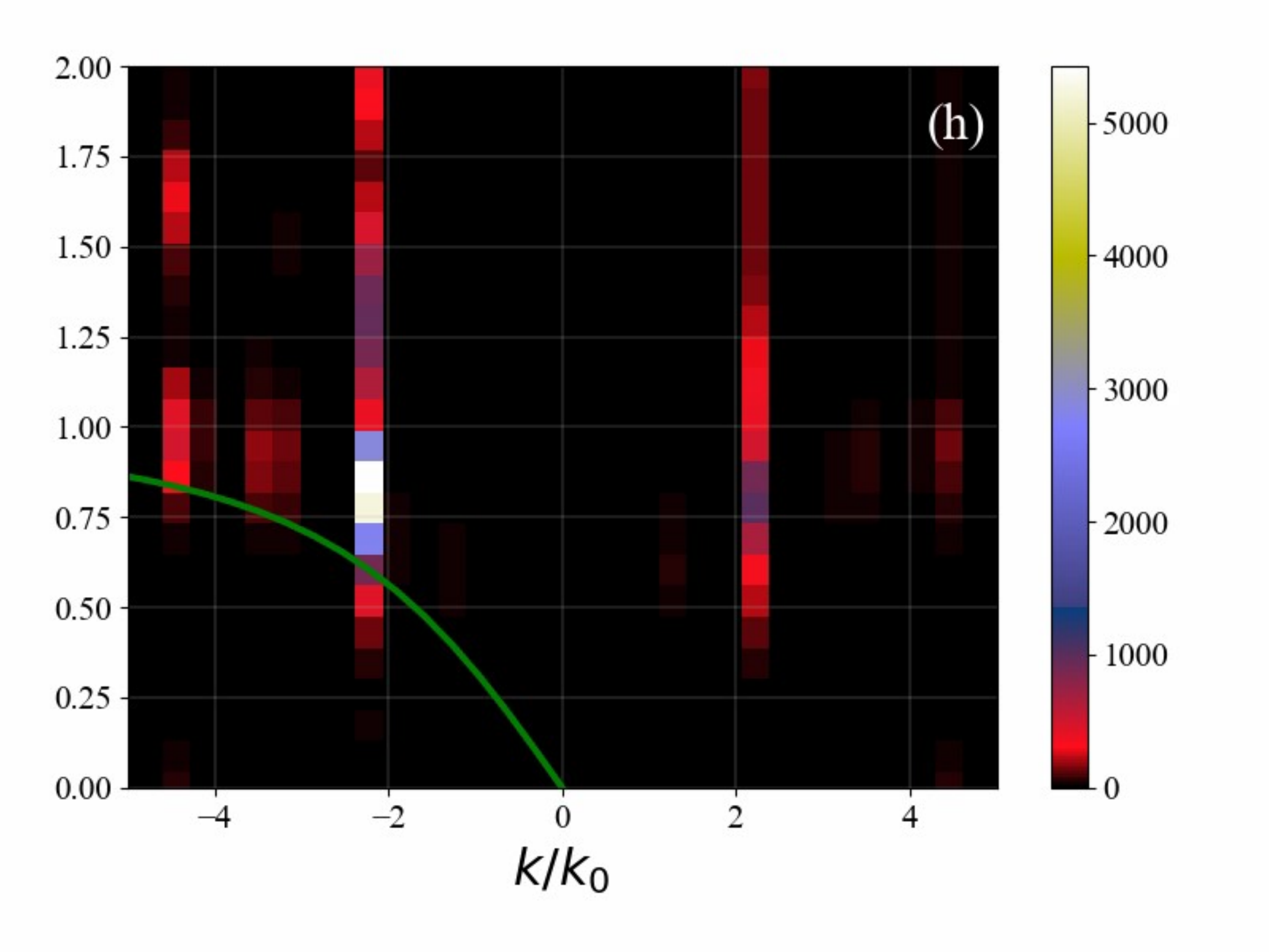}}
\vspace{-1.4cm}
\caption{The frequency spectrum of the electric field. (a), (c), (e),  and (g) are the PIC results ($N_{ppc}=10^4$). (b), (d), (f), and (h) are the Vlasov results. $L=627/k_0$,  $156.8/k_0$, $39.6/k_0$, and $19.8/k_0$ in the first, second, third, and fourth rows respectively. Green lines show the ion-sound dispersion, calculated with the mean temperature of each simulation.\label{fig:Ekw}}  
\end{figure}

\Cref{fig:Ekw} shows the nonlinear frequency spectrum of the PIC and Vlasov simulations for different azimuthal lengths using the two-dimensional FFT. In all cases, the dominant frequency is around $\omega \approx \omega_{pi}$, which is consistent with the observations of Ref.~\onlinecite{janhunen2018nonlinear}. Similar to \Cref{fig:Ek}, the dominant wave vector is around $k\approx k_0$ for $L=627/k_0$ and $L=156.8/k_0$, whereas it is around $k=2k_0$ for $L=39.6/k_0$. For all azimuthal lengths, the amplitude of the dominant mode is much higher in the PIC simulation than the Vlasov simulation. An important difference between the frequency spectra of the PIC and Vlasov simulations is that the backward waves  are clearly seen in the spectra of Vlasov simulations for all lengths. The backward waves propagate in the direction opposite to the initial electron drift and usually result from the strong flattening of the electron velocity distribution function \cite{tavassoli2021backward}. Experimental observations also confirm the existence of these waves in the Hall thruster plasma \cite{tsikata2009dispersion,tsikata2010three}.  These waves are, however, barely noticeable in the spectra of the PIC simulations. The reason for this may be the low amplitude of backward waves makes them more susceptible to being lost in the noise of the PIC simulations. 
In \Cref{fig:Ekw}, the ion-sound dispersion is also shown in all sub-figures. We can see that, in contrast to the ion-sound dispersion, in all simulations with $L\ge 39.6/k_0$, the frequency spectrum remains discrete. Nevertheless, in simulations with $L=39.6/k_0$, some smoothing of the cyclotron peaks is observed. In \Cref{fig:Ekw_L005Vl,fig:Ekw_L005_PIC_1e4}, the azimuthal length is further reduced to $L=19.8/k_0$. In \Cref{fig:Ekw_L005Vl}, the mode $k\approx k_0$ has essentially disappeared, and mode $k\approx 2k_0$ is dominant.  {We note that the first three cyclotron peaks are still visible in \Cref{fig:Ekw_L005Vl}.} Although the frequency spectrum of the Vlasov simulation remains quite discrete, the spectrum of the PIC simulation in \Cref{fig:Ekw_L005_PIC_1e4} shows a significant smoothing and similarity with the ion-sound dispersion. Moreover, the dominant mode in this simulation has a $k\approx 1/(\sqrt{2}\lambda_D)\approx 2.7k_0$, which belongs to the maximum-growth-rate mode of the ion-sound instability. Therefore, it is likely that the combination of a small azimuthal length with the noise of the PIC simulations can induce the transition to  ion-sound turbulence.

\section{The inverse cascade and phase space structures in  the PIC and Vlasov simulations}\label{sec:bunching}

\begin{figure}[htbp]
\centering
\captionsetup[subfigure]{labelformat=empty}
\subcaptionbox{\label{fig:PIC10000_fxvxe_238}}{\includegraphics[width=0.49\linewidth]{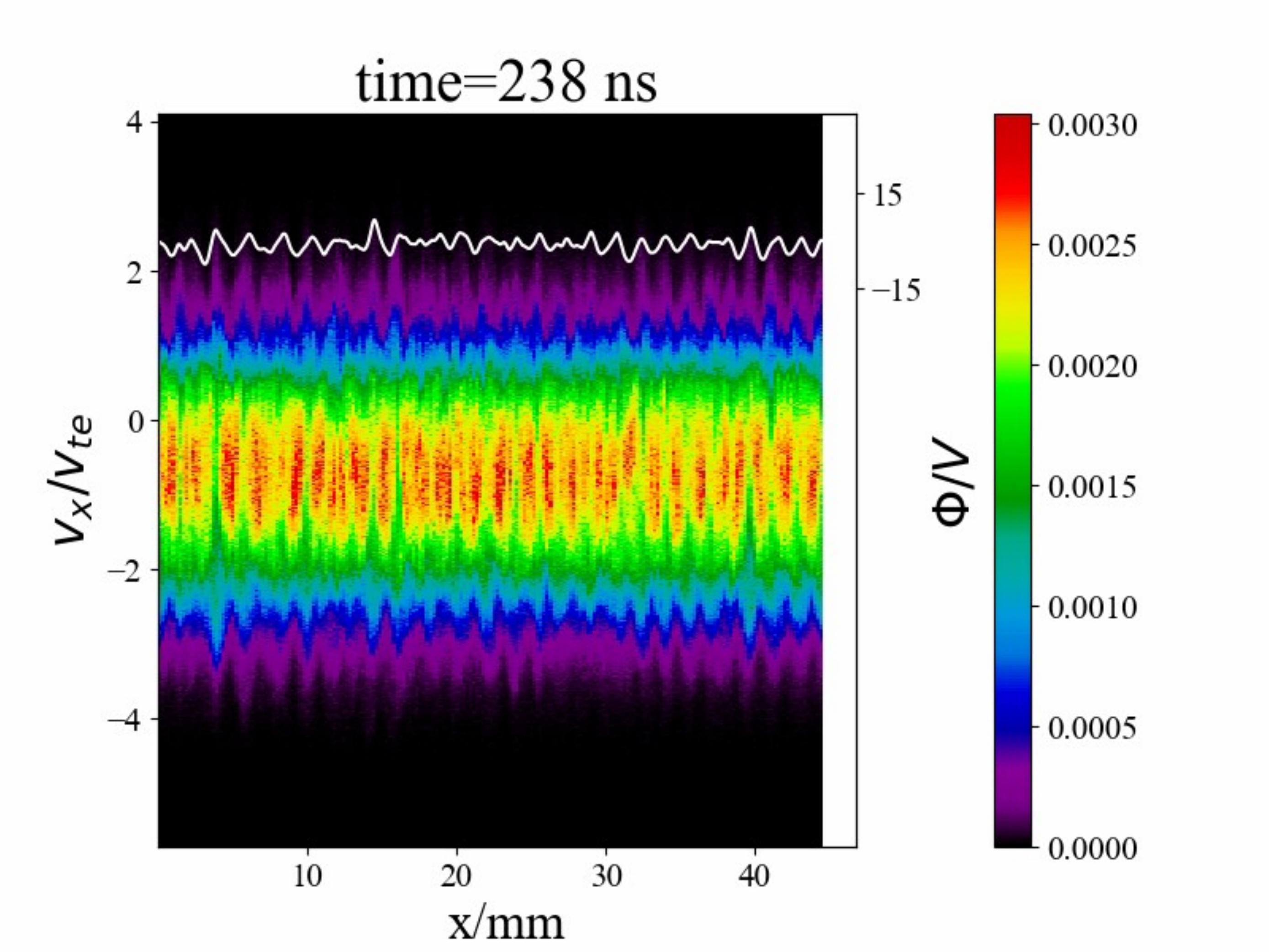}}
\subcaptionbox{\label{fig:PIC10000_fxvxe_797}}{\includegraphics[width=0.49\linewidth]{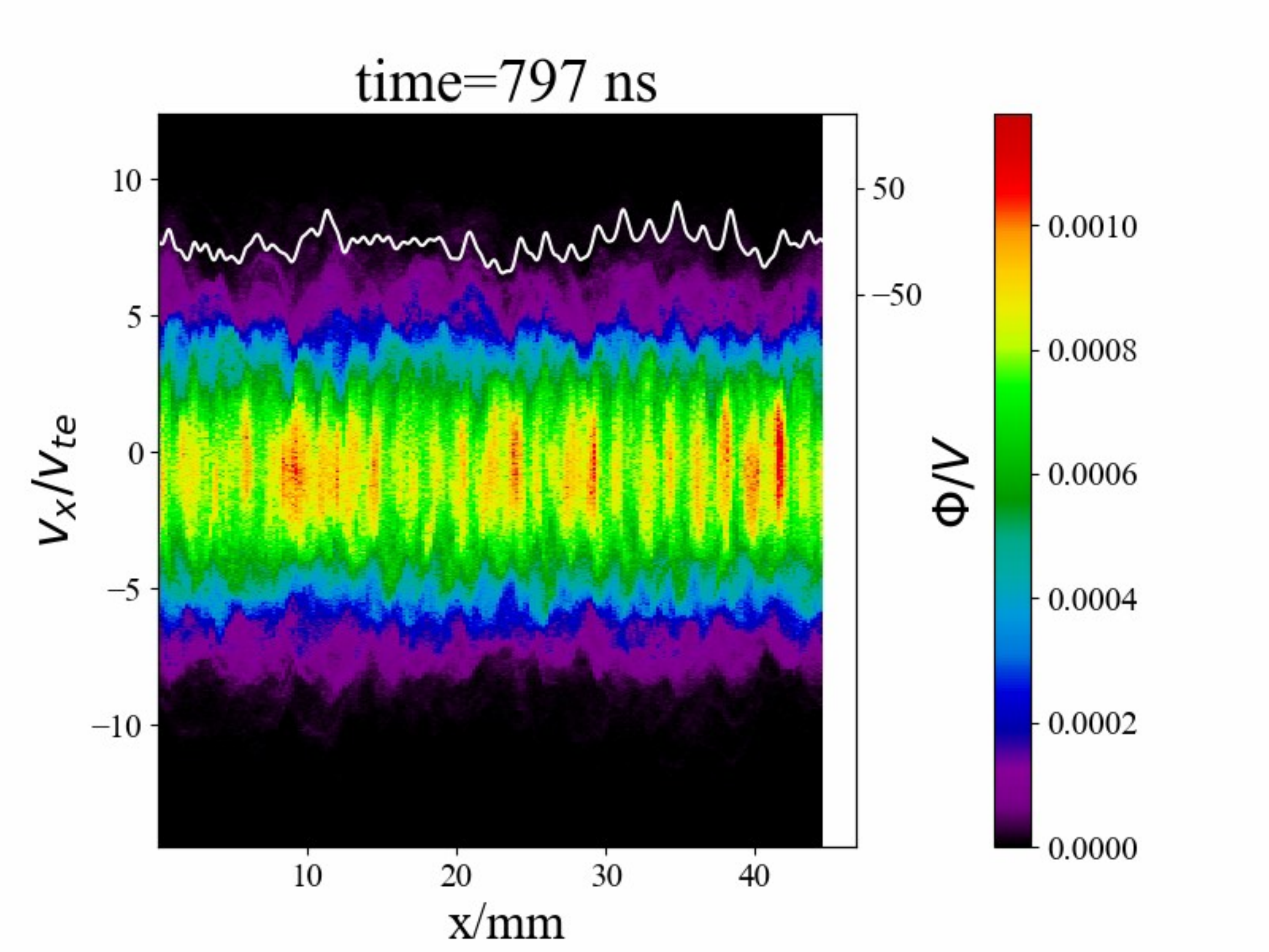}}
\vspace{-1cm}
\subcaptionbox{\label{fig:PIC10000_fxvxe_1325}}{\includegraphics[width=0.49\linewidth]{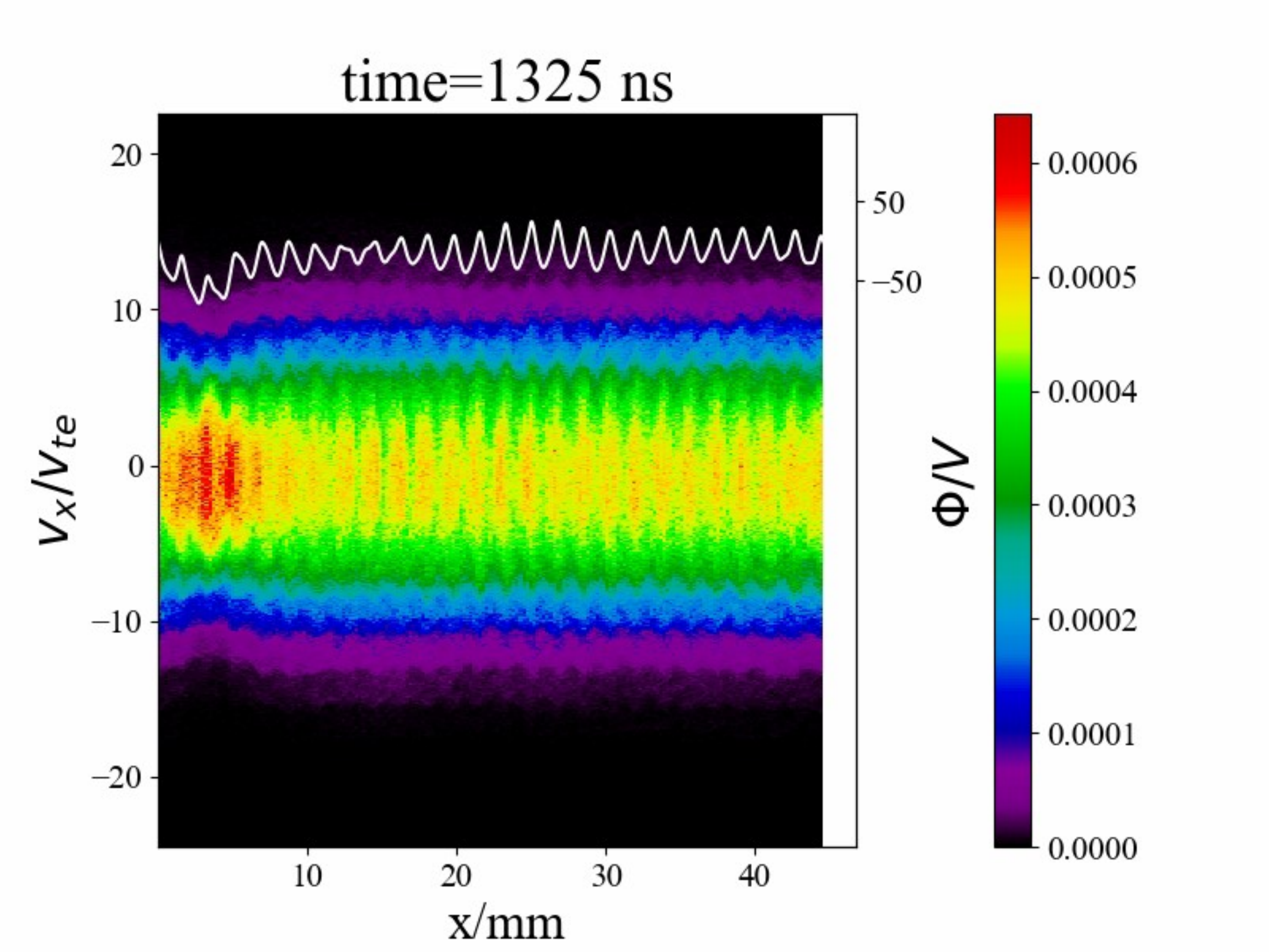}}
\subcaptionbox{\label{fig:PIC10000_fxvxe_1743}}{\includegraphics[width=0.5\linewidth]{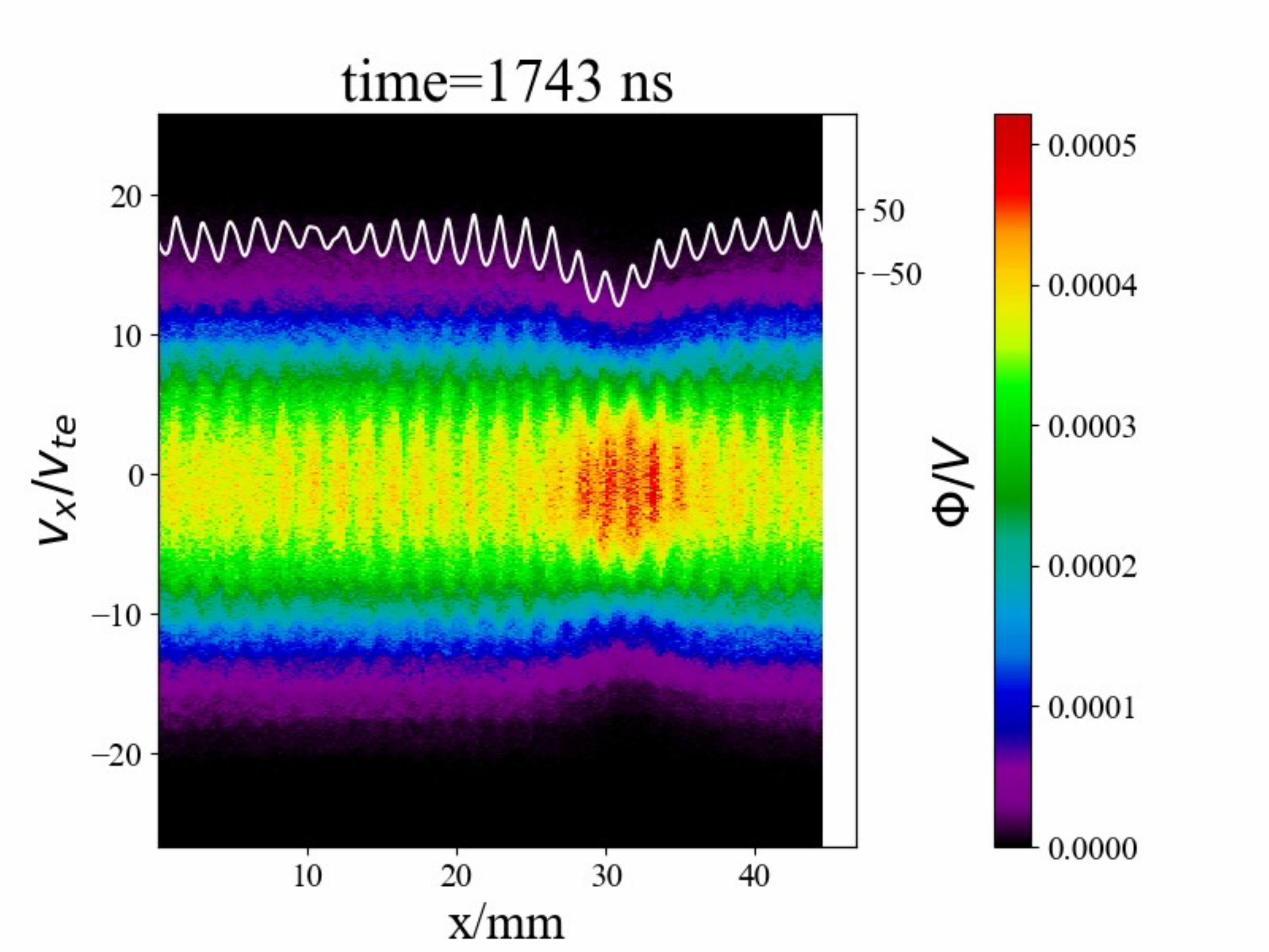}}
\caption{Electron bunches in phase space of the PIC simulation with $L=156.8/k_0=4.456$ cm and $N_{ppc}=10^4$. The white line shows the electrostatic potential ($\phi$).\label{fig:PIC_bunches}}
\end{figure}

Concentration of electrons in particular regions of the phase space form structures (bunches) that are seen in both PIC and Vlasov simulations. In the nonlinear regime of the ECDI, the electron bunches are likely the result of cnoidal waves that are formed in the nonlinear regime (see Refs.~\onlinecite{janhunen2018nonlinear,janhunen2018evolution,tavassoli2022nonlinear}). In contrast to  electron holes, which are associated with local maxima of the electrostatic potential, bunches are associated with local minima. \Cref{fig:PIC_bunches} shows the dynamics of electron bunches in the phase space of the PIC simulation. In \Cref{fig:PIC10000_fxvxe_238}, we see that several electron bunches co-exist, and the electrostatic potential has a cnoidal wave shape at $t=238$ ns. At time around $t=797$ ns, some of these bunches merge, and larger bunches appear in the phase space (e.g.,~around $x\approx 10$ mm). The formation of larger bunches from the smaller ones coincides with the inverse cascade seen in \Cref{fig:Ek_L4_PIC_1e4} and therefore might explain this phenomenon. At $t=1325$ ns (\Cref{fig:PIC10000_fxvxe_1325}), the process of merging has finished, and a ``solitary" bunch is formed around $x=5$ mm. Around this location, a clear global minimum in the shape of a soliton can be seen in the spectrum of electrostatic potential. The length of this soliton is about 8 mm, and  after $t=1325$ ns, it moves with an average velocity of about $ 26\;c_s\approx 7\;\omega_{pi}/k_0\approx0.06\;v_{te}$. 

\begin{figure}[htbp]
\centering
\captionsetup[subfigure]{labelformat=empty}
\subcaptionbox{\label{fig:vl_fxvxe_768}}{\includegraphics[width=0.49\linewidth]{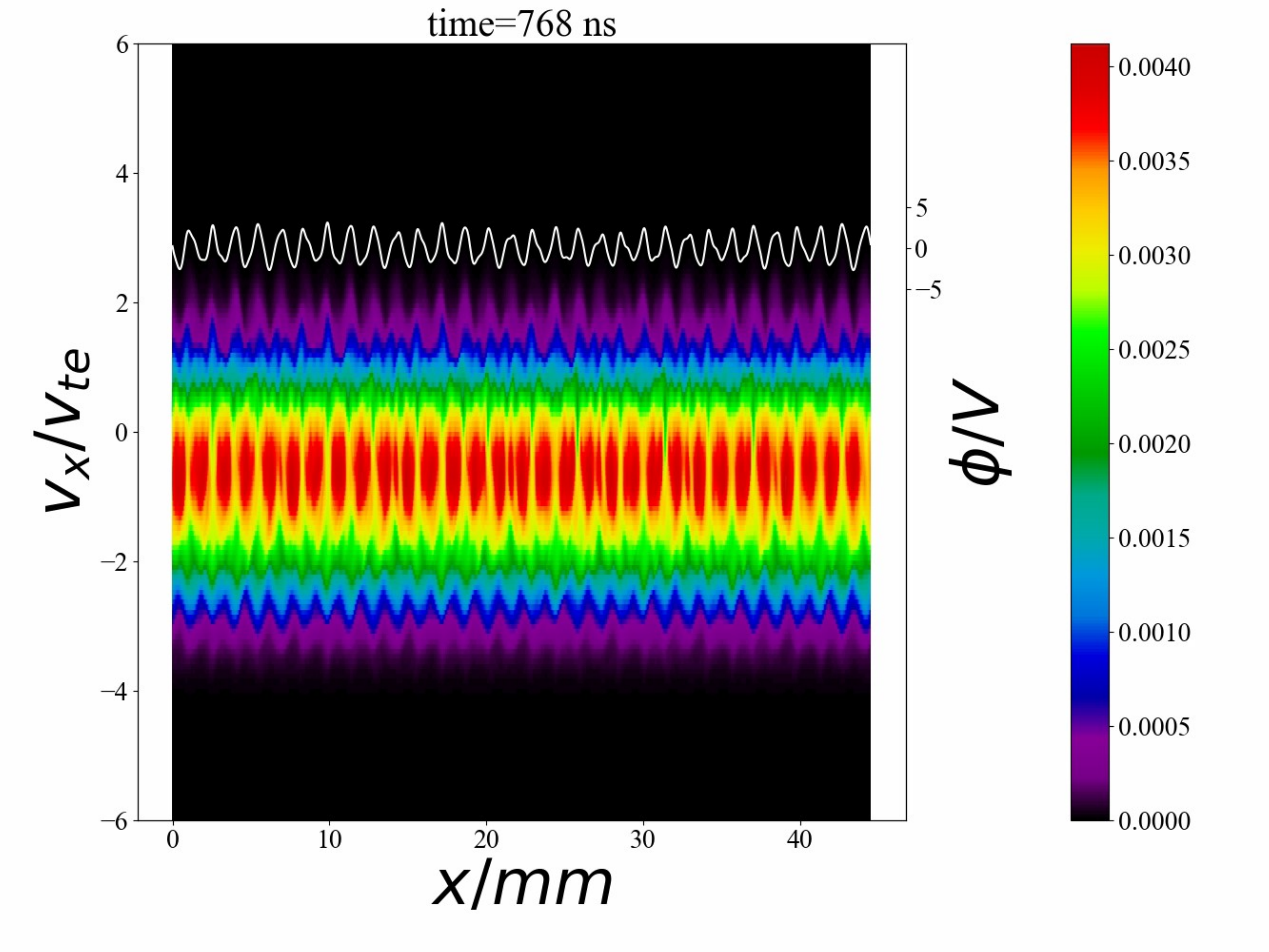}}
\subcaptionbox{\label{fig:vl_fxvxe1196}}{\includegraphics[width=0.49\linewidth]{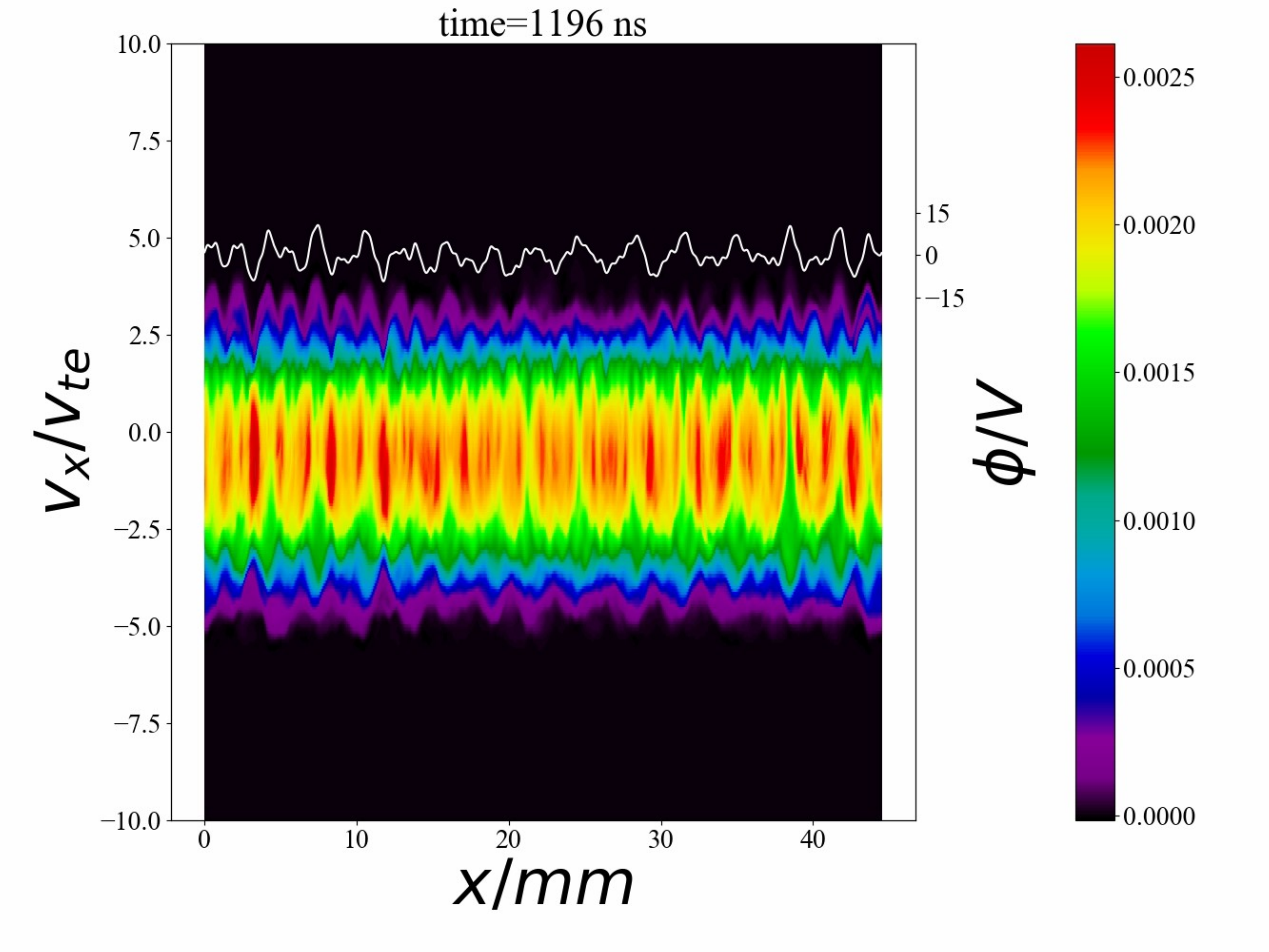}}
\subcaptionbox{\label{fig:vl_fxvxe1974}}{\includegraphics[width=0.49\linewidth]{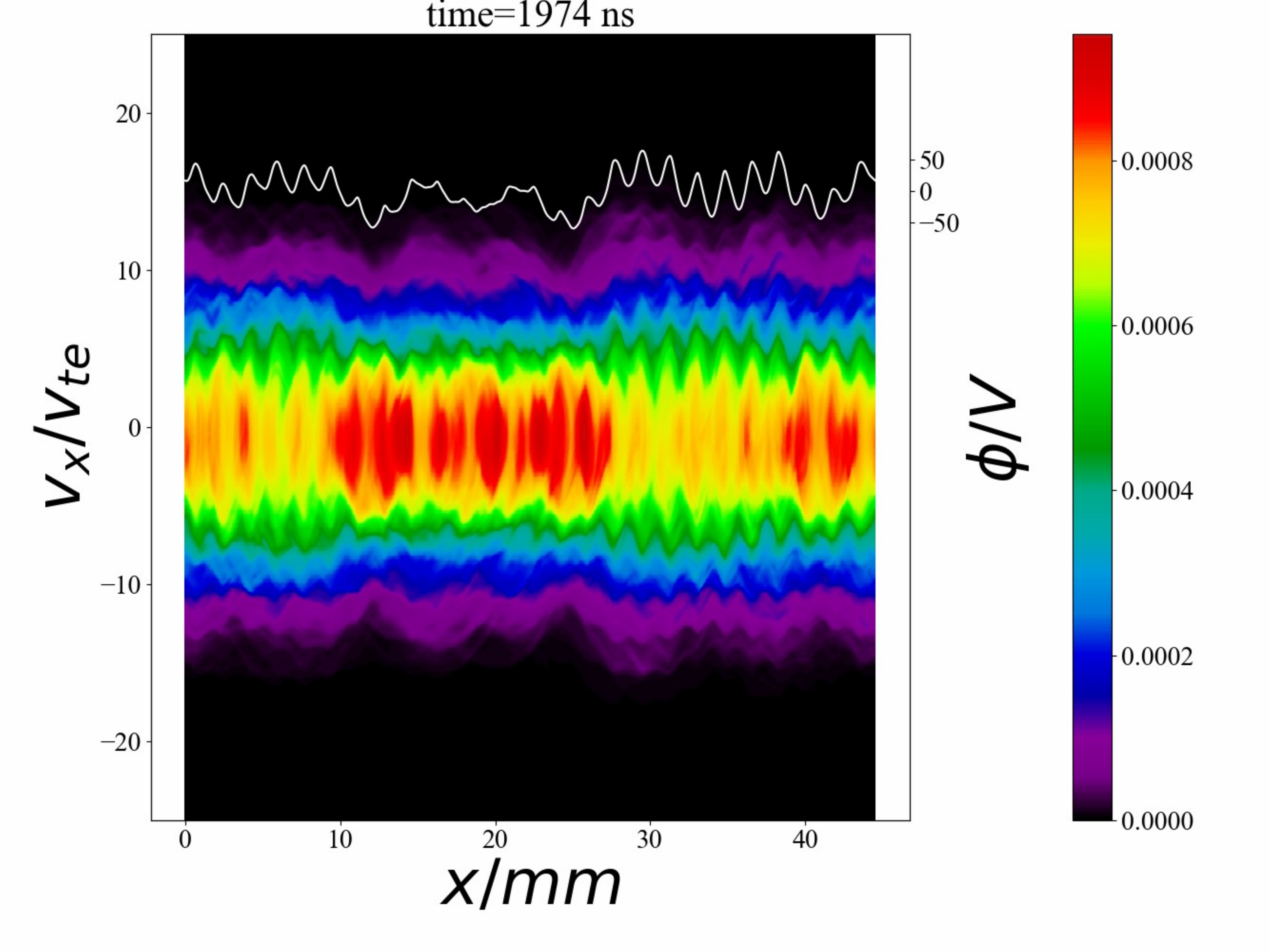}}
\caption{Electron bunches in phase space of the Vlasov simulation with $L=156.8/k_0=4.456$ cm. The white line shows the electrostatic potential ($\phi$).\label{fig:vl_bunches}} 
\end{figure}

Similar to the PIC simulations, electron bunching is observed in the Vlasov simulations during the time that the inverse cascade in happening in \Cref{fig:Ek_L4Vl}. \Cref{fig:vl_fxvxe_768} shows the electron phase space at $t=768$ ns. At this time, many electron bunches are observed, some of which merge by $t=1196$ ns (\Cref{fig:vl_fxvxe1974}). \Cref{fig:vl_fxvxe1974} shows that the process of bunching continues up to the last stage of the simulation. However, in contrast to the PIC simulation, no clear soliton is formed, and the wave has its cnoidal shape with several bunches until the last stage of the simulation. This discrepancy between the PIC and Vlasov simulations can be also explained based on the spectra in \Cref{fig:Ek}. In the $k$-spectrum of the PIC simulation in \Cref{fig:Ek_L4_PIC_1e4}, we can see a gap between $k\approx 0.2k_0$ to $k\approx 0.9k_0$ after $t\approx 1000$ ns. With no significant amplitudes in this gap, the small-$k$ modes of the PIC spectrum become similar to the spectrum of solitary waves. In contrast, this gap is not observed in the spectrum of the corresponding Vlasov simulation (\Cref{fig:Ek_L4Vl}). 

\section{The electrostatic energy and heating in PIC and Vlasov simulations}\label{sec:es_heating}

\begin{figure}[htbp]
\centering
\captionsetup[subfigure]{labelformat=empty}
\subcaptionbox{\label{fig:Ep_PIC_VL}}{\includegraphics[width=0.49\textwidth]{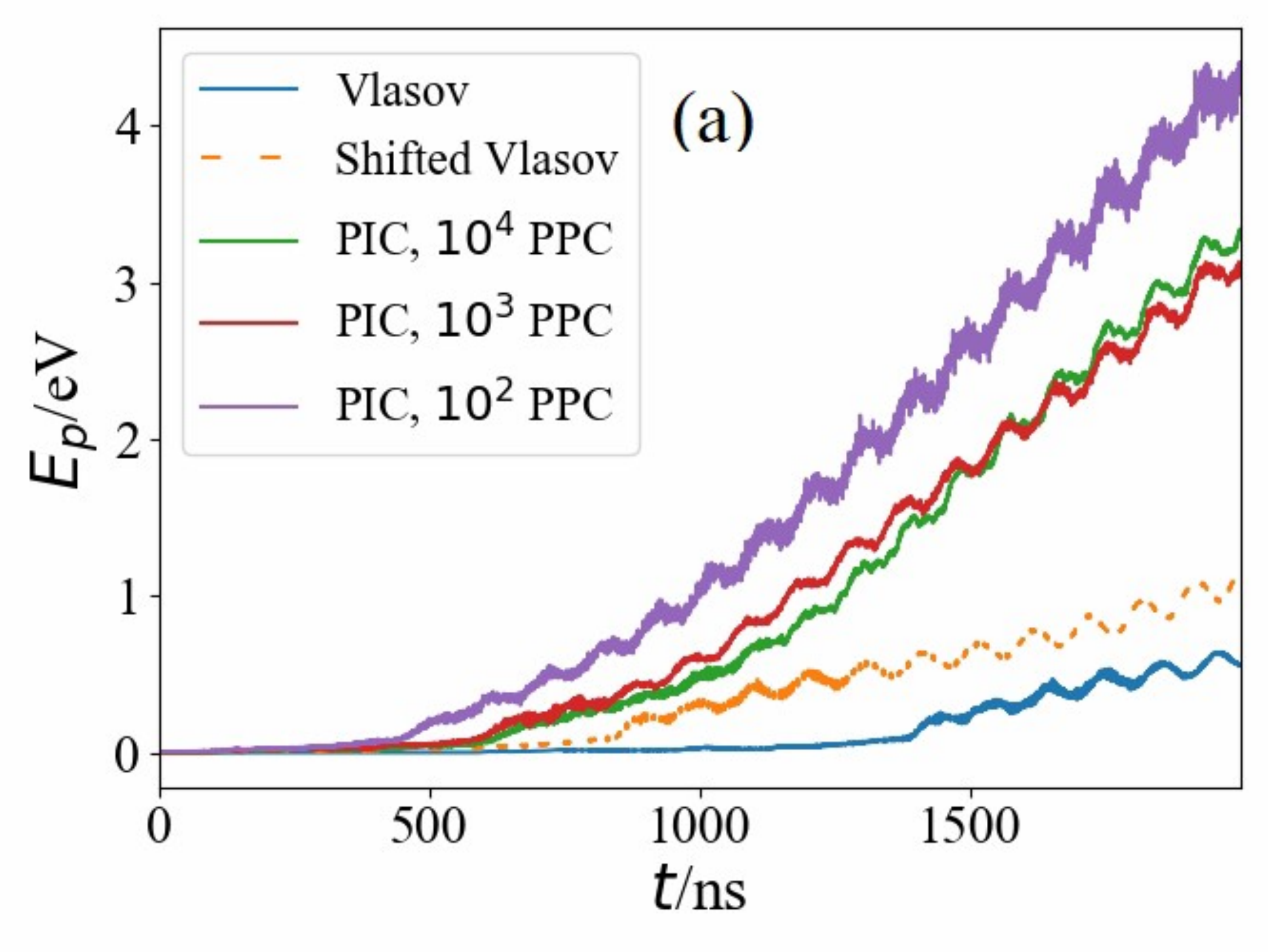}}
\subcaptionbox{\label{fig:avg_Txe_PIC_VL}}{\includegraphics[width=0.49\textwidth]{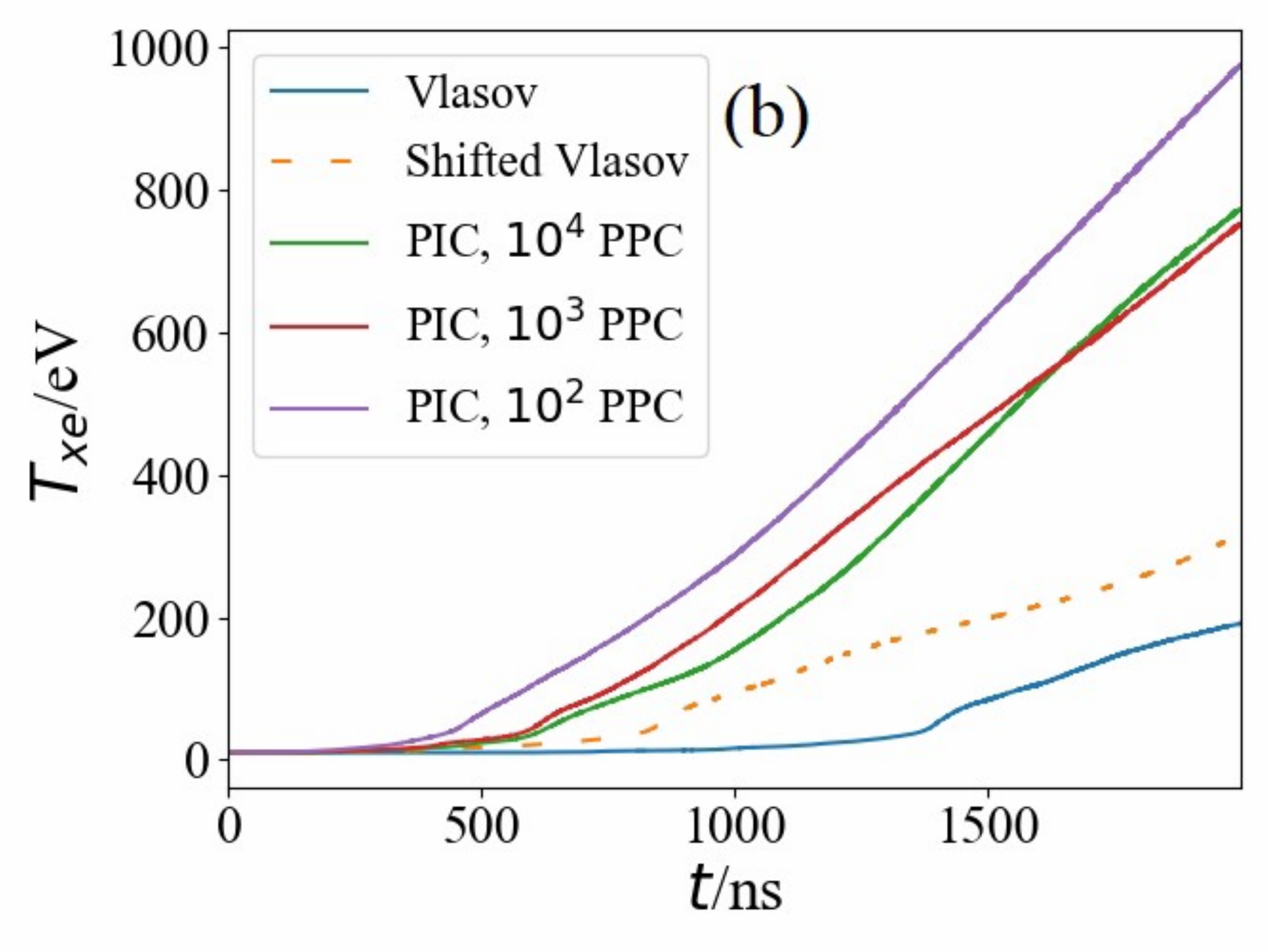}}
\subcaptionbox{\label{fig:Ew_PIC_VL}}{\includegraphics[width=0.49\textwidth]{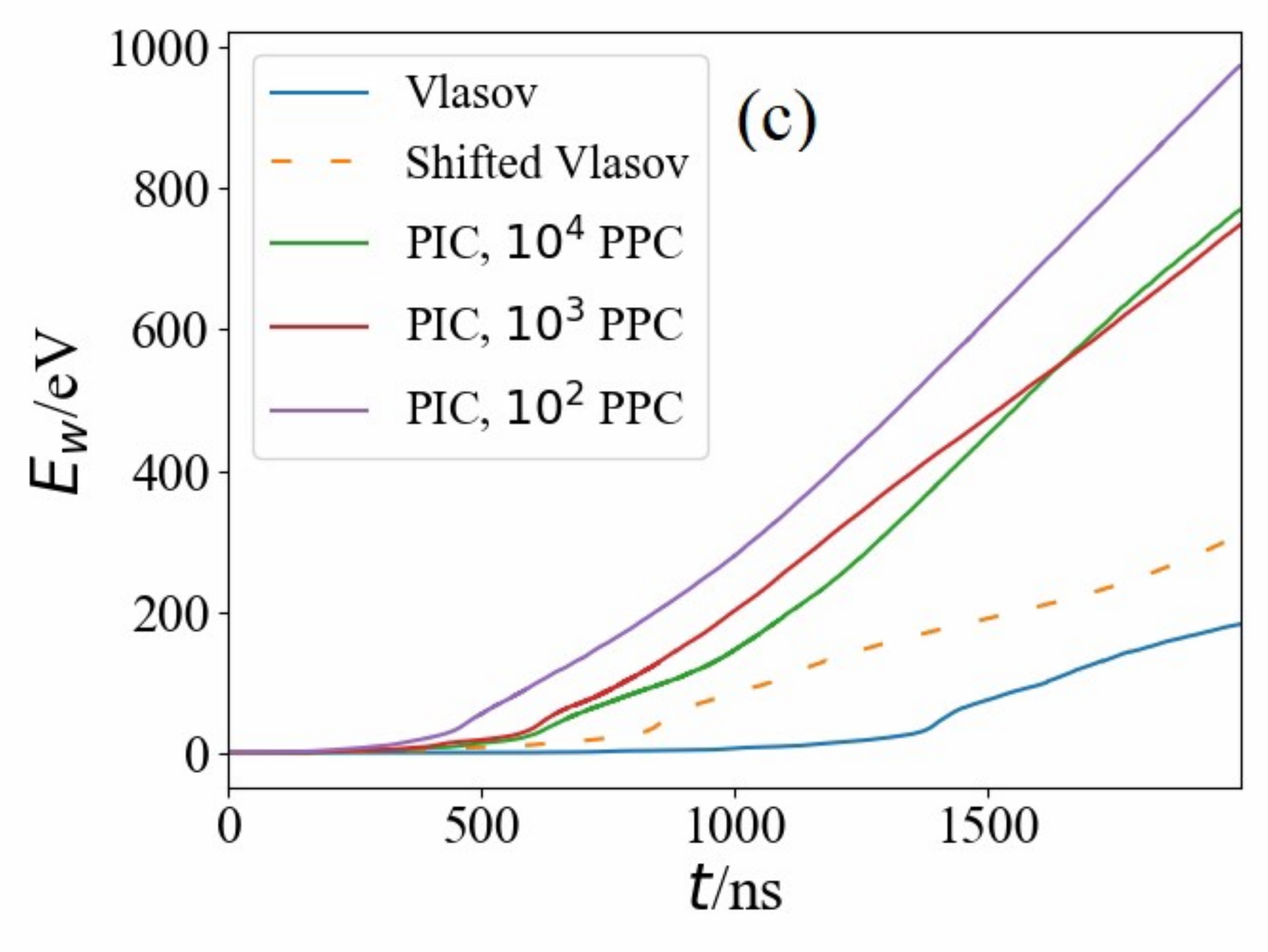}}
\caption{a) The electrostatic energy b) The electron temperature c) $E_w$ as defined by \Cref{eq:ew} in PIC and Vlasov simulations, using $L=158.4/k_0$.\label{fig:Txe_ES_PIC_Vl}}  
\end{figure}

  In our simulations, the electrostatic energy is calculated as $E_p=\frac{\epsilon_0}{2n_0L}\int_0^LE_x^2\dd x$, and the (spatially averaged) temperature in the $x$-direction is calculated as $T_{xe}=\frac{m_e}{n_eL}\int_0^L \int (v_x-V_x)^2f_e(x,v_x,v_z)\dd v_x\dd v_z \dd x$. In all simulations, we observed that the temperatures in the $x$- and $z$-directions are basically the same, and therefore, we only report $T_{xe}$. \Cref{fig:Ep_PIC_VL,fig:avg_Txe_PIC_VL} show the electrostatic energy and the temperature, respectively. In these figures, we do not use the logarithmic scale as in \Cref{fig:PIC10000_t_Ek,fig:vlasov_t_Ek,fig:Ep_PIC_VL_Log} in order to clearly show the differences in the nonlinear regime.  In each figure, the results of the PIC simulations with different $N_{ppc}$ are plotted along with the results of the Vlasov simulation. \Cref{fig:Ep_PIC_VL,fig:avg_Txe_PIC_VL} show that in the linear regime and early nonlinear regime (up to $t\approx 600$ ns), the PIC and Vlasov results remain in good agreement. After this time, however, the electrostatic energy and the temperature in the Vlasov simulation are much smaller than in the PIC simulations even when $N_{ppc}=10^4$. Because the Vlasov simulation starts from a much lower initial fluctuation amplitude, it does not go as far into the nonlinear regime as the PIC simulations. Therefore, to reduce confounds in the comparisons, we also plotted a ``shifted Vlasov" in \Cref{fig:Ep_PIC_VL,fig:avg_Txe_PIC_VL}. The shifted Vlasov is the same Vlasov simulation but with the time axis shifted by $t=546.259$ ns. The size of this shift is the smallest required to make the initial electrostatic energy of the Vlasov the same as the PIC with $N_{ppc}=10^4$. Therefore, the PIC and the shifted Vlasov start from the same level of fluctuation. Although this shift slightly raises the electrostatic energy and the temperature, they remain much less that in the PIC simulations for the most of the nonlinear regime. We also tried the same shift on the PIC simulations with the higher $N_{ppc}$ to make their initial amplitude similar to the ones with lower $N_{ppc}$; however, no significant change was observed in the results and hence are omitted.
  
  In \Cref{fig:Ep_PIC_VL,fig:avg_Txe_PIC_VL}, we also see that when $N_{ppc}$ in the PIC simulation increases from $10^2$ to $10^3$, both electrostatic energy and electron temperature significantly decrease. When $N_{ppc}$ increases further to $10^4$, however, these quantities converge to approximately the same values.  Another observation in \Cref{fig:Ep_PIC_VL,fig:avg_Txe_PIC_VL} is that the rate of  increase in the electron temperature closely tracks the electrostatic energy. This observation is at odds with the predictions of the ion-sound turbulence theory in Refs.~\onlinecite{lampe1971nonlinear,lampe1972theory}. We note that this discrepancy might be because of the difference in the setup of problems. Unlike Refs.~\onlinecite{lampe1972theory}, we have an energy input that prevents the full saturation of the growth. Also, the ions in our case are heavy Xe ions, for which we did not observe significant trapping or heating during the time of our simulations \footnote{For the ions, however, we observed a significant modification of the distribution function and density spikes similar to what is observed in Ref.~\onlinecite{janhunen2018nonlinear}.}.

\begin{figure}[htbp]
\centering
\captionsetup[subfigure]{labelformat=empty}
\subcaptionbox{\label{fig:PIC-t_Ep_Lcomp}}{\includegraphics[width=0.49\textwidth]{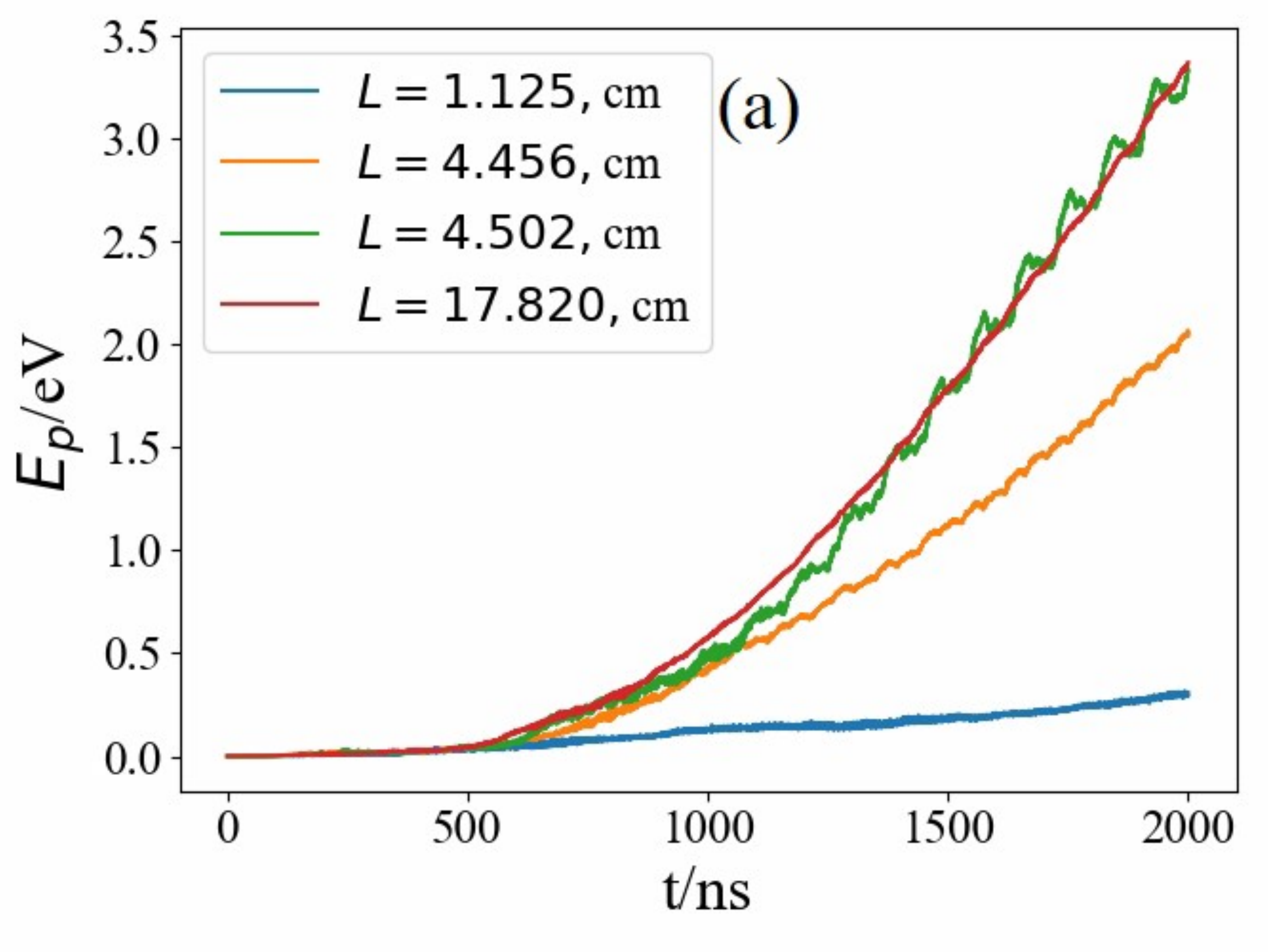}}
\vspace{-1cm}
\subcaptionbox{\label{fig:vl_t_ES_Lcomp}}{\includegraphics[width=0.49\textwidth]{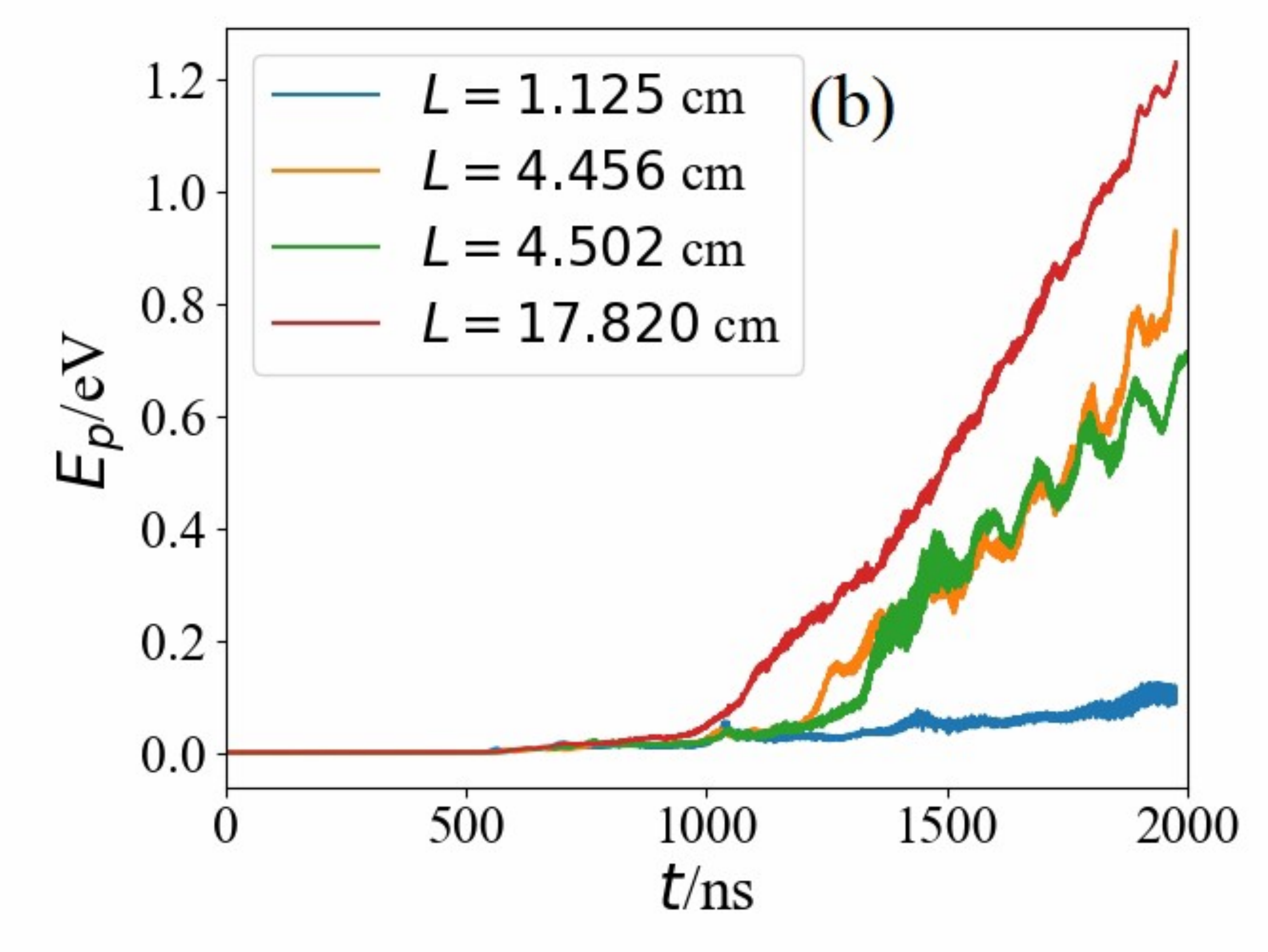}}
\vspace{-1cm}
\subcaptionbox{\label{fig:PIC-t_Tex_Lcomp}}{\includegraphics[width=0.49\textwidth]{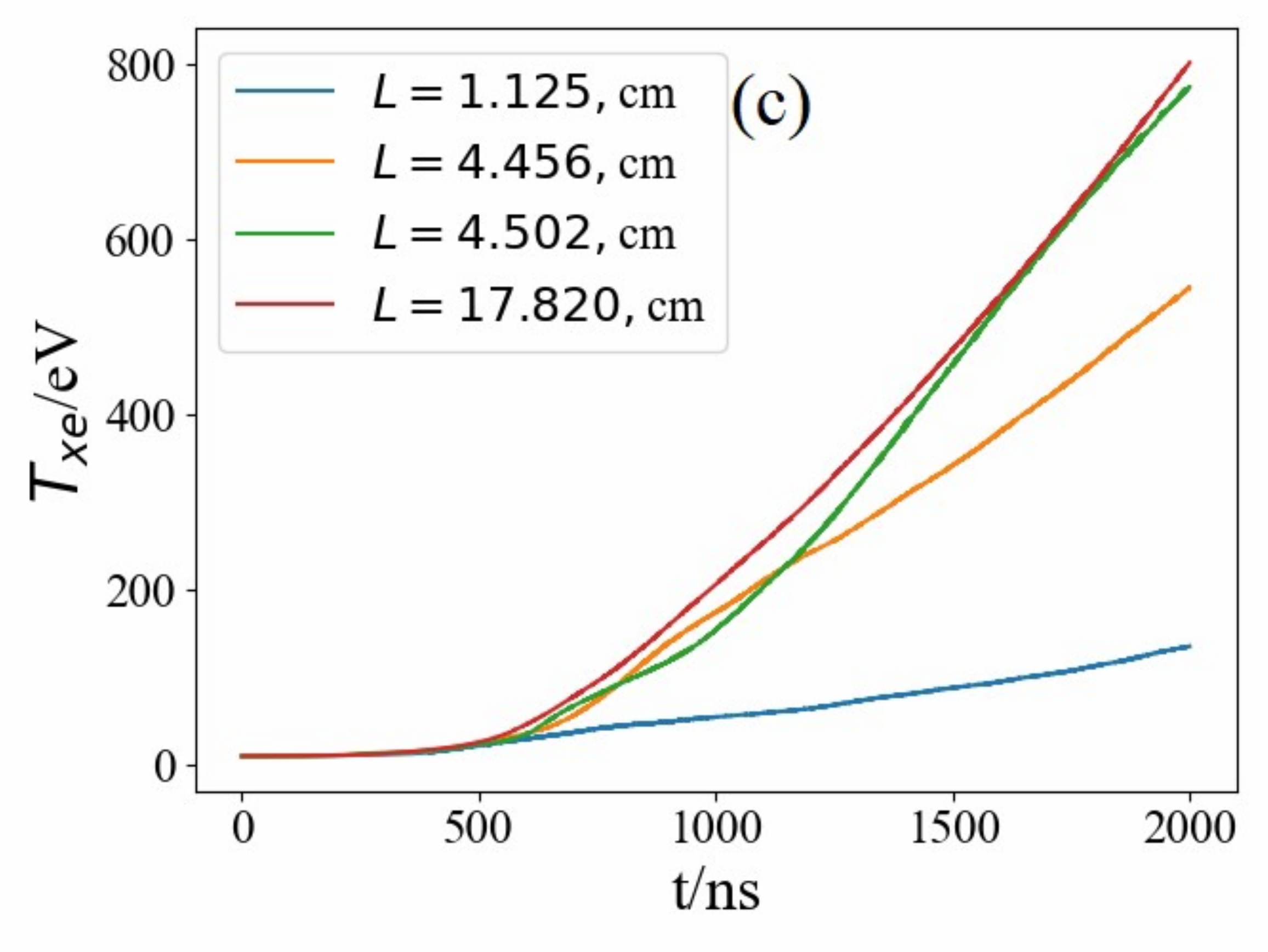}}
\subcaptionbox{\label{fig:vl_t_Txe_Lcomp}}{\includegraphics[width=0.49\textwidth]{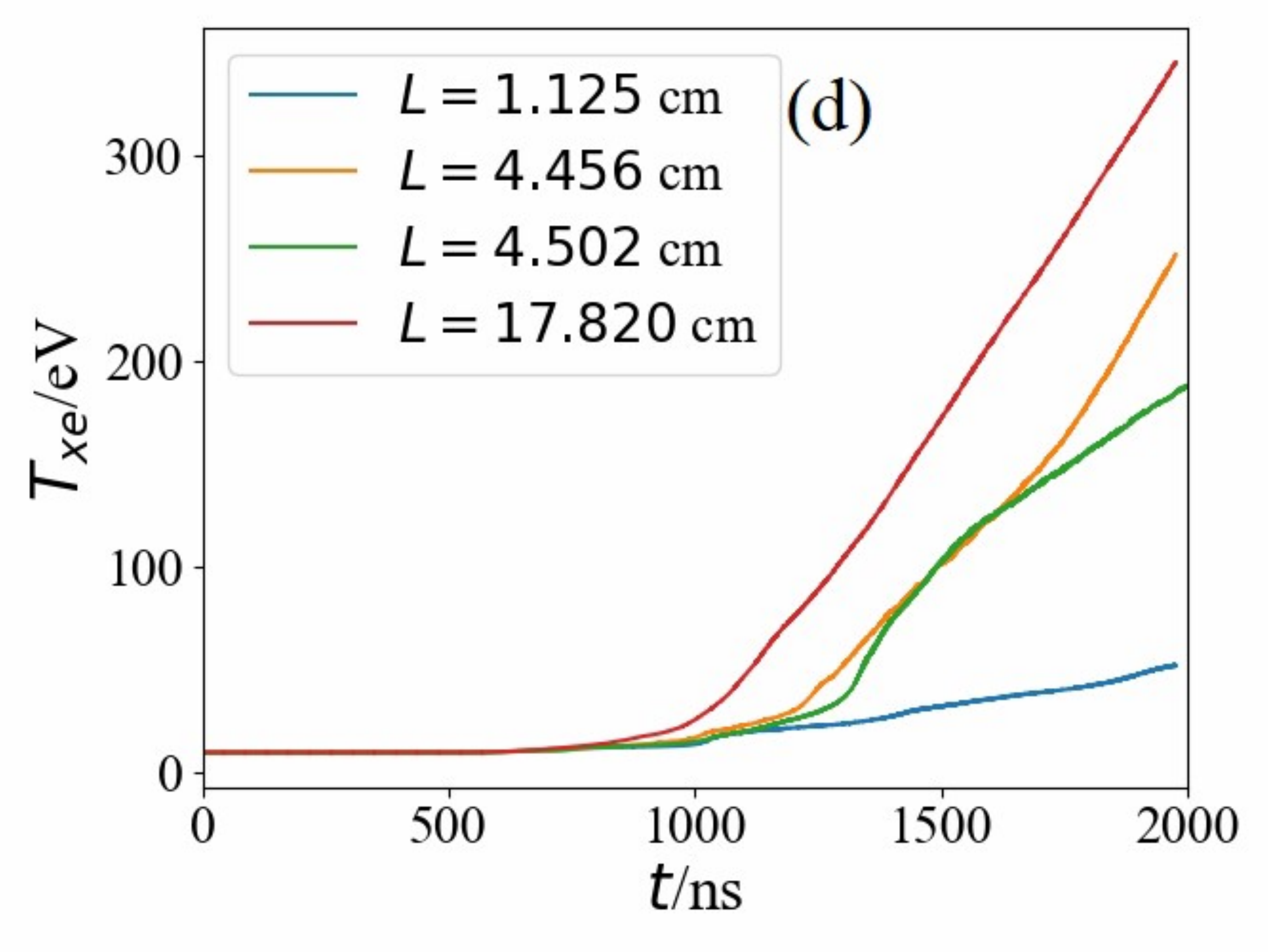}}
\caption{The effect of the azimuthal length on the electrostatic energy ($E_p$) and the electron temperature ($T_{xe}$). (a) and (c) are the PIC results ($N_{ppc}=10^4$). (b) and (d) are the Vlasov results. \label{fig:Txe_ES_Lcomp}} 
\end{figure}

 The effect of the azimuthal length on the electrostatic energy and temperature is shown in \Cref{fig:Txe_ES_Lcomp}. We can see that both quantities increase with increasing azimuthal length in both PIC and Vlasov simulations.  {This can be partly due to the increase in the number of the unstable resonant modes in the simulations with longer azimuthal length (increased resolution in the Fourier space), as shown in \Cref{sec:linear_regime}, that leads to an increase in the total electrostatic energy. Another reason for this increase can be the further extension of the spectrum to long wavelengths that is made possible by increasing $L$.} An important observation in \Cref{fig:Txe_ES_Lcomp} is the sensitivity of the heating and electrostatic energy to the azimuthal length. \Cref{fig:PIC-t_Ep_Lcomp,fig:PIC-t_Tex_Lcomp} show that, in the PIC simulation, a small variation of azimuthal length from $L=156.8/k_0$ to $L=158.4/k_0$ leads to a significant increase in electrostatic energy and electron temperature in the deep nonlinear regime. This sensitivity is, however, much less in the Vlasov simulations, as seen in \Cref{fig:vl_t_ES_Lcomp,fig:vl_t_Txe_Lcomp}. In \Cref{fig:Txe_ES_Lcomp}, we can also see that for all azimuthal lengths used, the electrostatic energy and electron heating in the PIC simulations are much greater than in the Vlasov simulations.
 
 In \Cref{fig:Txe_ES_PIC_Vl,fig:Txe_ES_Lcomp}, we see that neither the electrostatic energy nor the temperature completely saturate in the nonlinear regime. This observation is consistent with other PIC simulations of ECDI \cite{janhunen2018nonlinear,janhunen2018evolution,lafleur2016theory1}. In Ref.~\onlinecite{lafleur2016theory1},
 complete saturation is achieved due to the ion trapping only when the ``virtual axial length model" is used. In this model, particles are replaced when they are displaced beyond a given length in the axial direction. In the absence of the virtual axial length model, the constant axial electric field provides an energy reservoir for unlimited growth of the electrostatic energy and the temperature. The energy provided from this reservoir between the time 0 and $t$ is 
 \begin{equation}
    E_w= E_0\int_0^t\int_0^L J_{ze}(x,t)\dd x \dd t,
    \label{eq:ew}
 \end{equation}
 where $J_{ze}=-e\int v_zf_e(x,v_x,v_z) \dd v_x\dd v_z$ is the axial electron current (the anomalous current). $E_w$ is essentially wasted in the Hall thrusters because it does not contribute to the produced thrust.  {$E_w$ is shown in \Cref{fig:Ew_PIC_VL} for the PIC and Vlasov simulations. We see that in all simulations the $E_w$ remains relatively close to the electron temperature. This is in fact expected from the conservation of energy because, in the deep nonlinear regime, the electron heating is the dominant energy output of the system and $E_w$ is the only energy input.}

\section{The anomalous electron transport}\label{sec:anomalous}

\begin{figure}[htbp]
\centering
\captionsetup[subfigure]{labelformat=empty}
\subcaptionbox{\label{fig:PIC10000_Jze}}{\includegraphics[width=0.49\linewidth]{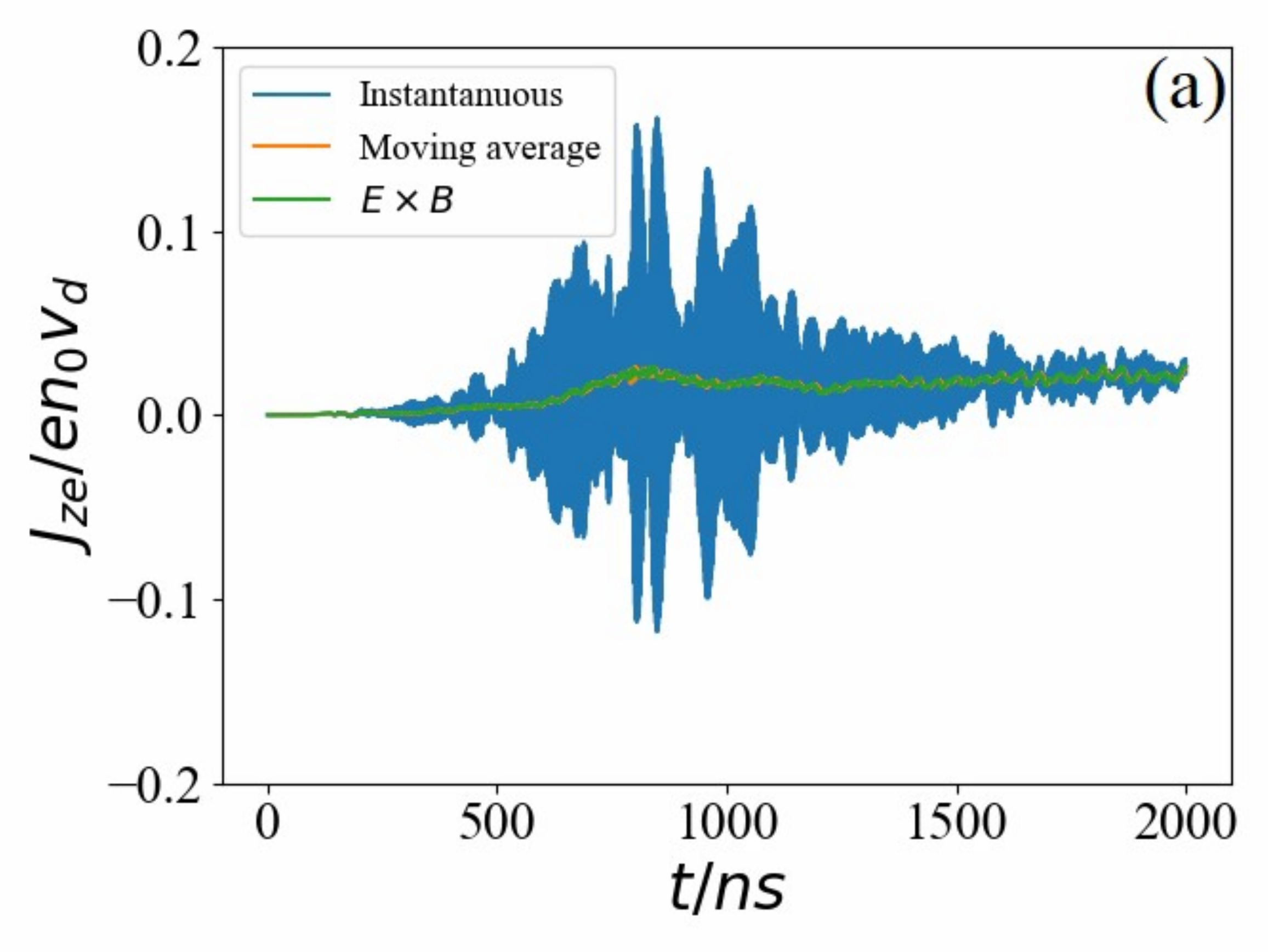}}
\subcaptionbox{\label{fig:Vlasov_Jze}}{\includegraphics[width=0.49\linewidth]{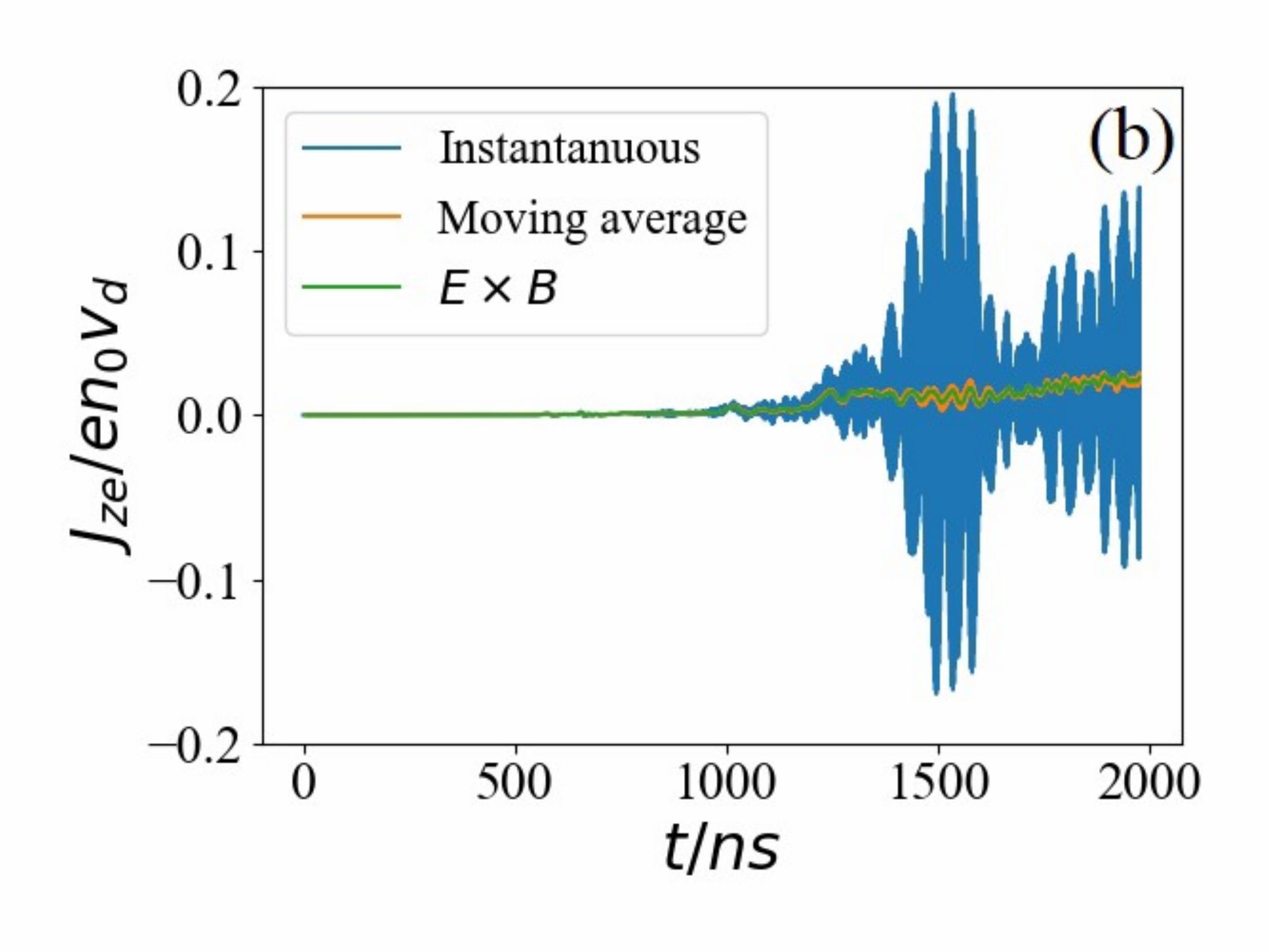}}
\caption{The electron anomalous current using $L=156.8/k_0$ in a) PIC with $10^4$ particles per cell b) Vlasov.}
\label{fig:jze_PIC_VL}
\end{figure}

\begin{figure}[htbp]
\centering
\includegraphics[width=0.49\textwidth]{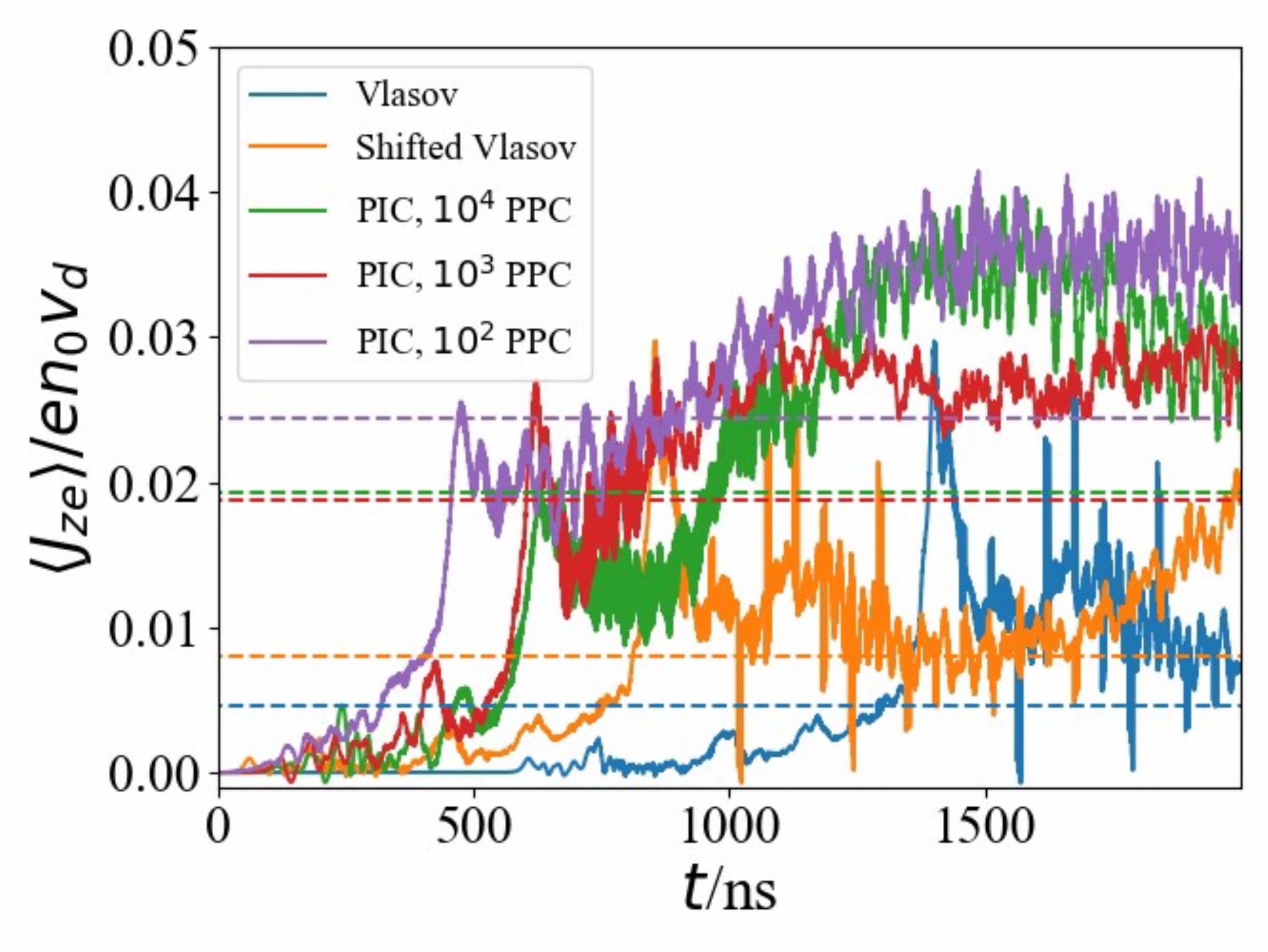}
\caption{The moving average of the electron anomalous current, $L=158.4/k_0$. The horizontal dashed lines show the total time average of each curve.} 
\label{fig:mvavg_Jze_PIC_VL}
\end{figure}

 \Cref{fig:Vlasov_Jze,fig:PIC10000_Jze} show the spatially averaged anomalous current for the Vlasov and PIC simulations, respectively. In these figures, the $E\times B$ current is calculated as $J_{E\times B}=\frac{-e\expval{n_eE_x}}{B_0}$, where the averaging is over the azimuthal length and a large time window. The $E\times B$ current is quite close to the moving average of the anomalous current $\expval{J_{ze}}$. This effect was also observed in all other simulations performed in this study with both PIC and Vlasov methods \cite{lafleur2016theory1,lafleur2016theory2}. \Cref{fig:mvavg_Jze_PIC_VL} shows $\expval{J_{ze}}$ for the PIC and Vlasov simulations.  Because the wasted energy is proportional to the total time average of $\expval{J_{ze}}$ (\Cref{eq:ew}), this quantity is also shown as horizontal lines in  \Cref{fig:mvavg_Jze_PIC_VL}. Another reason for plotting these lines is that, in the absence of a complete saturation, they help us better compare the results of different simulations. In \Cref{fig:mvavg_Jze_PIC_VL}, we see that similar to the electrostatic energy and electron temperature, the total time average of the $\expval{J_{ze}}$ is much smaller in the Vlasov simulation than the PIC simulations. During the nonlinear regime, the $\expval{J_{ze}}$ in the Vlasov simulation also  generally remains much smaller than in the PIC simulations (except temporarily at $t\approx 1450$ ns). The value of $\expval{J_{ze}}$ of the PIC simulations with $10^3$ and $10^4$ particles per cell are close to each other, whereas the anomalous current of the simulation with $10^2$ particles per cell is slightly larger than the other two. Similar to \Cref{fig:Ep_PIC_VL,fig:avg_Txe_PIC_VL}, the shifted anomalous current in the Vlasov is also shown on \Cref{fig:mvavg_Jze_PIC_VL}, but it does not significantly affect the  mentioned comparison of the PIC and Vlasov simulations.
 
\begin{figure}[htbp]
\centering
\captionsetup[subfigure]{labelformat=empty}
\subcaptionbox{\label{fig:PIC-t_mave_Jze_Lcomp}}{\includegraphics[width=0.49\linewidth]{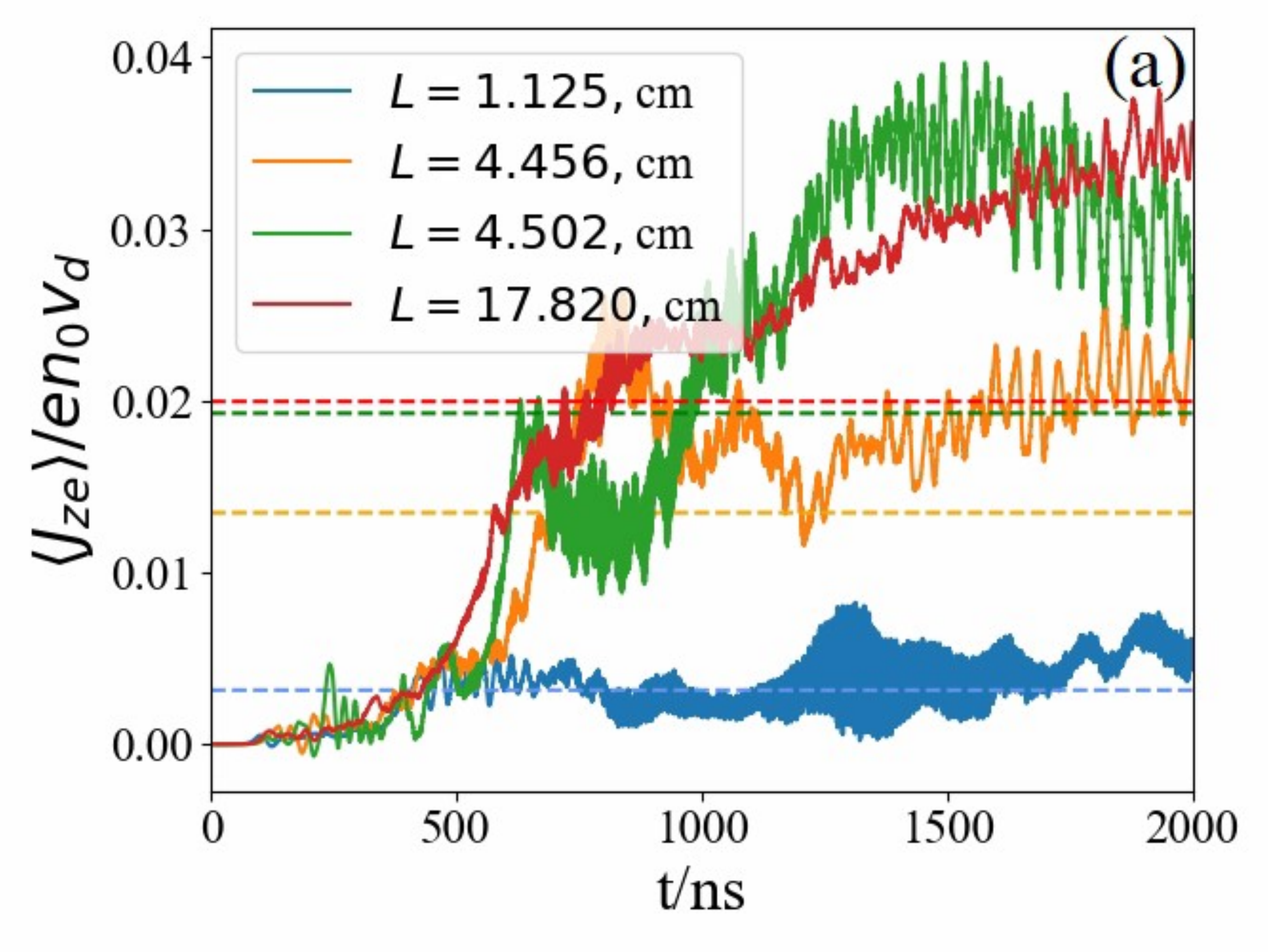}}
\subcaptionbox{\label{fig:vl_t_mvavg_Jze_Lcomp}}{\includegraphics[width=0.49\linewidth]{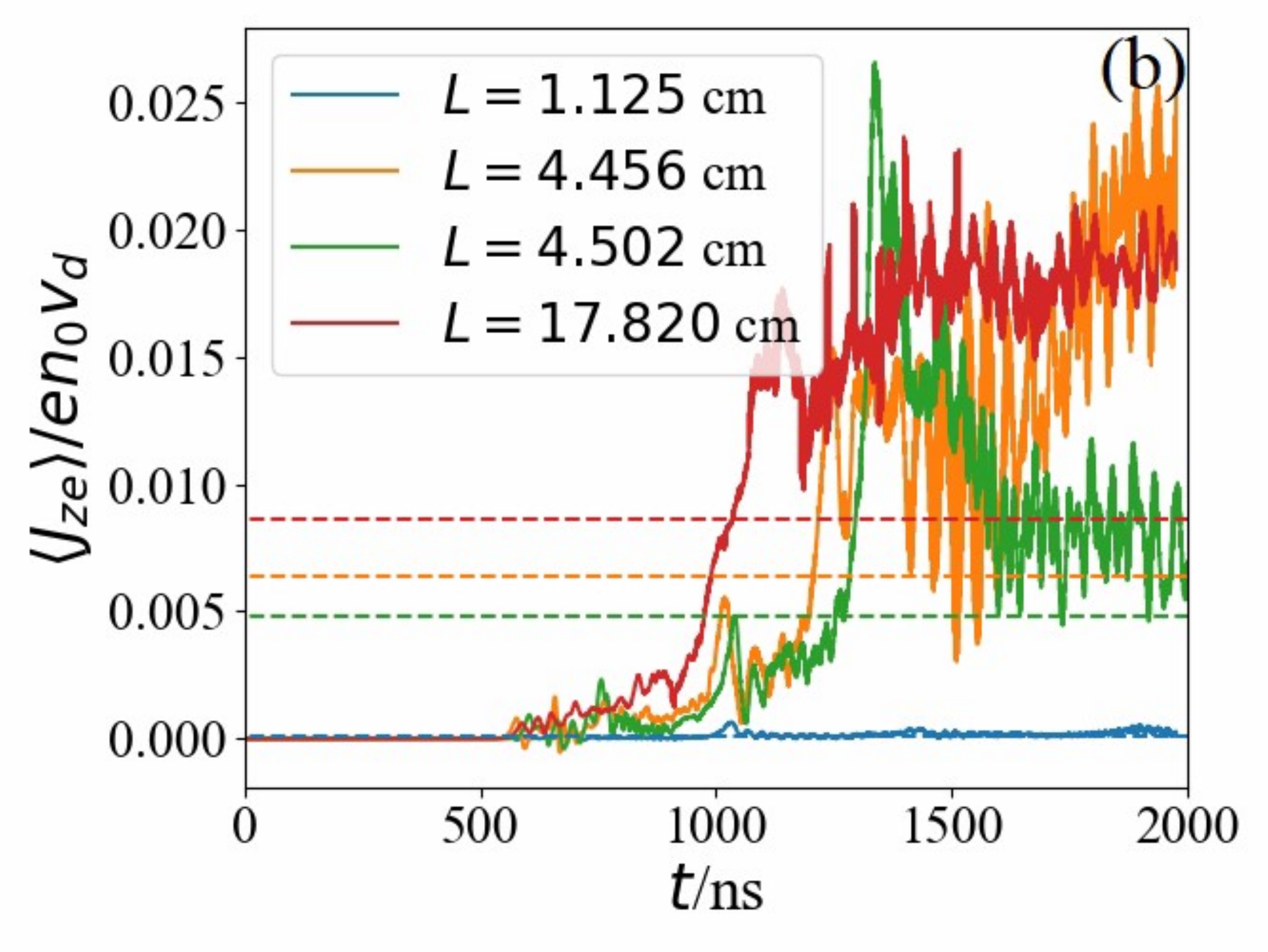}}
\caption{The effect of length on the $\expval{J_{ze}}$ in a) PIC simulation with $N_{ppc}=10^4$ and b) Vlasov simulation.  The horizontal dashed lines show the total time average of each curve.\label{fig:Vl_PIC_mvavg_Jze}} 
\end{figure}

 \Cref{fig:Vl_PIC_mvavg_Jze} shows $\expval{J_{ze}}$ for different azimuthal lengths for both PIC and Vlasov simulations. We can see that, similar to the temperature and the electrostatic energy, the total time average of $\expval{J_{ze}}$ increases with the azimuthal length in both PIC and Vlasov simulations. Also, the transient behaviour of the moving average generally shows the same trend, except for some short times in the nonlinear regime. The total time average of $\expval{J_{ze}}$ for all azimuthal lengths is larger in the PIC simulations than in the Vlasov simulation.  
 
 To study the sensitivity of the $\expval{J_{ze}}$ to a small variation of the azimuthal length, this quantity is shown for $L=156.8/k_0$ and $L=158.4/k_0$ in \Cref{fig:Vl_PIC_mvavg_Jze}. In \Cref{fig:vl_t_mvavg_Jze_Lcomp},  we see that although this variation has a small effect on the total time average of $\expval{J_{ze}}$, it changes the transient behaviour. In this figure, the maximum $\expval{J_{ze}}\approx 0.025$ occurs at $t\approx 1450$ ns for $L=158.4/k_0$, whereas it occurs at $t\approx 2000$ ns for $L=156.8/k_0$. In \Cref{fig:PIC-t_mave_Jze_Lcomp}, we see that the total time average of $\expval{J_{ze}}$ in the PIC simulations is more sensitive to this small variation. In PIC simulations, although for $L=158.4/k_0$ $\expval{J_{ze}}$ is generally much larger than when $L=156.8/k_0$, they come close to each other in the last few nanoseconds. 

\section{Discussion and conclusion}\label{sec:conclusion}

In summary, physical characteristics of the ECDI are compared using the PIC and the Vlasov simulation methods. The compared characteristics include the observed growth rates in the linear regime,  nonlinear saturation level, electron bunching, electron heating, the anomalous current, and the effect of the azimuthal length on the saturation of the instability and the inverse cascade. 

In the linear regime, resolving large variations of the growth rates with the wave vector requires a particularly high Fourier resolution, i.e.,~a long azimuthal length. Although the linear regime of PIC simulations with $156.8/k_0$ shows linear growth rates close to the theory, this was not true for the case of $L=627/k_0$. For the latter case, the Vlasov simulation showed a better consistency with theory, especially for the low-growth rate modes that did not exist in the simulations with $L=156.8/k_0$. This means that the low resolution in Fourier space might overshadow the adverse effect of the noise of the PIC simulation on the linear growth rates. 

 {In this study, we referred to the appearance of modes with $k\lesssim k_0$ in the nonlinear regime that have no significant linear growth rate as the inverse cascade. The inverse cascade was clearly seen in the nonlinear regime of PIC and Vlasov simulations when $L$ was sufficiently large. The inverse cascade happens along with the growth of the electrostatic energy in the first and other cyclotron peaks. An apparently similar process is discussed in the electromagnetic study of Ref.~\onlinecite{muschietti2013microturbulence}, where it is shown that the cyclotron peaks with $k>k_0$ are damped out while the first cyclotron peak grows. This process is not exactly the same as what we see in \Cref{fig:Ek_L17_PIC_1e4,fig:Ek_L17Vl,fig:Ek_L4_PIC_1e4,fig:Ek_L4Vl} of this study; because although we see the growth of the first cyclotron peak in that figure, other cyclotron peaks are also growing in the nonlinear regime. Therefore, the spectrum of the electric field in the late nonlinear regime remains discrete together with a notable energy in the $k\lesssim k_0$ region.} 

 {The inverse cascade has also been studied in the context of two-dimensional fluids\cite{kraichnan1967inertial,chen2006physical} and plasmas \cite{hasegawa1979nonlinear,xia2003inverse}. The inverse cascade, i.e. energy flow toward the larger scales, is a result of the additional conserved quantity (enstrophy integral) in two-dimensional systems,  such as in magnetized plasmas, where the motion in the plane perpendicular to the magnetic field is two-dimensional \cite{hasegawa1979nonlinear}. We believe that the inverse energy cascade, observed in our simulations, also occurs due to the two-dimensional nature of the electron dynamics, which remains strongly magnetized.  It is interesting that in our simulations, the inverse cascade in real space is accompanied by the formation and merging of electron structures in phase space, and therefore, these two effects might be related. Studies of  phase-space structures \cite{Dupree1982} have a long history; e.g., see reviews in Refs.\onlinecite{EliassonPR2006,hutchinson2017electron}. It is noteworthy that, because of the observed differences in the dynamics of the bunches in the PIC and Vlasov simulations,  the merging of bunches might have been impacted by the numerical noise. In the PIC simulations, the merging eventually leads to a large solitary structure \cite{NuwalPoP2022}, whereas  it  terminates in a few co-existing bunches in the Vlasov simulations. The exact nature of these bunches and their probable impact on the inverse cascade are left for future studies.}

The discrete nature of $n k_0$ modes that we observed in our simulations is at odds with the prediction of the ion-sound theory and suggests that the electrons remain magnetized in the nonlinear regime. Another characteristic of the Vlasov simulations that is not predicted by the ion-sound transition theory is the robust appearance of the backward waves over all the values of $L$ considered. The backward waves are, however, not observed in the PIC simulations of this study.
 {Since its introduction in Refs.~\onlinecite{lampe1971nonlinear,lampe1972theory}, the validity of the ion-sound turbulence theory has been a subject of debate\cite{forslund1972anomalous,janhunen2018nonlinear,lampe1972anomalous,muschietti2013microturbulence}. Ref.~\onlinecite{muschietti2013microturbulence} shows that this theory is only valid for the large wave numbers satisfying $k\rho_e\gtrsim 1$. This conclusion is, however, not supported by our results because in our case, several behaviors are at odds with predictions of ion-sound turbulence theory despite $k_0\rho_e\approx 1.3$ initially. This suggests that the ion-sound theory might not be valid even for large $k$ values. The discrepancy in the setup of problems and the parameter regime of the two studies makes a direct comparison of the behaviors difficult.}
It is known that the noise in the PIC simulations can facilitate the transition to the ion-sound regime by imitating the collisional effects of electrons \cite{janhunen2018nonlinear}. In this study, we have also shown that reducing the azimuthal length $L$ to $19.8/k_0$ can make the frequency spectrum of the PIC simulations similar to the ion-sound dispersion relation while hampering the inverse cascade. This effect, however, was not seen in the frequency spectrum of the Vlasov simulation with the same $L$, probably due to the absence of statistical noise.

In the nonlinear regime of PIC simulations, the electrostatic energy first remains close to the Vlasov results, but eventually it grows much faster than in the Vlasov simulations. The same trend was also observed for the electron temperature for all the values of $L$ considered. Therefore, our study suggests that the statistical noise of the PIC simulations might contribute to the extreme electron heating that is observed in some simulations\cite{lafleur2016theory1,janhunen2018nonlinear}. Also, when the fluctuations of the anomalous current are smoothed by applying a moving average on it, the anomalous current of the PIC simulations remains higher than Vlasov except for some transient times.

 {An important conclusion of our study of the effect of azimuthal length is that decreasing this length can seriously impact the physics of the simulations. In many previous two-dimensional studies of the typical Hall thruster parameter regime, an azimuthal length of about 0.5 cm to 1.5 cm is used \cite{villafana20212d,charoy20192d,croes2018effect,lafleur2018anomalous,croes20172d,janhunen2018evolution}, which our study finds inadequate for capturing the full physics of the ECDI.} The effect of the azimuthal length $L$ on the electrostatic energy, heating, and anomalous current was considered in two ways. First, we show that increasing the $L$ increases these quantities in both PIC and Vlasov simulations. This is in fact expected from the linear spectrum because a significant increase in $L$ leads to many more unstable resonant modes being resolved in the simulation. Second, we have shown that the electrostatic energy, the electron temperature, and the anomalous current can be sensitive to a small variation in $L$. This sensitivity is  generally much higher  in the PIC simulations than in the Vlasov simulations. At present, we do not have a conclusive answer as to why this sensitivity exists. From the steep spectrum of linear growth rates, it is evident that small variations of $L$ can drastically change the growth rate of the unstable modes. This means that this sensitivity can be partially due to the underlying physics of the ECDI problem. On the other hand, because this sensitivity is different in the PIC and Vlasov simulations, one can conclude that in addition to the physics of ECDI, numerical effects are partially responsible for this sensitivity. 

It is possible to reduce the initial noise in the PIC simulations by using the quiet-start initialization \cite{dawson1983particle}. However, this method only removes the noise in the initial time-step, and after this time, the PIC simulations are affected by the noise. In fact, if the initial amplitudes are very small, the effect of the noise on them can be even amplified in the quiet start simulations\cite{tavassoli2021role}.  Filtering can be used to  reduce the noise, albeit with increased computational cost. There are many low-pass filters proposed in the literature, each with pros and cons\cite{vay2011numerical}.  Another method for reducing the noise of the PIC simulations is the delta-f method\cite{aydemir1994unified,brunner1999collisional,allfrey2003revised}. In this method, the distribution function is assumed to be a known function (usually Maxwellian) that is added to a small $\delta f$. The macroparticles are then only used to calculate the $\delta f$ part, and the noise is reduced. We note that the filtering and the delta-f methods can also be used in the Vlasov simulations to reduce the discretization error in the velocity subspace \cite{mehrenberger2013vlasov,klimas2018absence}. Another useful method for reducing the noise of PIC simulations is remapping. In this method, the macroparticles are frequently mapped onto a phase space grid where the distribution function is calculated \cite{wang2011particle,myers20174th}. Investigating the effect of quiet-start, filtering, delta-f, and remapping on the ECDI simulations is beyond the scope of this study.

Many similarities and discrepancies in the PIC and Vlasov simulations discussed in this study need to be addressed in the broader context of 2D and 3D effects \cite{charoy20192d,villafana20212d} but are left for future studies.  {Future work can consider the electromagnetic effects using the full Maxwell equation in the simulations \cite{muschietti2013microturbulence}. The electromagnetic effects might be particularly important in the study of the long-wavelength regions of the spectrum \cite{callen1973electromagnetic}.}

\section*{Acknowledgment}
The authors acknowledge illuminating discussions with S. Janhunen. This work is partially supported in part by funding from the US Air Force Office of Scientific Research FA9550-15-1-0226 and the Natural Sciences and  Engineering Council of Canada (NSERC) as well as computational resources from the Digital Research Alliance of Canada (the Alliance).  

 \section*{Author Declarations}
 \subsection*{Conflict of interest}
 The authors have no conflicts to disclose.
 
 \section*{Data Availability Statement}
 The data that support the findings of this study are available from the corresponding author
upon reasonable request.

\bibliographystyle{apsrev4-2}
\bibliography{Refs.bib}

\end{document}